\documentclass[manuscript]{acmart}
\AtBeginDocument{%
  }
\usepackage{listings}
\usepackage{fancyvrb}
\usepackage[most]{tcolorbox}
\tcbuselibrary{skins,breakable}
\usepackage{booktabs}
\usepackage{multirow}
\usepackage[table,xcdraw]{xcolor}
\usepackage{threeparttable}
\tcbuselibrary{listings,breakable}
\usepackage{algorithm}
\usepackage{algpseudocode}
\usepackage{xcolor}
\usepackage{caption}
\usepackage{subcaption}
\usepackage{soul}
\usepackage[nameinlink,noabbrev]{cleveref}
\lstdefinestyle{jsonverbatim}{
  basicstyle=\ttfamily\footnotesize,
  columns=fullflexible,
  keepspaces=true,
  showstringspaces=false,
  breaklines=true,
  breakatwhitespace=false,
  frame=none,
  numbers=left,
  numberstyle=\tiny,
  numbersep=6pt
}
\lstdefinestyle{jsonpretty}{
  basicstyle=\ttfamily\footnotesize,
  breaklines=true,
  breakatwhitespace=true,
  columns=fullflexible,
  keepspaces=true,
  showstringspaces=false,
  tabsize=2,
  frame=none,
  postbreak={},                 
  literate={\\n}{{\newline}}2 
}
\newtcblisting{jsonbox}{
  breakable,
  colback=gray!5,       
  colframe=black!60,
  boxrule=0.4pt,
  arc=1mm,
  left=1.5mm,right=1.5mm,top=1mm,bottom=1mm,
  listing only,
  listing options={
       style=jsonpretty, 
       numbers=none,
       frame=none
    }
}
\lstdefinestyle{mdcli}{
  basicstyle=\small, 
  breaklines=true,
  breakatwhitespace=true,
  columns=fullflexible,
  keepspaces=true,
  showstringspaces=false,
  tabsize=2,
  frame=none,
  numbers=left,
  numberstyle=\tiny,
  numbersep=5pt,
  breakautoindent=false,
  breakindent=0pt,
  xleftmargin=1.5em,
  framexleftmargin=1.5em,
  xrightmargin=0.5em,
  postbreak={},
  morecomment=[l]{####},
  commentstyle=\bfseries
}

\newtcblisting{mdbox}{
  breakable,
  colback=gray!5,       
  colframe=black!60,
  boxrule=0.5pt,
  arc=1.5mm,
  width=0.92\textwidth,
  left=0.8em,
  right=0.8em,
  top=1.5mm,
  bottom=1.5mm,
  listing only,
  listing options={style=mdcli}
}

\setcopyright{acmlicensed}
\copyrightyear{2018}
\acmYear{2018}
\acmDOI{XXXXXXX.XXXXXXX}
\acmConference[Conference acronym 'XX]{Make sure to enter the correct
  conference title from your rights confirmation email}{June 03--05,
  2018}{Woodstock, NY}
\acmISBN{978-1-4503-XXXX-X/2018/06}

\begin{document}

\title{Supporting System Testing with a Multi-Agent LLM-based Framework for Knowledge Graph Extraction: A Case Study with Ethernet Switch Systems}
\author{Rongqi Pan}
\affiliation{
    \institution{University of Ottawa}
    \city{Ottawa}
    \country{Canada}}
\email{rpan@uottawa.ca}
\orcid{0000-0002-9096-6241}

\author{Mahboubeh Dadkhah}
\affiliation{
    \institution{University of Ottawa}
    \city{Ottawa}
    \country{Canada}}
\email{mdadkhah@uottawa.ca}
\orcid{0000-0002-0436-8369}

\author{Jean Baptiste Minani}
\affiliation{
    \institution{University of Ottawa}
    \city{Ottawa}
    \country{Canada}}
\email{baptiste2k8@gmail.com}
\orcid{0000-0002-9164-6645}

\author{Hussein Al Osman}
\affiliation{
    \institution{University of Ottawa}
    \city{Ottawa}
    \country{Canada}}
\email{hussein.alosman@uottawa.ca}
\orcid{0000-0002-7189-5644}

\author{Lionel Briand}
\affiliation{
    \institution{University of Ottawa}
    \city{Ottawa}
    \country{Canada}}
\affiliation{
  \institution{Research Ireland Lero Centre for Software and University of Limerick}
    \city{Limerick}
    \country{Ireland}
    }
\orcid{0000-0002-1393-1010}
\email{Lbriand@uottawa.ca}

\author{Haiwei Dong}
\affiliation{
    \institution{Huawei Canada}
    \city{Ottawa}
    \country{Canada}}
\email{haiwei.dong@ieee.org}
\orcid{0000-0003-1437-7805}

\renewcommand{\shortauthors}{Pan et al.}

\begin{abstract}
Technical documents contain rich domain knowledge for automating downstream tasks such as system testing. 
While this paper focuses on Ethernet switch configuration manuals (ESCMs), we propose a general framework that can be adapted to different industrial contexts.
ESCMs provide valuable domain knowledge for Ethernet switch testing, but their semi-structured format, implicit step attributes, and complex section dependencies make them difficult to directly leverage for test automation.
To address this, we generate knowledge graphs (KGs) that capture configuration knowledge from ESCM in a structured form.
We propose a multi-agent LLM-based framework that extracts, evaluates, and improves KGs from ESCMs using a fine-grained KG schema and an iterative Extract-Evaluate-Improve (EEI) loop.
Our evaluation on 50 real-world ESCMs shows that our framework achieves high extraction correctness using the original prompts, with average correctness scores ranging from 0.97 to 0.99 across three extraction tasks. For challenging ESCMs, the EEI loop further improves correctness through manual-specific prompt refinement.
Moreover, the LLM judgments and human evaluations show substantial agreement, with Cohen's kappa of at least 0.72 across all extraction tasks. 
Finally, feedback from industry testers indicates that the generated KGs can support the generation of useful and correct test case specifications (TCSs) for downstream testing.
\end{abstract}



\maketitle

\section{Introduction}

Knowledge Graph (KG) generation from technical documents has gained increasing attention in recent years. These documents, mainly in text format, contain rich, often tacit knowledge that can be leveraged to automate downstream software engineering tasks. In this work, we focus on system testing and KG generation from Ethernet Switch Configuration Manuals (ESCMs), a class of semi-structured documents that describe configuration steps, supported commands, and expected device behavior under various settings. Similar to many available technical documents, ESCMs are designed to guide end users rather than support automation. However, KGs generated from these manuals can support the automation of downstream tasks, particularly Ethernet switch testing.
Testing Ethernet switches is largely a manual process, requiring testers to write test scripts and execute test commands. Consequently, deriving comprehensive and diverse test cases is a challenging, labor-intensive, and highly expensive task. While ESCMs represent a valuable, rich, yet underexplored source for system testing and configuration validation, they cannot be directly leveraged for automated test generation due to their inherent complexities, including implicit attributes of configuration steps and dependencies among configuration steps that are difficult to extract and verify. In contrast, KGs are widely used to represent structured knowledge and have proven effective in supporting automated testing~\cite{su2024enhancing,su2025automated}. Early KG generation approaches, primarily based on rule-based methods or traditional NLP techniques, often struggle with variations in document structure and the presence of long, complex sentences. Recent advances in Large Language Models (LLMs), however, offer new opportunities to address these challenges. 
When integrated into a well-designed framework, LLMs can effectively interpret complex technical content and help build accurate KGs. In this paper, we propose a multi-agent LLM-based framework for knowledge extraction from technical documents with high accuracy, enabling the automation of downstream testing activities, such as test case specification (TCS) generation. We evaluate our approach using Ethernet switches as a case study. Beyond this important industrial application, our goal is to provide a more general framework that can be adapted to other tasks and contexts, as we discuss in Section~\ref{sec:adapt_framework}.

KGs capture structured knowledge from large-scale, diverse data sources using a graph-based data model, where nodes denote entities and edges represent the relations between them~\cite{hogan2021knowledge,peng2023knowledge,zhang2024extract}. KGs have been widely adopted to represent extracted knowledge across various application scenarios and have demonstrated effectiveness in automating downstream tasks, such as test generation, issue resolution, and decision-making~\cite {chen2025prometheus,su2024enhancing,su2025automated,guo2021automatic}. 
However, KG construction remains challenging as it requires accurately capturing both syntactic and semantic information from data sources to generate a high-accuracy KG~\cite{zhang2024extract,ye2022generative}. This challenge is particularly pronounced in technical documents, such as configuration manuals, which are inherently more complex than general text and therefore demand more precise and robust methods for knowledge extraction and KG generation. Achieving a high level of correctness is essential in this context to effectively handle such complexities and enable the generated KGs to support downstream tasks, such as test generation.

KG generation from technical documents has traditionally been studied using rule-based, ontology-based, and traditional NLP approaches~\cite{li2014configuration, rizvi2018ontology, su2022constructing}. Advances in LLMs have led to their adoption to automate knowledge extraction and KG generation, mostly from general texts~\cite{zhang2024extract, kommineni2024human}. 
Given the diversity of general texts, a wide range of entities and relationships are required to represent the extracted knowledge in KG format. As a result, the existing approaches primarily focus on dynamic KG schema construction. 
On the other hand, KG generation from technical documents presents distinct challenges, with the primary focus on the correctness and precision of the generated KGs, while the underlying KG schema is often predefined. This is particularly important to ensure reliable and consistent KG construction.
While LLM-based methods have shown promising performance on general text~\cite{zhang2024extract, kommineni2024human}, their effectiveness on technical documents, particularly in real-world settings such as ESCM, remains limited. ESCMs provide highly detailed configuration instructions for human users in a semi-structured format. Consequently, they are inherently more complex than general text inputs and thus necessitate more precise and robust methods for effective knowledge extraction and KG generation.

To address these challenges and motivated by LLMs' ability to interpret complex technical documents and capture implicit dependencies~\cite{pan2023large}, we propose a multi-agent LLM-based framework for KG extraction, evaluation, and improvement to support system testing. Although we focused on KG generation from ESCMs in this paper, the proposed framework is general and broadly applicable to other types of technical documents in different industrial contexts, as we discuss in Section~\ref{sec:adapt_framework}.
This framework is built on a well-defined, fine-grained KG schema designed specifically for the ESCM context, enabling precise and consistent knowledge representation.
Based on this KG schema, we design three specialized Extraction Agents (ExtrAgents) in the framework, each corresponding to a specific extraction task that identifies and extracts different types of entities and dependencies. 
ExtrAgents are guided by carefully designed prompts, with a set of representative task-specific examples included only where they effectively enhance performance.
In addition to the well-designed extraction prompts, we introduce an iterative mechanism, called the Knowledge Graph Extract-Evaluate-Improve (EEI) Loop, to iteratively improve the generated KGs. This mechanism is designed to mitigate challenges arising from structural and semantic variations across ESCMs and ensure a high level of correctness in the generated KGs. 
The EEI loop involves two additional LLM agents: an Extraction Evaluation Agent (EvalAgent) that evaluates the KG generated by ExtrAgent against the corresponding source ESCM using a set of task-specific evaluation guidelines and produces detailed feedback for that KG, and an Extraction Improvement Agent (ImprovAgent), which refines the extraction prompt based on the provided feedback. The ExtrAgents are then re-invoked with the refined prompt to generate an improved KG. This process continues until the overall correctness score of the generated KG exceeds a predefined threshold or the maximum number of iterations is reached. The EvalAgent is designed using the LLM-as-a-Judge paradigm and is guided by carefully crafted evaluation guidelines tailored to each extraction task. This design enables us to automatically evaluate KGs without requiring a ground-truth reference. Traditional NLP metrics (e.g., BLEU, ROUGE, METEOR) are often unreliable for reasoning-intensive tasks and depend on ground truth, which is costly to collect~\cite{liu2023g}. In contrast, LLM-based evaluation has been shown to align well with human judgment when clear criteria are provided~\cite{zheng2023judging,liu2023g}, offering a more practical, scalable, and cost-effective solution.

To comprehensively evaluate our approach, we conducted an empirical study on 50 ESCMs provided by Huawei, enabling us to assess it on realistic industrial documentation. Our results show that using the original prompts, i.e., the initial extraction prompts before any refinement by the EEI loop, our approach achieves high extraction correctness scores, with average correctness scores ranging from 0.97 to 0.99 across all three extraction tasks and all ESCMs. However, for a small subset of ESCMs, particularly when extracting dependencies between configuration steps, the KGs generated using original prompts do not meet the predefined correctness threshold and therefore require further refinement. In such cases, the EEI loop is triggered. Our findings demonstrate that the EEI loop effectively refines extraction prompts for these manuals, leading to consistently high-quality KG generation.
Furthermore, we evaluated the consistency between LLM and human judgment on extraction correctness. Two authors independently evaluated the extracted KGs across all three extraction tasks and the full set of 50 ESCMs. The results indicate substantial agreement between LLM and human judgments across three extraction tasks, with Cohen’s kappa of at least 0.72. Notably, most discrepancies are due to minor categorization differences or formatting issues, rather than errors that impact the correctness or usefulness of the generated KGs.

We also investigated the usefulness of generated KGs for automating downstream tasks, particularly test generation. For this purpose, we selected five representative KGs, spanning a range of complexity, size, and structure, and generated their corresponding TCSs. We then asked five Huawei testers with a wide range of experience to evaluate the usefulness, correctness, and completeness of these TCSs for test case generation using a structured questionnaire. The results, based on Likert-scale responses, indicate consistently high ratings, suggesting that the KGs generated and verified by our approach effectively support and guide test case generation. All the code, the 50 ESCMs used in our study, the original and refined prompts, the generated KGs, and the complete experimental results, including the questionnaire and detailed responses, will be made publicly available upon acceptance.
Furthermore, our proposed framework is modular and can be adapted to configuration manuals in other domains and to broader classes of technical documents. Adapting the framework primarily involves reusing or defining a KG schema and tailoring the two agent types, i.e., ExtrAgents and EvalAgent, to the target domain and task, requiring only revisions to their prompts, evaluation guidelines, and task-specific example sets for few-shot prompting, rather than redesigning the overall architecture. This is discussed in detail in Section ~\ref{sec:adapt_framework}. 
 
To summarize, the key contributions of this paper are as follows:
\begin{itemize}
    \item We introduce a multi-agent LLM-based framework for KG extraction, evaluation, and improvement. Our framework comprises specialized agents for knowledge extraction, KG evaluation, and prompt refinement, each guided by a well-designed prompt. It further incorporates an iterative Extract-Evaluate-Improve (EEI) mechanism to address structural and semantic expression variations across documents by refining extraction prompts. This mechanism leverages the LLM-as-a-Judge paradigm, supported by carefully designed evaluation guidelines tailored to each extraction task. While the experiments in this study are conducted on subjects from the ESCM domain, the proposed framework is domain-adaptable and broadly applicable to other technical documents with revisions primarily to the KG schema, extraction prompts, and evaluation guidelines. Its modular, agent-based design supports adaptation to configuration manuals in other domains and extension to diverse technical document types with minimal modifications.
    \item We design a fine-grained KG schema tailored to ESCMs, encompassing key entities and relationships, enabling the generation of KGs with high granularity and precision.
    \item We empirically evaluate our approach on 50 real-world ESCMs obtained from our industry partner. The results show that our approach without the EEI mechanism already achieves high average correctness scores of 0.97-0.99 across all extraction tasks and ESCMs. For the KGs that do not meet the required correctness score threshold, the EEI loop effectively refines the extraction prompt and improves the quality of generated KGs.
    
    \item We further investigate the usefulness of generated KGs in automating system test generation through a structured questionnaire and responses from five experienced testers. The results, based on Likert-scale responses, show consistently high ratings, indicating that the KGs generated by our approach effectively support and facilitate test case generation.
\end{itemize}

The remainder of this paper is organized as follows: Section 2 motivates this study and provides a real ESCM example. Section 3 introduces our proposed framework. Section 4 describes the experiments we performed to evaluate our framework. Section 5 discusses the results for each research question. Section 6 discusses the potential threats to the validity of our study and the actions taken to mitigate them. Section 7 discusses related work, and Section 8 concludes the paper.

\section{Motivation}
In this section, we first outline the real-world industry need for accurate knowledge extraction from ESCMs and for representing the extracted knowledge in a well-structured format to reduce manual testing effort, which motivates our study. We then discuss the main challenges and explain how our approach addresses them. In addition, we provide a real ESCM example to illustrate the typical structure of these manuals and the challenges associated with extracting knowledge from them. This example is concise enough to explain in detail while still covering most of the sections and information types observed in our ESCM dataset. Thus, we use it throughout the paper as a running example to demonstrate our approach.

\subsection{Ethernet Switches and Manual Testing}
\noindent
Ethernet switch systems are commonly deployed in enterprise and campus networks to provide essential networking functionalities, including network monitoring, fault management, and traffic statistics collection. 
Ensuring the correctness of these functionalities requires extensive testing, which in practice is often conducted manually by following the product documentation. Specifically, given a testing requirement, test engineers first identify the Ethernet switch function to be tested and select relevant testing scenarios based on their domain expertise. They then manually write test cases for these functions and scenarios by referring to the product documentation, where each test case consists of natural-language test steps. Finally, these test steps are converted into device commands and automatically executed on the target devices. However, given that the Ethernet Switch testing is often performed manually in the industry, it remains one of the most time-consuming and labor-intensive tasks in Ethernet Switch production and the verification and validation (V\&V) process. This bottleneck motivates our study to develop an effective approach for KG generation from ESCMs, which serve as a valuable and available source of testing knowledge for supporting automated testing.

ESCMs provide highly detailed configuration instructions to guide users in deploying, configuring, and validating Ethernet switch functionalities across different networking scenarios. These manuals typically contain multiple sections, including \emph{Configuration Roadmap}, \emph{Procedure}, \emph{Networking Requirements}, and \emph{Configuration Files}. Consequently, they are valuable resources for automating Ethernet Switch testing. The ESCMs are inherently semi-structured and contain extensive information, and important information is not explicitly documented, including the implicit attributes of configuration steps and many-to-many mappings between roadmap steps and procedure steps. As a result, their potential to automate downstream engineering tasks has remained underexplored, primarily due to challenges in extracting knowledge from these semi-structured documents. In this study, we address this challenge by leveraging LLMs with dynamic prompt refinement to effectively extract knowledge from ESCMs, particularly to extract the key attributes of the configuration steps and the implicit dependencies between them, and present them as KGs in a structured and machine-readable format.

\subsection{Challenges in extracting knowledge from configuration steps}
ESCMs often include detailed instructions, providing step-by-step guides for configuring switches. While these steps are usually presented in dedicated sections of the ESCMs, their implicit attributes, such as the goal and notes associated with each step, as well as the dependencies between steps, make knowledge extraction challenging. Specifically, in our study, configuration steps are often provided in two sections, i.e., the \emph{Configuration Roadmap} and \emph{Procedure} sections, at different levels of abstraction. The \emph{Configuration Roadmap} section provides abstract, high-level instructions, while the \emph{Procedure} section provides very detailed actions and commands for configuring switches, without explicitly specifying how these steps correspond to those in the \emph{Configuration Roadmap} section.

Listing~\ref{lst:configuration-md} presents an example ESCM taken from the \emph{S300, S500, S2700, S5700, and S6700 Series Ethernet Switches Product Documentation}~\footnote{\url{https://support.huawei.com/hedex/hdx.do?docid=EDOC1100333029&id=index}}, provided by our industry partner, which is used as input for generating the KG. The example ESCM provides detailed configuration information for a specific scenario. Specifically, it contains the following sections \emph{Overview},\emph{Configuration Notes},\emph{Networking Requirements}, \emph{Configuration Roadmap}, \emph{Procedure}, and \emph{Configuration Files}:
\begin{itemize}
  \item \textbf{\emph{Overview}} introduces the background context and explains the purpose and underlying principles of this ESCM.
  \item \textbf{\emph{Configuration Notes}} specifies platform applicability, including supported and unsupported device models, versions, or constraints.
  \item \textbf{\emph{Networking Requirements}} describes the target network environment, including topology, configuration objectives, and expected operational outcomes.
  \item \textbf{\emph{Configuration Roadmap}} outlines the high-level sequence of configuration steps needed to achieve the intended functionality.
  \item \textbf{\emph{Procedure}} provides detailed, step-by-step configuration actions and commands to be executed on the involved devices.
  \item \textbf{\emph{Configuration Files}} presents the complete device configurations to support deployment and reproducibility.
\end{itemize}

In this example, the first step in the \emph{Configuration Roadmap} section includes two sub-actions: (1) enabling LBDT on the interfaces and (2) configuring the Switch to detect loops in VLAN 100. These sub-actions are implemented through three detailed procedure steps: Procedure Step 1 enables LBDT on the interfaces, Procedure Step 2 configures the VLAN ID of LBDT packets, and Procedure Step 4 verifies that the LBDT configuration is successful. Similarly, the second step in the \emph{Configuration Roadmap} section contains two sub-actions: (1) configuring the action to be taken after a loop is detected and (2) setting the recovery time, which are implemented in the \emph{Procedure} through two detailed steps: Procedure Step 3 configures the loop handling action and recovery time. Procedure Step 4 confirms that the configured blocking action is triggered when a loop is detected.

This example highlights two key challenges in constructing a KG from an ESCM. First, each configuration step includes rich step-level attributes that must be explicitly extracted and represented. For roadmap steps, such attributes include their configuration goals and accompanying notes. For procedure steps, important attributes include commands, expected outputs, and notes that provide additional clarification or explanation. Detailed definitions of these attributes are provided in Section~\ref{sec:entity_extraction}. These attributes are valuable for downstream tasks, such as test generation. Representing this knowledge in a structured KG can further facilitate knowledge retrieval and reuse, and may reduce the risk of hallucination when LLMs are later used for downstream tasks. However, these attributes are often expressed in various formats and writing styles across different manuals. For example, commands and expected outputs may appear as separate code blocks, be embedded within step descriptions, or be mixed with explanatory text. Similarly, goals and notes for roadmap steps may be stated differently. Such variation makes it difficult for rule-based methods or traditional NLP techniques to accurately and completely extract these attributes.

Second, there are implicit dependencies between roadmap steps and procedure steps. A roadmap step may correspond to multiple procedure steps, while a procedure step may support multiple roadmap goals. Extracting these dependencies is particularly challenging because it requires the semantic understanding of the configuration intent, the actions performed by each procedure step, and the verification logic used to confirm the configuration outcome. Moreover, these mappings are not explicitly documented in the manual and often involve many-to-many relationships across sections. As a result, simple keyword matching, rule-based heuristics, or traditional NLP methods are insufficient to reliably identify the complete set of dependencies between roadmap and procedure steps. LLMs combined with our dynamic prompt refinement mechanism enable us to precisely extract the attributes of the configuration steps and their dependencies, as described in Section~\ref{sec:approach}.

\vspace{2pt}
\captionsetup{type=lstlisting}
\caption{An Example of ESCM}
\label{lst:configuration-md}
\vspace{-5pt}
\begin{mdbox}
# Example for Configuring LBDT to Detect Loops on the Local Network

#### Overview
<text of the Overview>

#### Configuration Notes
<text of the Configuration Notes>

#### Networking Requirements
<text of the Networking Requirements>

#### Configuration Roadmap
To detect loops on the network where the Switch is deployed, configure LBDT on GE1/0/1 and GE1/0/2 of the Switch. In this example, untagged LBDT packets sent by the Switch will be discarded by other switches on the network. As a result, the packets cannot be sent back to the Switch, and LBDT fails. Therefore, LBDT is configured in a specified VLAN. The configuration roadmap is as follows:
  1. Enable LBDT on interfaces and configure the Switch to detect loops in VLAN 100 to implement LBDT on the network where the Switch is located.
  2. Configure an action to be taken after a loop is detected and set the recovery time. After a loop is detected, the Switch blocks the interface to reduce the impact of the loop on the network. 

#### Procedure
  1. Enable LBDT on interfaces.
         <HUAWEI> **system-view**
         [HUAWEI] **sysname Switch**
         [Switch] **interface gigabitethernet 1/0/1**
         [Switch-GigabitEthernet1/0/1] **loopback-detect enable**  //Enable LBDT on ...
         ...

  2. Specify the VLAN ID of LBDT packets.
         [Switch] **vlan 100**
         [Switch-vlan100] **quit**
         [Switch] **interface gigabitethernet 1/0/1**
         [Switch-GigabitEthernet1/0/1]**port link-type hybrid**  //In V200R005C00 and ...
         ...

  3. Configure an action to be taken after a loop is detected and set the recovery time.
         [Switch] **interface gigabitethernet 1/0/1**
         [Switch-GigabitEthernet1/0/1] **loopback-detect action block**  //Configure ...
         [Switch-GigabitEthernet1/0/1] **loopback-detect recovery-time 30**  //Set the ...
         ...

  4. Verify the configuration.
     1. Run the **display loopback-detect** command to check the LBDT configuration.           
            [Switch] **display loopback-detect**
            Loopback-detect sending-packet interval:  5                                                                  
            --------------------------------------------------------------------------                                   
            Interface                     RecoverTime  Action     Status                                                 
            --------------------------------------------------------------------------                                   
            GigabitEthernet1/0/1          30           block      NORMAL                                                 
            GigabitEthernet1/0/2          30           block      NORMAL 
            --------------------------------------------------------------------------
The preceding command output shows that the LBDT configuration is successful.

     2. After about 5s, run the **display loopback-detect** command to check whether GE1/0/2 is blocked.         
            [Switch] **display loopback-detect**
            Loopback-detect sending-packet interval:  5                                                                  
            --------------------------------------------------------------------------                                   
            Interface                     RecoverTime  Action     Status                                                 
            --------------------------------------------------------------------------                                   
            GigabitEthernet1/0/1          30           block      NORMAL                                                 
            GigabitEthernet1/0/2          30           block      **BLOCK(Loopback detected)** 
            --------------------------------------------------------------------------
The preceding command output shows that GE1/0/2 is blocked.

#### Configuration Files
<text of the Configuration Files>
\end{mdbox}

\section{Approach}
\label{sec:approach}

In this section, we present our multi-agent framework for extracting, evaluating, and improving knowledge graphs (KGs) from ESCMs. We first introduce the KG schema tailored to the Ethernet switch domain, enabling a precise representation of extracted knowledge in KG format. We then describe each component of our framework, which extracts, evaluates and improves the KG according to the schema.

\subsection{Knowledge Graph Schema}

\begin{figure*}[htbp]
\centering
\includegraphics[width=1.0\textwidth]{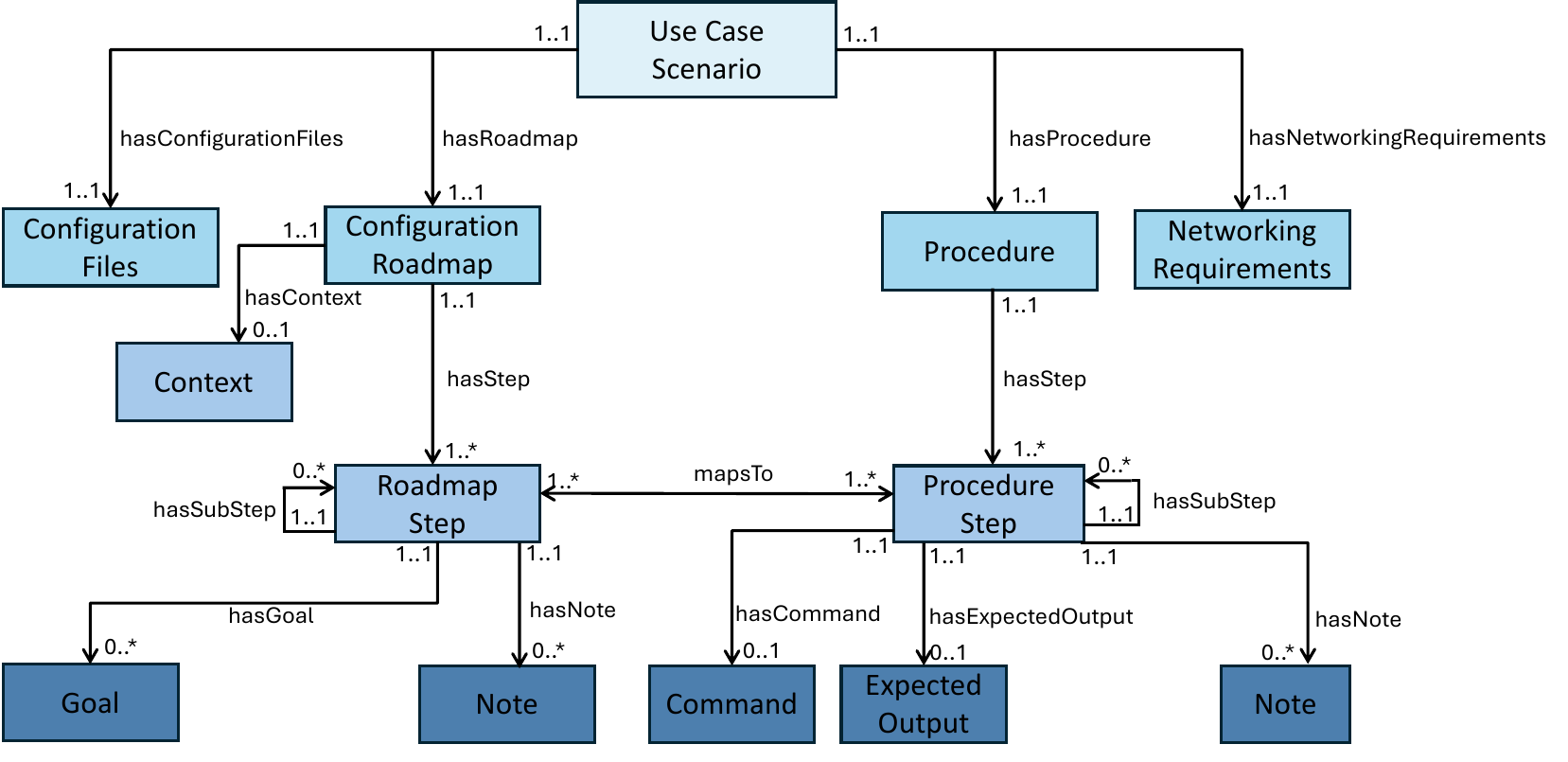}
\Description{}
\caption{Overview of the proposed knowledge graph (KG) schema. }
\label{fig:kg_schema}
\end{figure*}

The product documentations of Ethernet switches provide highly detailed, semi-structured configuration manuals, typically organized into multiple sections such as the \emph{Configuration Roadmap}, \emph{Procedure}, \emph{Networking Requirements}, and \emph{Configuration Files}. Although these sections are individually well organized, the overall information is extensive, and dependencies across sections—especially the implicit many-to-many mappings, where one roadmap step may correspond to multiple procedure steps and one procedure step may also relate to multiple roadmap steps—are not explicitly documented. 
Moreover, important attributes of configuration steps, such as goals, notes, commands, and expected outputs, are not explicitly labeled in a structured form. To unify information scattered across sections in the ESCM into a single structured representation that explicitly captures both cross-section dependencies and step-level attributes while supporting efficient querying, reasoning, and downstream automation, we transform each ESCM into a Knowledge Graph (KG).

Figure~\ref{fig:kg_schema} provides an overview of our KG schema, showing all entity and relation types it contains. For readability, some low-level entities and relations are omitted from the figure. For example, \emph{Roadmap Step} and \emph{Procedure Step} entities are linked to their corresponding textual content through the \textit{hasContent} relation, which is not shown in the figure.
Specifically, the KG contains the following entity types: \emph{Use Case Scenario}, \emph{Networking Requirements}, \emph{Configuration Files}, \emph{Configuration Roadmap} (including \emph{Context} and a hierarchy of \emph{Roadmap Steps/Substeps} annotated with \emph{Goal} and \emph{Note}), and \emph{Procedure} (including a hierarchy of \emph{Procedure Steps/Substeps} annotated with \emph{Note}, \emph{Command}, and \emph{Expected Output}). These entities are connected through relations that (i) link the major manual sections (e.g., \textit{hasConfigurationFiles}, \textit{hasNetworkingRequirements}, \textit{hasRoadmap}, \textit{hasProcedure}), (ii) encode hierarchical structures (e.g., \textit{hasStep}, \textit{hasSubstep}), (iii) attach step descriptive information (e.g., \textit{hasGoal}, \textit{hasNote}), (iv) capture execution details (e.g., \textit{hasNote}, \textit{hasCommand}, \textit{hasExpectedOutput}), and (v) explicitly align roadmap and procedure steps via step-level mappings (i.e., \textit{mapsTo}).

Overall, the KG captures (i) the hierarchical structures and step-level attributes of both the \emph{Configuration Roadmap} and the \emph{Procedure}, (ii) explicit mappings between roadmap steps and their implementing procedure steps, and (iii) associated \emph{Use Case Scenario}, \emph{Networking Requirements}, and \emph{Configuration Files}.

\subsection{Multi-Agent Framework}

\begin{figure*}[htbp]
\centering
\includegraphics[width=1.0\textwidth]{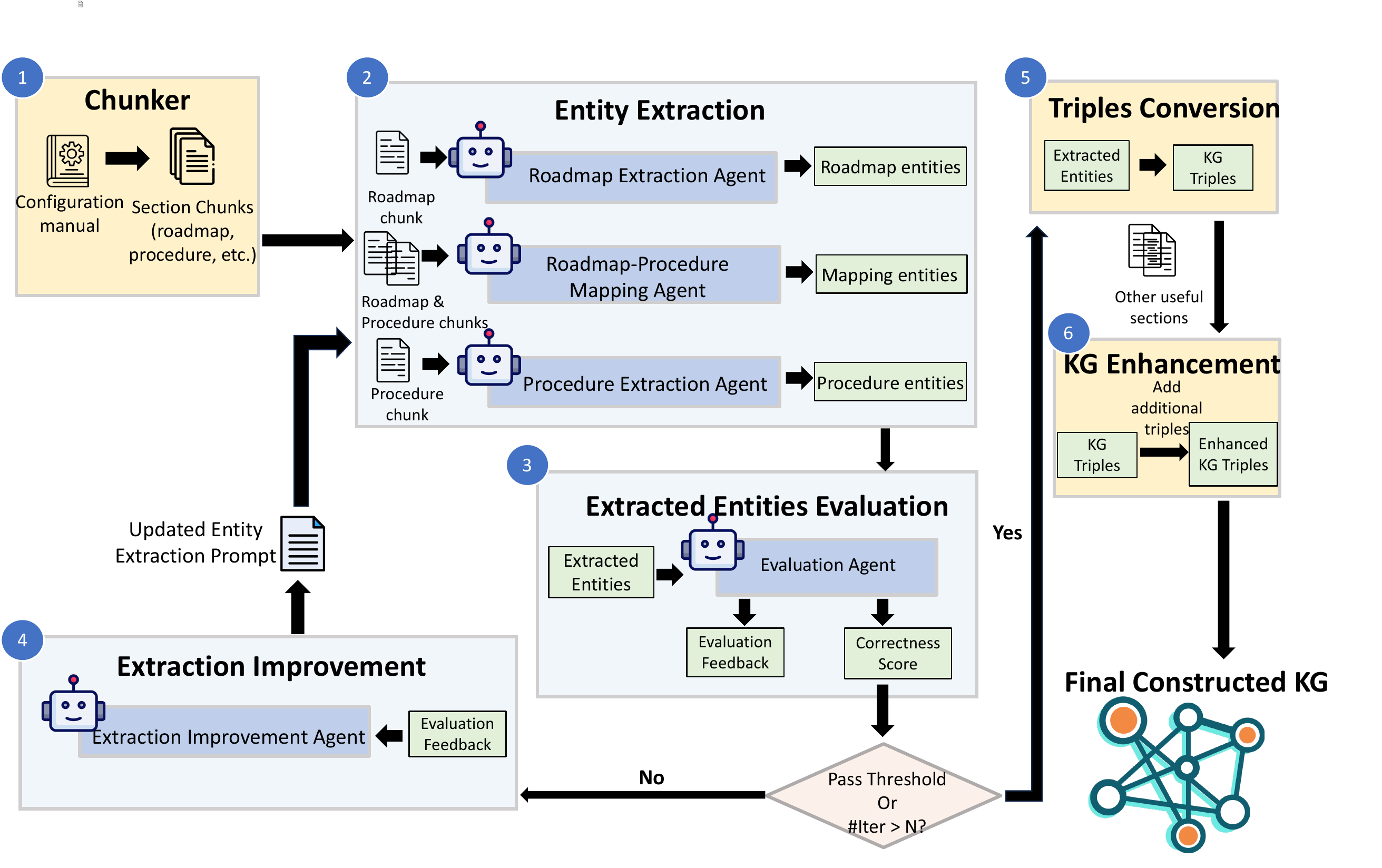}
\caption{An overview of our multi-agent framework for knowledge graph extraction, evaluation, and improvement. }
\label{fig:multi-agent-framework}
\end{figure*}

To facilitate the automation of KG extraction, evaluation, and improvement, we introduce a multi-agent framework. 
As shown in Figure~\ref{fig:multi-agent-framework}, the framework comprises three task-specific LLM agents for entity extraction (ExtrAgents) in roadmap extraction, roadmap-procedure mapping, and procedure extraction (Component 2), together with an EvalAgent for assessing the extracted entities (Component 3), and an ImprovAgent for refining the extraction prompt based on evaluation feedback (Component 4). Given an ESCM, we first split it into section-based chunks (e.g., roadmap and procedure chunks). The three ExtrAgents then (i) extract structured roadmap entities from the roadmap chunk, (ii) map each roadmap main step in the roadmap chunk to one or more corresponding procedure main steps in the procedure chunk to make cross-section dependencies explicit, and (iii) extract structured procedure entities from the procedure chunk.
To improve the quality of extracted entities, the EvalAgent assesses the output of each ExtrAgent against predefined evaluation guidelines and provides structured feedback. Based on this feedback, the ImprovAgent iteratively refines the extraction prompts until the overall quality of the extracted entities, as measured by correctness score, exceeds a predefined threshold or a maximum number of iterations is reached.
Finally, we apply entity-to-triple conversion (Component 5) and KG enhancement (Component 6) by adding other useful sections (e.g., \emph{Networking Requirements} and \emph{Configuration Notes}) as additional entities to obtain the final constructed KG. The details of each component are presented in the following sections.

\subsubsection{Chunker}
ESCMs are often lengthy and may exceed LLM input token limits. Moreover, they are organized into heterogeneous sections (e.g., \emph{Overview}, \emph{Networking Requirements}, \emph{Configuration Roadmap}, \emph{Procedure}, and \emph{Configuration Files}), each containing different types of information and following distinct structures. To enable targeted processing by task-specialized agents, we first decompose each ESCM into section-based chunks. 

As shown in Listing~\ref{lst:configuration-md}, each ESCM in our dataset is provided in Markdown format, where major sections (e.g.,\emph{Configuration Roadmap}, \emph{Procedure}, \emph{Networking Requirements}, and \emph{Configuration Notes}) are separated by explicit structural markers (i.e., \texttt{\#\#\#\#}). We use a regular-expression-based parser to segment each ESCM into section-level chunks and extract the raw text for each section. 
We also remove text that is not part of the technical documentation, such as navigation links and footer metadata (e.g., ``Parent Topic'', copyright notices, and ``Previous topic'' links), to avoid introducing noise into subsequent extraction tasks.
Finally, each cleaned section chunk is provided as input to the corresponding LLM agent for the entity extraction task.

\subsubsection{Entity Extraction.}
\label{sec:entity_extraction}

As shown in Figure~\ref{fig:multi-agent-framework}, the entity extraction component is implemented by three task-specific ExtrAgents, each responsible for extracting a different type of information from the ESCM. The first agent is the roadmap extraction agent, which extracts entities from the \emph{Configuration Roadmap} section, including the configuration context, roadmap steps and substeps, and their associated goals and notes. The second agent is the roadmap--procedure mapping agent, which identifies the semantic mappings between roadmap main steps and procedure main steps. The third agent is the procedure extraction agent, which extracts entities from the \emph{Procedure} section, including procedure steps and substeps, and their associated commands, expected outputs, and notes.

We design these three ExtrAgents separately because the corresponding extraction tasks involve different input sections, information structures, and reasoning requirements. For example, roadmap extraction focuses on high-level configuration intent, roadmap--procedure mapping requires cross-section dependency reasoning, and procedure extraction focuses on concrete configuration operations and outputs. The details of each agent are provided below.

\noindent\textbf{Roadmap Extraction Agent. } 
ESCMs typically include a \emph{Configuration Roadmap} section that outlines the high-level workflow for completing a configuration task for a specific use case. As shown in Figure~\ref{fig:roadmap}, a \emph{Configuration Roadmap} section typically contains the following information:

\begin{figure*}[htbp]
\centering
\includegraphics[width=0.9\textwidth]{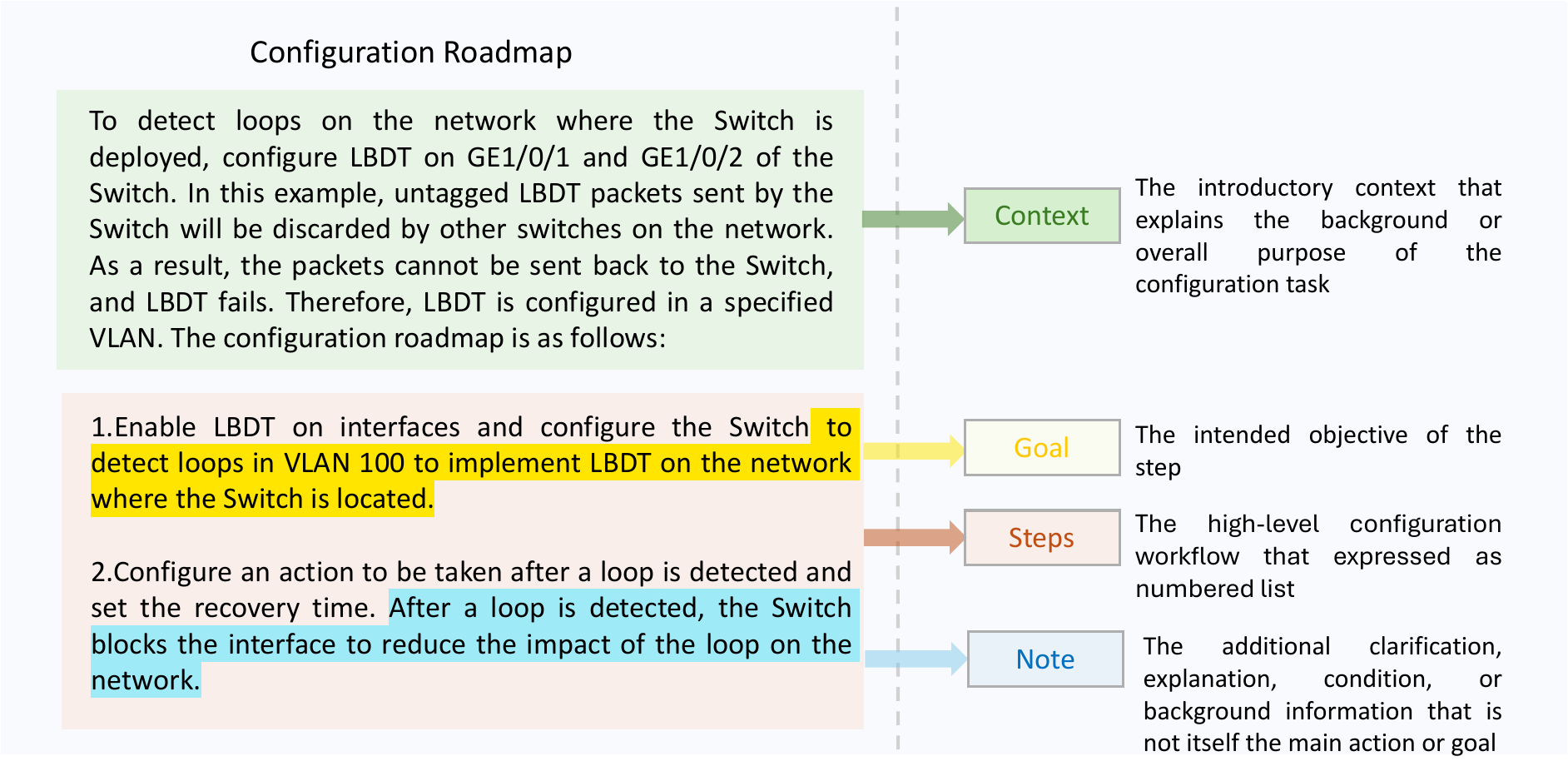}
\caption{An example of the \emph{Configuration Roadmap} section in the ESCM. }
\label{fig:roadmap}
\end{figure*}

\begin{itemize}
\item \textbf{Context (optional)}: The background and/or overall purpose of the configuration task before any explicit steps are listed.
\item \textbf{Main steps}: A sequence of high-level configuration steps, typically expressed as a numbered list.
\item \textbf{Substeps and deeper-level steps (optional)}: Finer-grained steps under a main step, indicated by explicit structural markers such as hierarchical numbers, letters, or bullet points.
\item \textbf{Goal (optional)}: The descriptive text within a step that explicitly states the step objective (e.g., introduced by phrases such as ``to detect'' or  ``to implement’’). 
\item \textbf{Note (optional)}: The additional clarification, explanation, condition, or background information that is not itself the main action or goal.
\end{itemize}

The roadmap extraction agent aims to extract structured roadmap entities from the \emph{Configuration Roadmap} section. Specifically, the agent aims to identify: (1) the context of the roadmap, if present; (2) the hierarchical structure of configuration steps, including main steps, substeps, and any deeper-level steps; and (3) for each step, the explicitly stated goal and any associated notes, if present. Capturing these elements allows the KG to represent not only the sequence of configuration actions, but also the configuration intent and constraints that help interpret the purpose of each step and support downstream reasoning.

\begin{figure*}[htbp]
\centering
\includegraphics[width=0.85\textwidth]{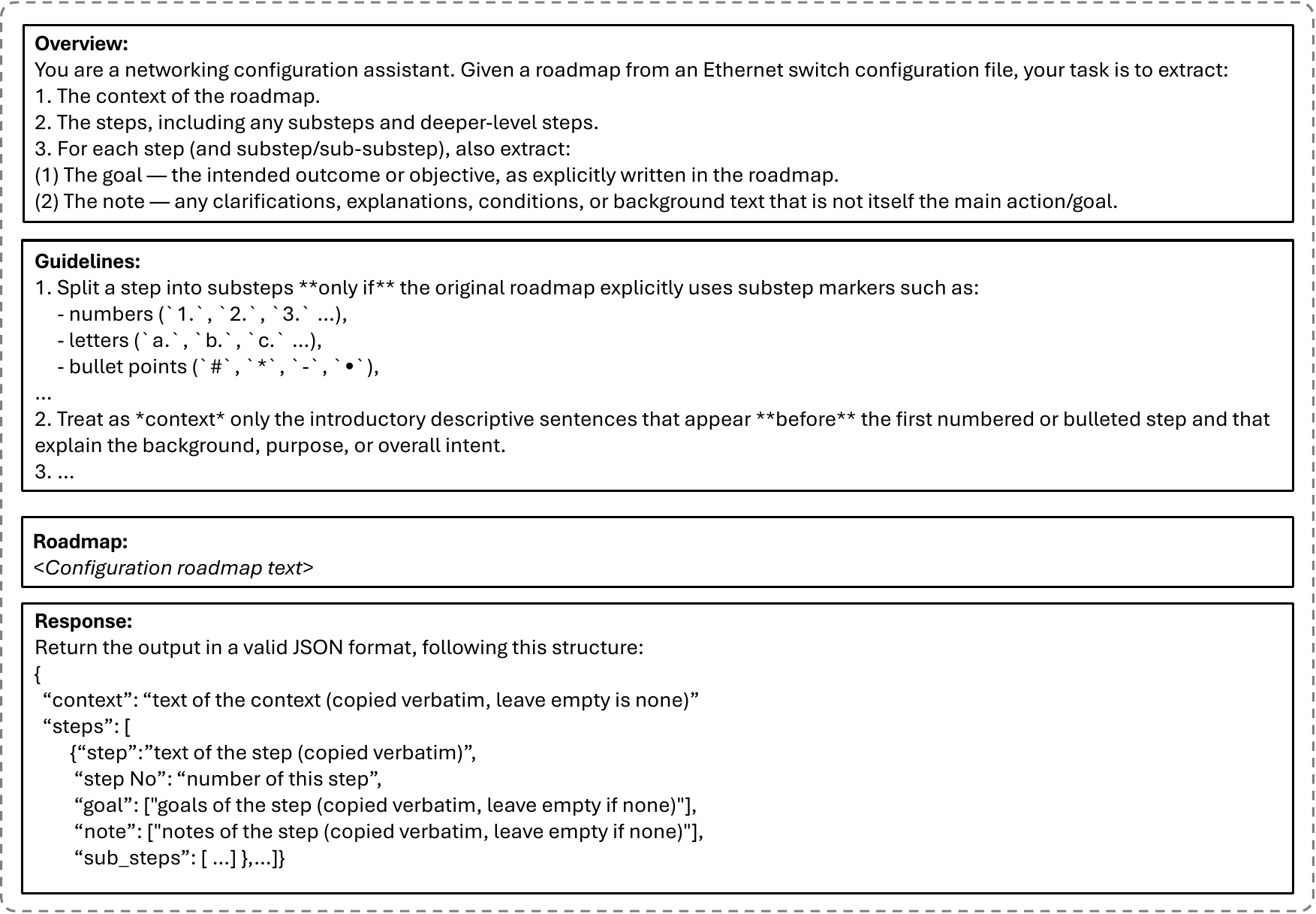}
\Description{}
\caption{The prompt for roadmap extraction agent. }
\label{fig:prompt_task1}
\end{figure*}

We adopt an LLM for this task because context, goals, and notes can be expressed in diverse ways across the \emph{Configuration Roadmap} sections, making them difficult to extract accurately using rule-based parsing. Moreover, step markers and hierarchy notations vary widely across ESCMs, and rule-based approaches may therefore fail to cover all marker patterns, resulting in incomplete or incorrect step extraction. Figure~\ref{fig:prompt_task1} presents a simplified version of the prompt used to guide the agent.
The prompt is organized into four sections:
\begin{itemize}
\item \textit{<Overview>}: This section defines the extraction task, i.e., extracting structured roadmap information (context, hierarchical steps, and per-step goals and notes) from the \emph{Configuration roadmap} section.
\item \textit{<Guidelines>}: This section specifies extraction rules, such as step segmentation, hierarchy numbering of steps, context identification, and verbatim extraction to mitigate LLM hallucinations.
\item \textit{<Roadmap>}: This section provides the original \emph{Configuration Roadmap} section text from the ESCM as the input for the extraction task.
\item \textit{<Response>}: This section constrains the output to a predefined hierarchical JSON schema.
\end{itemize}

As shown in Figure~\ref{fig:output_task1}, the roadmap extraction agent produces a structured JSON representation of the \emph{Configuration Roadmap} section, consisting of a \textit{Context} field and an ordered list of hierarchical \textit{Steps}. For each extracted step, the output preserves the step text verbatim and records its step number, together with any explicitly stated \textit{Goal} and \textit{Note} information, as well as nested \textit{Sub\_steps} when applicable. Note that when certain fields are not present in the roadmap, the agent returns them as empty lists (e.g., \textit{Note}: [], \textit{Sub\_steps}: []).

\begin{figure*}[t]
\centering

\begin{subfigure}[t]{0.53\textwidth}
\centering
\begin{tcolorbox}[
  title=Example Output of the Roadmap Extraction Task,
  width=\linewidth,
  colback=gray!5,
  colframe=black!40!white,
  fonttitle=\bfseries\footnotesize,
  boxsep=1pt,
  left=2pt,right=2pt,top=2pt,bottom=2pt,
  arc=1mm,
]
\begin{lstlisting}[frame=none]
{
  "context": "To detect loops on the network where the Switch is deployed, configure LBDT on GE1/0/1 and GE1/0/2 of the Switch. In this example, untagged LBDT packets sent by the Switch will be discarded by other switches on the network. As a result, the packets cannot be sent back to the Switch, and LBDT fails. Therefore, LBDT is configured in a specified VLAN.",
  "steps": [
    {
      "step": "Enable LBDT on interfaces and configure the Switch to detect loops in VLAN 100 to implement LBDT on the network where the Switch is located.",
      "step No": "1",
      "goal": [
        "detect loops in VLAN 100",
        "implement LBDT on the network where the Switch is located."
      ],
      "note": [],
      "sub_steps": []
    },
    {
      "step": "Configure an action to be taken after a loop is detected and set the recovery time. After a loop is detected, the Switch blocks the interface to reduce the impact of the loop on the network.",
      "step No": "2",
      "goal": [],
      "note": [
        "After a loop is detected, the Switch blocks the interface to reduce the impact of the loop on the network."
      ],
      "sub_steps": []
    }
  ]
}
\end{lstlisting}
\end{tcolorbox}
\caption{Roadmap extraction output.}
\label{fig:output_task1}
\end{subfigure}
\hfill
\begin{subfigure}[t]{0.43\textwidth}
\centering
\begin{tcolorbox}[
  title=Example Triples of the Roadmap Extraction Task,
  width=\linewidth,
  colback=gray!5,
  colframe=black!40!white,
  fonttitle=\bfseries\footnotesize,
  boxsep=1pt,
  left=2pt,right=2pt,top=2pt,bottom=2pt,
  arc=1mm,
]
\scriptsize
\begin{tabular}{@{}p{0.18\linewidth} p{0.20\linewidth} p{0.50\linewidth}@{}}
\textbf{subject} & \textbf{relation} & \textbf{object} \\ \hline
R\_1 & hasContent & Enable LBDT on interfaces ... \\
R\_1 & hasGoal & detect loops ... \\
R\_1 & hasGoal & implement LBDT on the network... \\
R\_2 & hasContent & Configure an action to be taken after a loop is detected... \\
R\_2 & hasNote & After a loop is detected, the Switch blocks the interface ... \\
Roadmap & hasContext & To detect loops on the network... \\
\end{tabular}
\end{tcolorbox}
\caption{Constructed roadmap triples.}
\label{fig:triples_task1}
\end{subfigure}

\caption{Example roadmap extraction output and the corresponding KG triples constructed from the output.}
\label{fig:roadmap_output_and_triples}
\end{figure*}
Finally, as shown in Figure~\ref{fig:triples_task1}, the output of the roadmap extraction agent is converted into a set of entity–relation–entity triples that encode the structured knowledge extracted from the \emph{Configuration Roadmap} section. Specifically, the context, the roadmap steps, and their associated attributes (i.e., content, goals, and notes) are transformed into triples according to the predefined KG schema shown in Figure~\ref{fig:kg_schema}. We assign identifiers with the prefix \textit{R} to roadmap steps based on their hierarchical indices (e.g., \(roadmap~step~1 \rightarrow \textit{R\_1}\), \(roadmap~substep~1.1 \rightarrow \textit{R\_1\_1}\)).
In this example, roadmap step $R\_1$ is connected to its original description through \textit{hasContent} and to multiple goals via \textit{hasGoal}. Similarly, step $R\_2$ is linked to a note using \textit{hasNote} relation.

\noindent\textbf{Roadmap--Procedure Mapping Agent.}
While the \emph{Configuration Roadmap} section provides a coarse-grained abstraction of the configuration workflow, the \emph{Procedure} section details fine-grained configuration steps with concrete commands, parameters, and verification actions. The roadmap--procedure mapping agent aims to establish an explicit mapping between coarse-grained roadmap steps and fine-grained procedure steps, bridging the gap between high-level configuration workflows and their concrete implementation details. These mappings provide structured knowledge that can be leveraged by downstream tasks, such as generating test cases at different levels of granularity.

\begin{figure*}[htbp]
\centering
\includegraphics[width=0.60\textwidth]{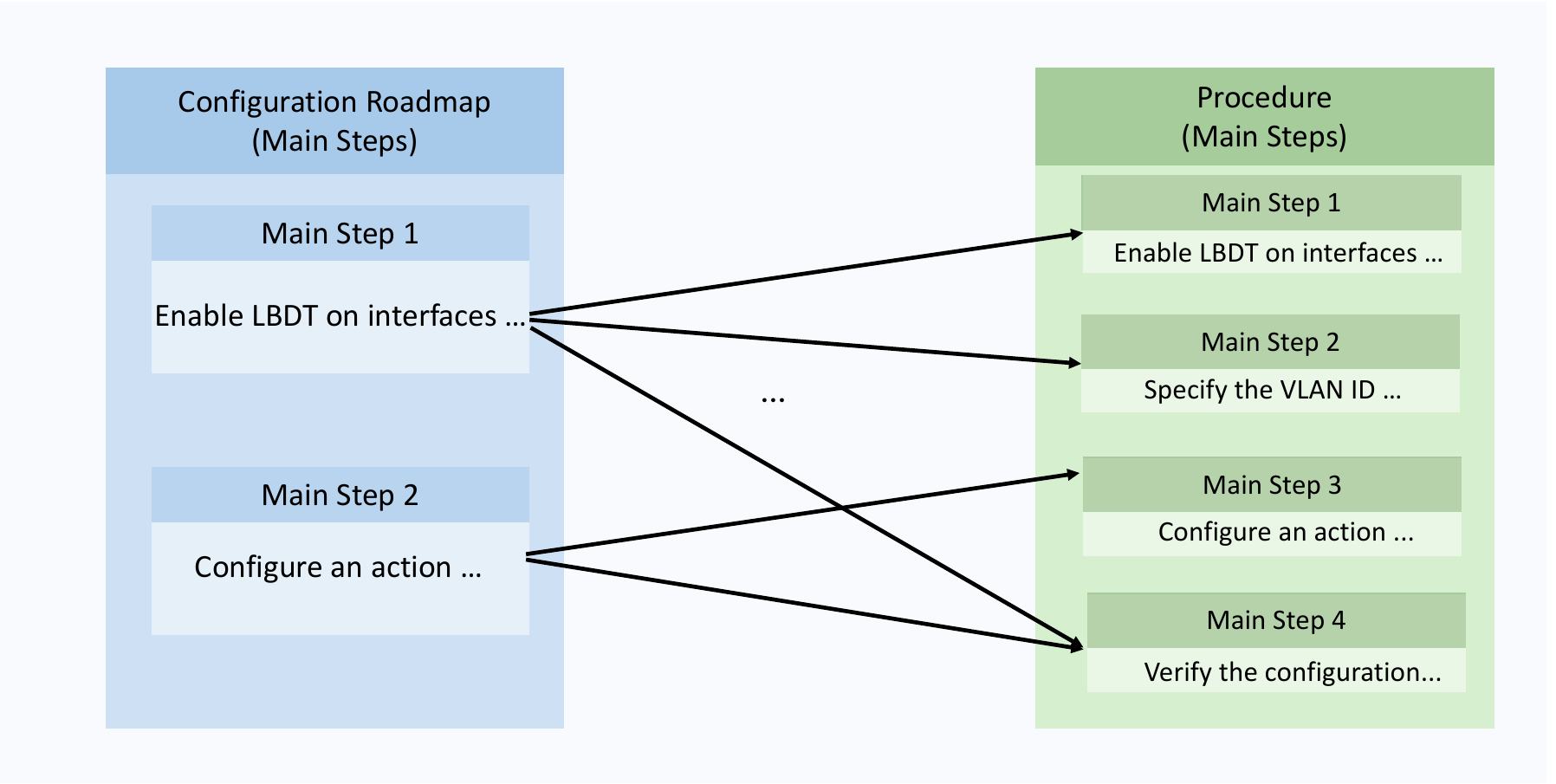}
\Description{}
\caption{An example illustrating the mapping between main steps in the \emph{Configuration Roadmap} section and the corresponding \emph{Procedure} section main steps. }
\label{fig:task2}
\end{figure*}

\begin{figure*}[htbp]
\centering
\includegraphics[width=0.7\textwidth]{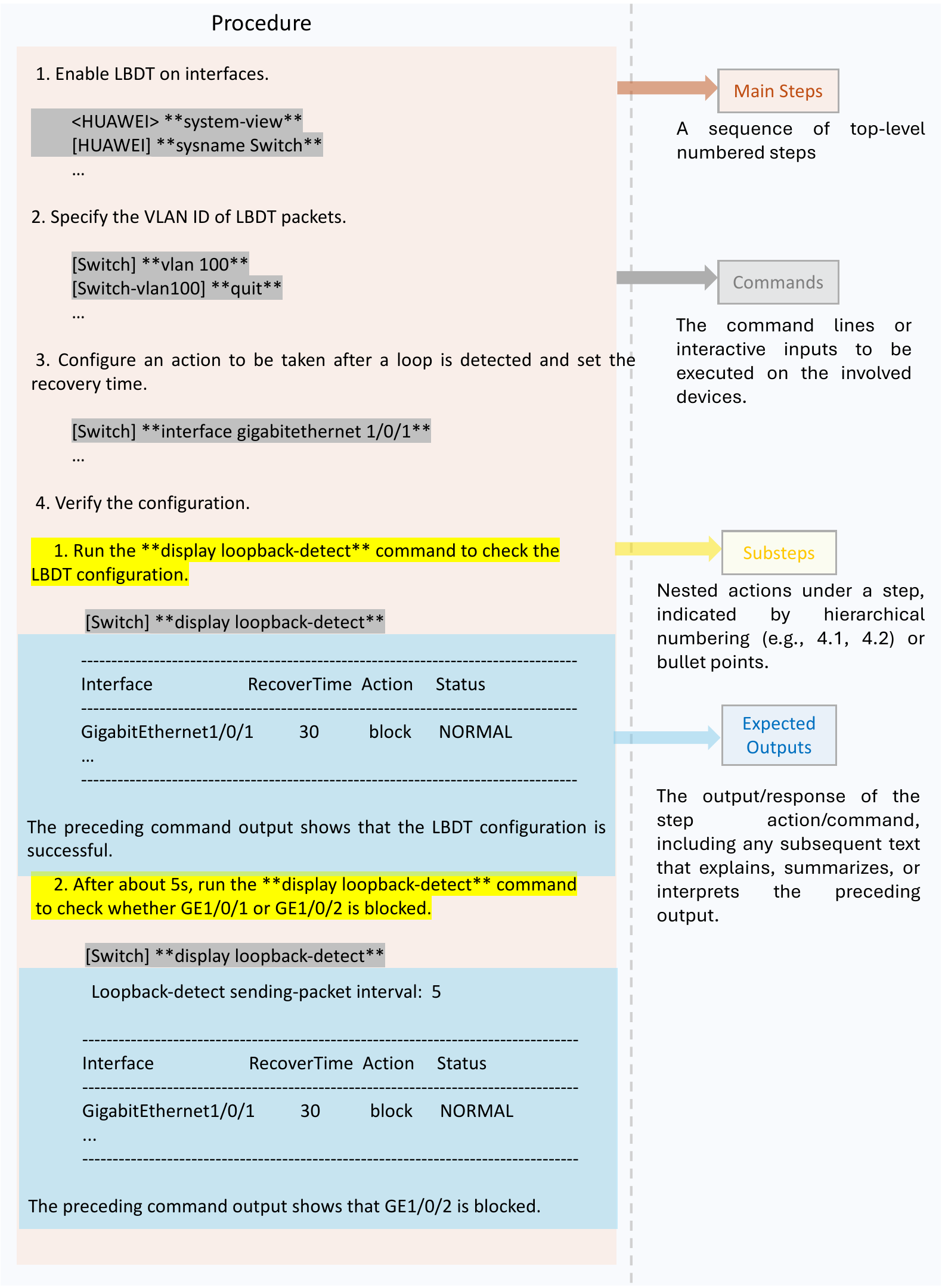}
\caption{An example of the \emph{Procedure} section in the ESCM. }
\label{fig:procedure_excerpt}
\end{figure*}

This task is also performed using an LLM because aligning roadmap steps with procedure steps is primarily a semantic matching problem that requires reasoning over the configuration intent, rather than a purely syntactic match. As shown in Figure~\ref{fig:task2}, the roadmap--procedure mapping agent maps the main steps between the \emph{Configuration Roadmap} section and \emph{Procedure} section. A main step in both sections is defined as a top-level numbered step that includes all associated content. For example, as illustrated in Figure~\ref{fig:roadmap}, the roadmap has two main steps, main step 1 includes the step content and its associated goal, while main step 2 consists of the step content accompanied by an explanatory note. Similarly, the \emph{Procedure} section shown in Figure~\ref{fig:procedure_excerpt} contains four main steps: main steps 1, 2, and 3 include the step contents and configuration commands, whereas main step 4 further contains substeps, commands, and corresponding expected outputs. For each roadmap main step, the agent identifies one or more corresponding procedure main steps that collectively implement and/or verify the intended configuration action described by that roadmap main step. As shown in Figure~\ref{fig:task2}, roadmap main step 1 is mapped to procedure main steps 1, 2, and 4, whereas roadmap main step 2 is mapped to procedure main steps 3 and 4. This many-to-many mapping arises because a high-level roadmap main step is typically decomposed into multiple finer-grained procedure main steps that specify concrete configuration actions and commands, while a single procedure main step, especially a verification step, may help validate the outcomes of multiple roadmap main steps.

\begin{figure*}[htbp]
\centering
\includegraphics[width=0.85\textwidth]{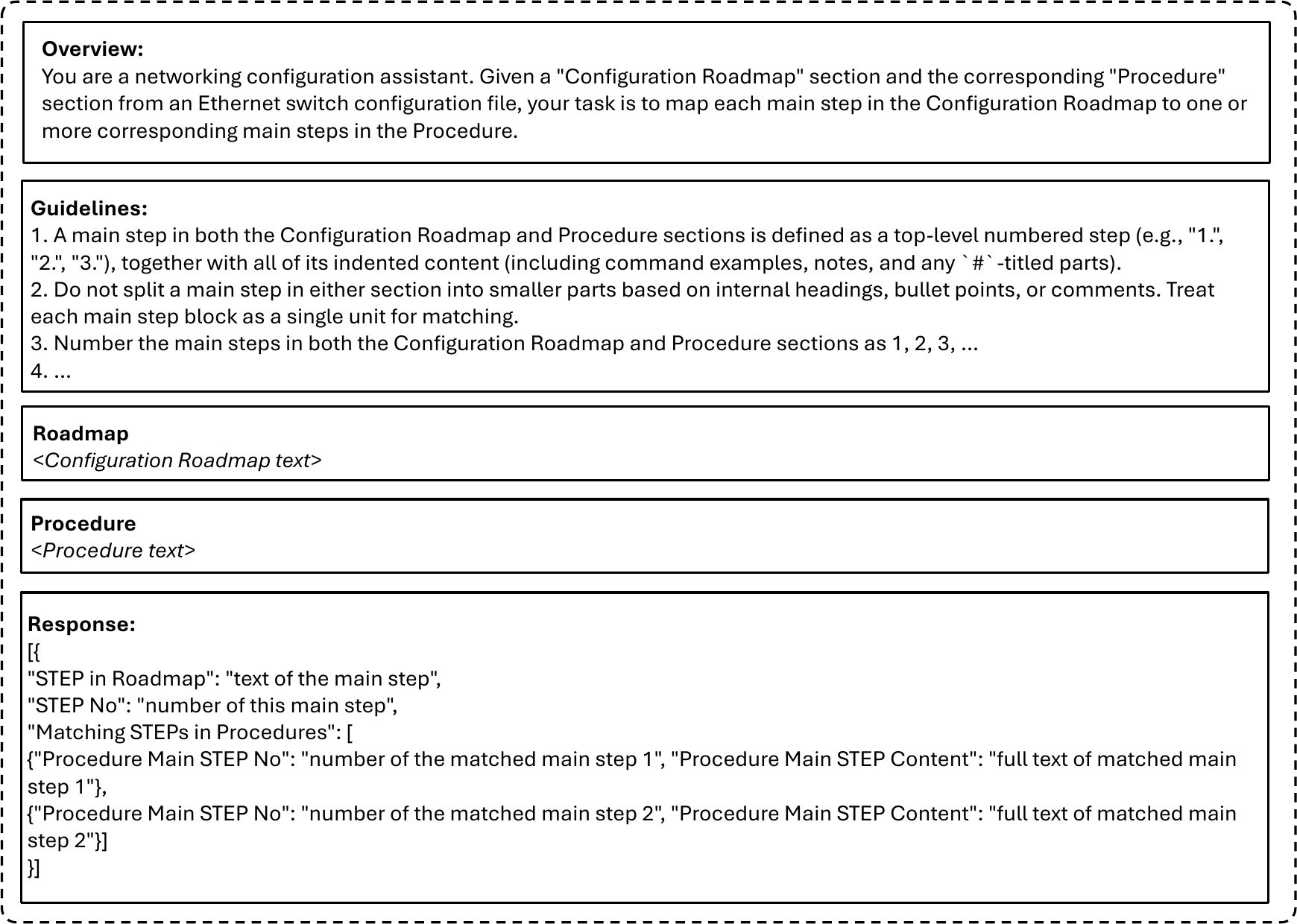}
\Description{}
\caption{The prompt for roadmap--procedure mapping Agent. }
\label{fig:prompt_task2}
\end{figure*}

Figure~\ref{fig:prompt_task2} presents a simplified version of the prompt used for guiding the roadmap--procedure mapping agent. The prompt is organized into five sections: 
\begin{itemize}
\item \textit{<Overview>}: This section defines the task of mapping each main step in the \emph{Configuration Roadmap} to one or more corresponding main steps in the \emph{Procedure} section.
\item \textit{<Guidelines>}: This section specifies the matching rules, such as treating main steps as atomic units, enforcing consistent numbering, and including relevant verification steps.
\item \textit{<Roadmap>}: This section provides the original \emph{Configuration Roadmap} section text as input.
\item \textit{<Procedure>}: This section provides the original \emph{Procedure} section text as input.
\item \textit{<Response>}: This section constrains the mapping results to be returned in a structured JSON format that explicitly captures the relationships between roadmap main steps and procedure main steps.
\end{itemize}

\begin{figure*}[t]
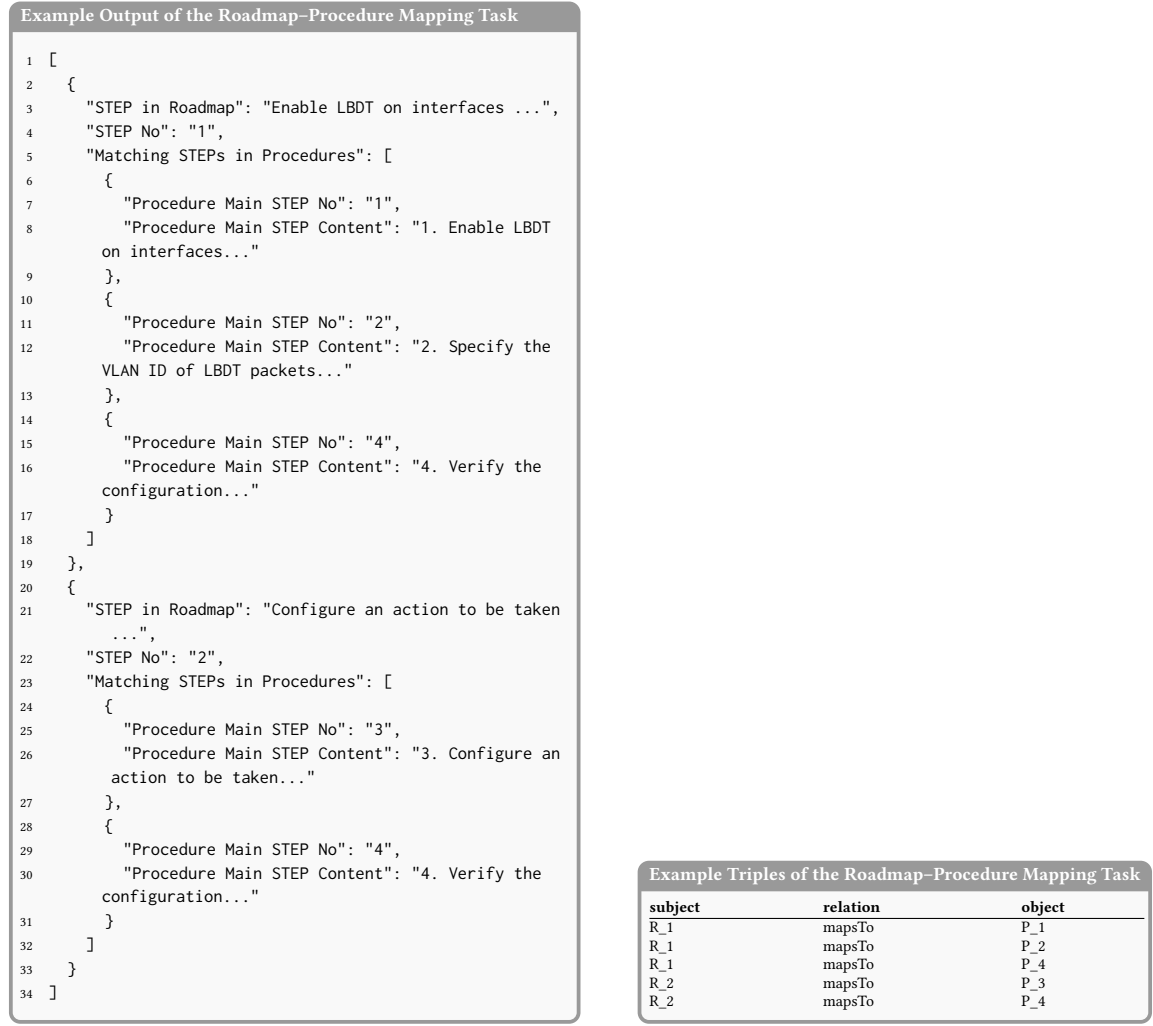

\centering

\begin{subfigure}[t]{0.50\textwidth}
\centering
\begin{tcolorbox}[
  title=Example Output of the Roadmap--Procedure Mapping Task,
  width=\linewidth,
  colback=gray!5,
  colframe=black!40!white,
  fonttitle=\bfseries\footnotesize,
  boxsep=1pt,
  left=2pt,right=2pt,top=2pt,bottom=2pt,
  arc=1mm,
]
\begin{lstlisting}[frame=none]
[
  {
    "STEP in Roadmap": "Enable LBDT on interfaces ...",
    "STEP No": "1",
    "Matching STEPs in Procedures": [
      {
        "Procedure Main STEP No": "1",
        "Procedure Main STEP Content": "1. Enable LBDT on interfaces..."
      },
      {
        "Procedure Main STEP No": "2",
        "Procedure Main STEP Content": "2. Specify the VLAN ID of LBDT packets..."
      },
      {
        "Procedure Main STEP No": "4",
        "Procedure Main STEP Content": "4. Verify the configuration..."
      }
    ]
  },
  {
    "STEP in Roadmap": "Configure an action to be taken ...",
    "STEP No": "2",
    "Matching STEPs in Procedures": [
      {
        "Procedure Main STEP No": "3",
        "Procedure Main STEP Content": "3. Configure an action to be taken..."
      },
      {
        "Procedure Main STEP No": "4",
        "Procedure Main STEP Content": "4. Verify the configuration..."
      }
    ]
  }
]
\end{lstlisting}
\end{tcolorbox}
\caption{Roadmap--procedure mapping output.}
\label{fig:output_task2}
\end{subfigure}
\hfill
\begin{subfigure}[t]{0.45\textwidth}
\centering
\begin{tcolorbox}[
  title=Example Triples of the Roadmap--Procedure Mapping Task,
  width=\linewidth,
  colback=gray!5,
  colframe=black!40!white,
  fonttitle=\bfseries\footnotesize,
  boxsep=1pt,
  left=2pt,right=2pt,top=2pt,bottom=2pt,
  arc=1mm,
]
\centering
\scriptsize
\begin{tabular}{@{}p{0.30\linewidth} p{0.35\linewidth} p{0.25\linewidth}@{}}
\textbf{subject} & \textbf{relation} & \textbf{object} \\ \hline
R\_1 & mapsTo & P\_1 \\
R\_1 & mapsTo & P\_2 \\
R\_1 & mapsTo & P\_4 \\
R\_2 & mapsTo & P\_3 \\
R\_2 & mapsTo & P\_4 \\
\end{tabular}
\end{tcolorbox}
\caption{Constructed mapping triples.}
\label{fig:triples_task2}
\end{subfigure}

\caption{Example roadmap--procedure mapping output and the corresponding KG triples constructed from the output.}
\label{fig:mapping_output_and_triples}
\end{figure*}

As shown in Figure~\ref{fig:output_task2}, the roadmap--procedure mapping agent produces a structured mapping that explicitly aligns each roadmap main step with one or more corresponding procedure main steps. This alignment bridges high-level configuration intent and low-level operational details, thereby enabling the construction of a coherent KG in which information is explicitly represented at multiple levels of granularity and can be effectively leveraged by downstream tasks.

Finally, the mapping results produced by the roadmap--procedure mapping agent are converted into a set of entity–relation–entity triples. As shown in Figure~\ref{fig:triples_task2}, each triple takes the form $<R\_i, mapsTo, P\_j>$, indicating that the roadmap main step $R\_i$ is aligned with the procedure main step $P\_j$. This representation naturally captures many-to-many relationships between the roadmap main steps and the procedure main steps. In this example, the triples $<R\_1, mapsTo, P\_1>$, $<R\_1, mapsTo, P\_2>$, and $<R\_1, mapsTo, P\_4>$indicate that roadmap main step $R\_1$ is operationalized through multiple procedure main steps, while $<R\_2, mapsTo, P\_3>$ and $<R\_2, mapsTo, P\_4>$ show that a verification step (i.e., $P\_4$) can verify both roadmap main steps $R\_1$ and $R\_2$.

\noindent\textbf{Procedure Extraction Agent.} 
ESCMs typically include a \emph{Procedure} section that describes how each high-level roadmap step is implemented through an ordered hierarchy of concrete actions. As shown in Figure~\ref{fig:procedure_excerpt}, a \emph{Procedure} section typically contains the following information:

\begin{itemize}
  \item \textbf{Steps}: A sequence of top-level numbered steps that organize the procedure into coarse-grained actions.
  \item \textbf{Substeps and deeper-level steps (optional)}: Nested actions under a step, indicated by hierarchical numbering (e.g., 1.1, 1.2) or bullet points.
  \item \textbf{Commands (optional)}: Commands used to implement the configuration steps on the involved device(s), typically presented as lines prefixed with device prompts such as \textit{[Switch]}.
  \item \textbf{Expected output (optional)}: The output/response of the step action/command, including any subsequent text that explains or summarizes the command output.
  \item \textbf{Note (optional)}: Text that provides background information, suggestions, or additional clarifications.
\end{itemize}

The procedure extraction agent aims to extract structured procedural information from the \emph{Procedure} section. Specifically, this agent aims to identify: (1) the hierarchical structure of configuration steps,
including main steps, substeps, and any deeper-level steps; (2) for each step level, the associated commands, notes, and expected outputs, if present.

We use an LLM for this task since the \emph{Procedure} section exhibits substantial variation in writing style and formatting across ESCMs, and key elements such as commands, notes, and expected outputs are often expressed implicitly or embedded in text or commands, making them difficult to extract accurately using rule-based parsing approaches.

The \emph{Procedure} section in ESCM is often long (typically 2.5K–4.5K tokens) and densely structured, containing many steps and substeps. Moreover, each step includes multiple types of knowledge to be extracted, such as commands, expected outputs, and notes. This increases inference cost and can degrade the LLM agent's extraction accuracy~\cite{liu2024lost,press2021train,zhang2025siren,ji2023survey}. Therefore, we further split each procedure section into top-level step chunks (i.e., main steps), extract the information from each step chunk independently, and then merge the results in the original step order to reconstruct the complete procedure representation.

\begin{figure*}[htbp]
\centering
\includegraphics[width=0.85\textwidth]{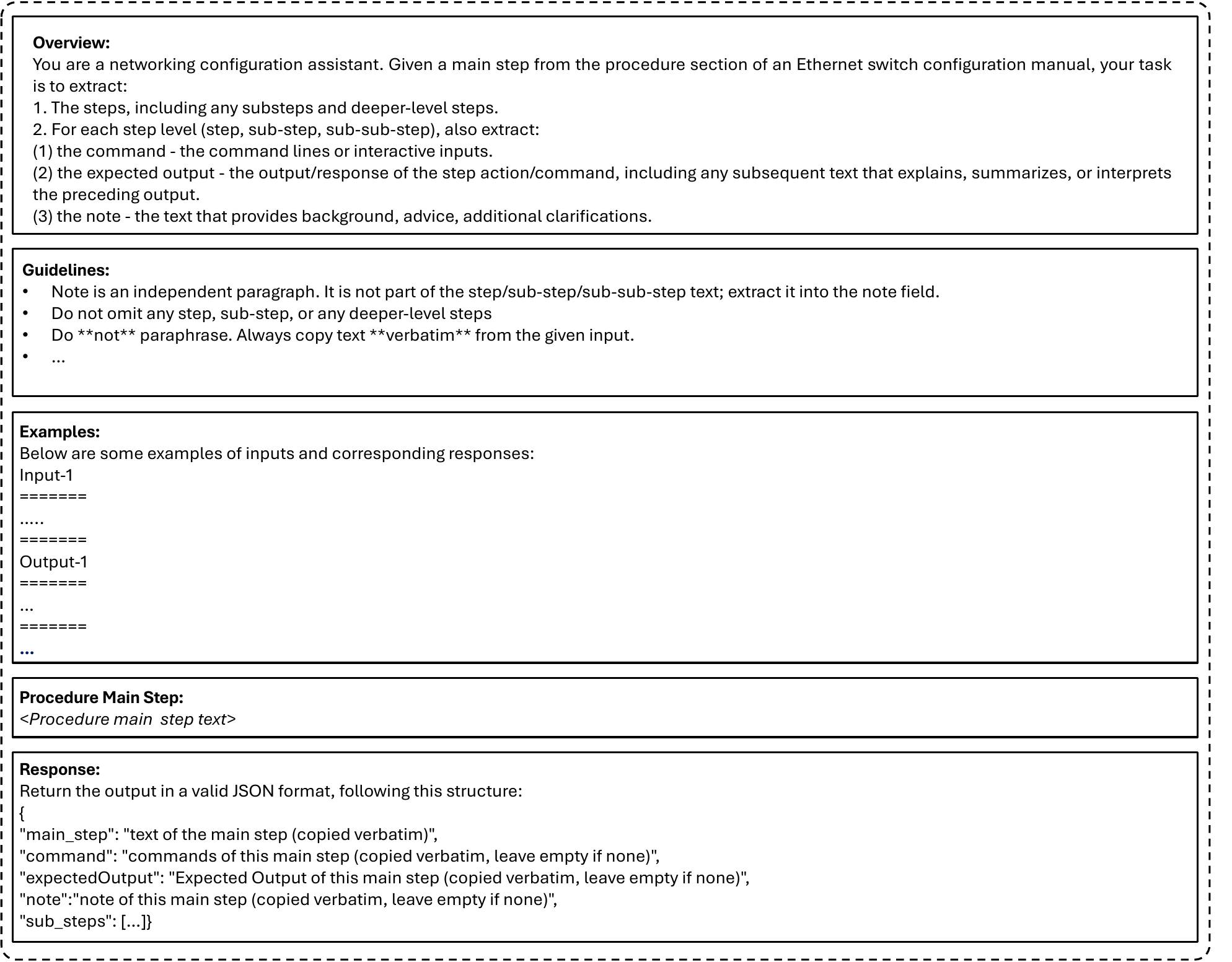}
\Description{}
\caption{The prompt for procedure extraction agent. }
\label{fig:prompt_task3}
\end{figure*}

Figure~\ref{fig:prompt_task3} shows the simplified prompt we used to guide the procedure extraction agent. The prompt is organized into five sections:
\begin{itemize}
  \item \textit{<Overview>}: This section defines the extraction task, i.e., extracting structured information from a procedure main step, including the hierarchical step structure and, for each step level, any associated commands, expected outputs, and notes when present.
  \item \textit{<Guidelines>}: This section specifies the extraction rules, including note identification and attachment, complete preservation of all hierarchical steps, expected output identification, hierarchical numbering of extracted steps, and verbatim copying requirements.
  \item \textit{<Examples>}: This section provides three representative examples, each consisting of an input procedure main step and its corresponding structured JSON output. These examples cover representative cases involving steps, substeps, notes, commands, and expected outputs. 
  Unlike the other two tasks, namely roadmap extraction and roadmap--procedure mapping, for which prompts without examples already achieved strong performance, procedure extraction was significantly improved by the inclusion of in-context examples. We found that prompts without examples were less effective for this task, likely because procedure steps exhibit greater structural and linguistic variation. We therefore included three representative examples in the final prompt to better capture these variations and guide the model toward more accurate extraction.
  \item \textit{<Procedure Main Step>}: This section provides the original procedure main step text from the ESCM for extraction.
  \item \textit{<Response>}: This section constrains the output to a predefined hierarchical JSON schema.
\end{itemize}

As shown in Figure~\ref{fig:output_task3}, the procedure extraction agent produces a structured JSON representation of a procedure main step (Step 4 shown in Figure~\ref{fig:procedure_excerpt}).
In this example, Step 4 comprises two substeps (Steps 4.1 and 4.2), each specifying the command to execute and the corresponding expected output. When certain fields are not present in the procedure main step, the procedure extraction agent returns them as empty lists (e.g., \textit{Command}: [], \textit{ExpectedOutput}: [], \textit{Sub\_sub\_steps}: []).

Finally, as shown in Figure \ref{fig:triples_task3}, the output of the procedure extraction agent is converted into a set of entity-relation-entity triples. Specifically, each procedural main step and its associated elements (nested substeps, commands, expected outputs, and notes) are transformed into triples according to the predefined KG schema shown in Figure \ref{fig:kg_schema}.
We assign identifiers with the prefix \textit{P} to procedure steps based on their hierarchical indices (e.g., \(procedure~main~step~4 \rightarrow \textit{P\_4}\), \(procedure~substep~4.1 \rightarrow \textit{P\_4\_1}\), \(procedure~subsubstep~4.1.1 \rightarrow \textit{P\_4\_1\_1}\)) to distinguish them from roadmap steps (prefixed with \textit{R}).
In this example, procedure step \textit{\(P\_4\)} is connected to its substeps via \textit{hasSubStep} relations, such as ⟨\textit{\(P\_4\)}, \textit{hasSubStep}, \textit{\(P\_4\_1\)}⟩, and ⟨\textit{\(P\_4\)}, \textit{hasSubStep}, \textit{\(P\_4\_2\)}⟩. Substep \textit{\(P\_4\_1\)} is linked to its content through the \textit{hasContent} relation and to its command through \textit{hasCommand} relation. Substep \textit{\(P\_4\_2\)} is linked to its expected output via the \textit{hasExpectedOutput} relation. 

\begin{figure*}[t]
\centering

\begin{subfigure}[t]{0.96\textwidth}
\centering
\begin{tcolorbox}[
  title=Example Output of the Procedure Extraction Task,
  width=\linewidth,
  colback=gray!5,
  colframe=black!40!white,
  fonttitle=\bfseries\footnotesize,
  boxsep=1pt,
  left=2pt,right=2pt,top=2pt,bottom=2pt,
  arc=1mm,
]
\begin{lstlisting}[basicstyle=\ttfamily\tiny,breaklines=true,columns=fullflexible,keepspaces=true,showstringspaces=false,frame=none]
{
  "main_step": "4. Verify the configuration.",
  "command": [],
  "expectedOutput": [],
  "note": [],
  "sub_steps": [
    {
      "sub_step_No": "4.1",
      "sub_step": "1. Run the **display loopback-detect** command to check the LBDT configuration.",
      "command": "[Switch] **display loopback-detect**",
      "expected_Output": "Loopback-detect sending-packet interval:  5
      ----------------------------------------------------------------------------------
      Interface                     RecoverTime  Action     Status
      ----------------------------------------------------------------------------------
      GigabitEthernet1/0/1          30           block      NORMAL
      GigabitEthernet1/0/2          30           block      NORMAL
      ----------------------------------------------------------------------------------
      The preceding command output shows that the LBDT configuration is successful.",
      "note": [],
      "sub_sub_steps": []
    },
    {
      "sub_step_No": "4.2",
      "sub_step": "2. After about 5s, run the **display loopback-detect** command to check whether GE1/0/1 or GE1/0/2 is blocked.",
      "command": "[Switch] **display loopback-detect**",
      "expected_Output": "Loopback-detect sending-packet interval:  5
      ----------------------------------------------------------------------------------
      Interface                     RecoverTime  Action     Status
      ----------------------------------------------------------------------------------
      GigabitEthernet1/0/1          30           block      NORMAL
      GigabitEthernet1/0/2          30           block      **BLOCK(Loopback detected)**
      ----------------------------------------------------------------------------------
      The preceding command output shows that GE1/0/2 is blocked.",
      "note": [],
      "sub_sub_steps": []
    }
  ]
}
\end{lstlisting}
\end{tcolorbox}
\caption{Procedure extraction output.}
\label{fig:output_task3}
\end{subfigure}

\vspace{0.5em}

\begin{subfigure}[t]{0.96\textwidth}
\centering
\begin{tcolorbox}[
  title=Example Triples of the Procedure Extraction Task,
  width=\linewidth,
  colback=gray!5,
  colframe=black!40!white,
  fonttitle=\bfseries\footnotesize,
  boxsep=1pt,
  left=2pt,right=2pt,top=2pt,bottom=2pt,
  arc=1mm,
]
\scriptsize
\begin{tabular}{@{}p{0.14\linewidth} p{0.25\linewidth} p{0.56\linewidth}@{}}
\textbf{subject} & \textbf{relation} & \textbf{object} \\ \hline
P\_4 & hasContent & 4. Verify the configuration. \\
P\_4 & hasSubStep & P\_4\_1 \\
P\_4\_1 & hasContent & 1. Run the **display loopback-detect** ... \\
P\_4\_1 & hasCommand & [Switch] **display loopback-detect** \\
P\_4\_1 & hasExpectedOutput & Loopback-detect sending-packet interval: 5... \\
P\_4 & hasSubStep & P\_4\_2 \\
P\_4\_2 & hasContent & 2. After about 5s ... \\
P\_4\_2 & hasCommand & [Switch] **display loopback-detect** \\
P\_4\_2 & hasExpectedOutput & Loopback-detect sending-packet interval: 5... \\
\end{tabular}
\end{tcolorbox}
\caption{Constructed procedure triples.}
\label{fig:triples_task3}
\end{subfigure}

\caption{Example procedure extraction output and the corresponding KG triples constructed from the output.}
\label{fig:procedure_output_and_triples}
\end{figure*}

\subsubsection{Knowledge Graph Extract-Evaluate-Improve (EEI) Loop}
\label{sec:eeiloop}
Although an LLM can effectively extract structured KG entities from ESCM when guided by a well-designed prompt, the extracted content may still contain errors. For example, the extraction guidelines in the prompt may need to be adapted for a specific ESCM to match the writing style and format of that particular ESCM. In addition, the LLM may generate inaccurate or incomplete extractions due to hallucinations. We therefore design an Extract-Evaluate-Improve (EEI) loop to systematically refine extraction prompts based on the evaluation feedback of extracted KG entities. This loop is applied to three extraction tasks, namely roadmap extraction, roadmap-procedure mapping, and procedure extraction, and iteratively improves entity extraction by regenerating KG entities with the revised prompts. In our preliminary experiments, we found that regenerating KG entities from the original ESCM with a revised prompt was more reliable than directly repairing erroneous KG entities from the previous extraction. A possible reason is that, when directly repairing a previous extraction, the LLM may still be influenced by the erroneous KG entities and keep making the same mistakes. In contrast, regenerating KG entities from the original ESCM can help avoid repeating errors from the previous extraction.

As shown in Figure~\ref{fig:multi-agent-framework}, in the EEI loop, the outputs of the ExtrAgents, namely the extracted entities, are fed into an EvalAgent, which scores the extraction quality according to the predefined task-specific evaluation guidelines and produces detailed feedback. We then compute an average correctness score across all the evaluation guidelines. If the average score falls below a predefined threshold, an ImprovAgent revises the extraction prompt based on the evaluation feedback, producing a revised prompt for the next iteration. The loop repeats until the predefined threshold is met or a maximum number of iterations is reached. In this study, we set the EEI loop correctness threshold to 0.9 and the maximum number of iterations to 3. This threshold was chosen because it represents a relatively high level of correctness. This allows the EEI loop to focus on outputs with clear deficiencies, while avoiding unnecessary refinement for outputs that are already of high quality. The maximum iteration number was set to 3 to bound the refinement cost and prevent the loop from running indefinitely. Both parameters are configurable and can be adjusted when applying the framework to other domains, depending on the required quality standard, available budget, and acceptable computational cost. In general, a higher threshold or a larger maximum iteration number may trigger more refinement attempts and thus require more time and computational resources.

\vspace{4pt}
\noindent\textbf{Extraction Evaluation Agent.}
Evaluating the quality of LLM-generated results remains challenging~\cite{liu2023g}. While there are traditional, widely used NLP evaluation metrics such as BLEU~\cite{papineni2002bleu}, ROUGE~\cite{lin2004rouge}, and METEOR~\cite{banerjee2005meteor}, they are particularly unreliable for evaluating reasoning-intensive natural language generation tasks~\cite {liu2023g}. Moreover, these metrics typically rely on ground-truth reference outputs, which are costly and time-consuming to collect~\cite {liu2023g}.
Therefore, we adopt an \emph{LLM-as-a-Judge}~\cite{fu2024gptscore,wang2023chatgpt,liu2023g, zheng2023judging,zhu2023judgelm,jung2024trust,gu2024survey} evaluation approach, which prompts an LLM to evaluate the generated text, to assess the quality of the entity extraction results. LLM-as-a-Judge methods have been shown to align reasonably well with human judgments, particularly when evaluation criteria are clearly specified~\cite{zheng2023judging,liu2023g}. Specifically, we provide (i) the extraction input (i.e., the \emph{Configuration Roadmap} and/or \emph{Procedure} section text), (ii) the extraction output (i.e., extracted entities), and (iii) a set of carefully-designed task-specific evaluation guidelines, (iv) a set of representative examples of input text and correctly extracted output, and ask the EvalAgent to assign a binary score ($1$ or $0$) to each evaluation guideline.  When an evaluation guideline receives a score of $0$, the EvalAgent also returns the failure reason and suggestions for improving the extraction prompt.

\begin{figure*}[htbp]
\centering
\includegraphics[width=0.85\textwidth]{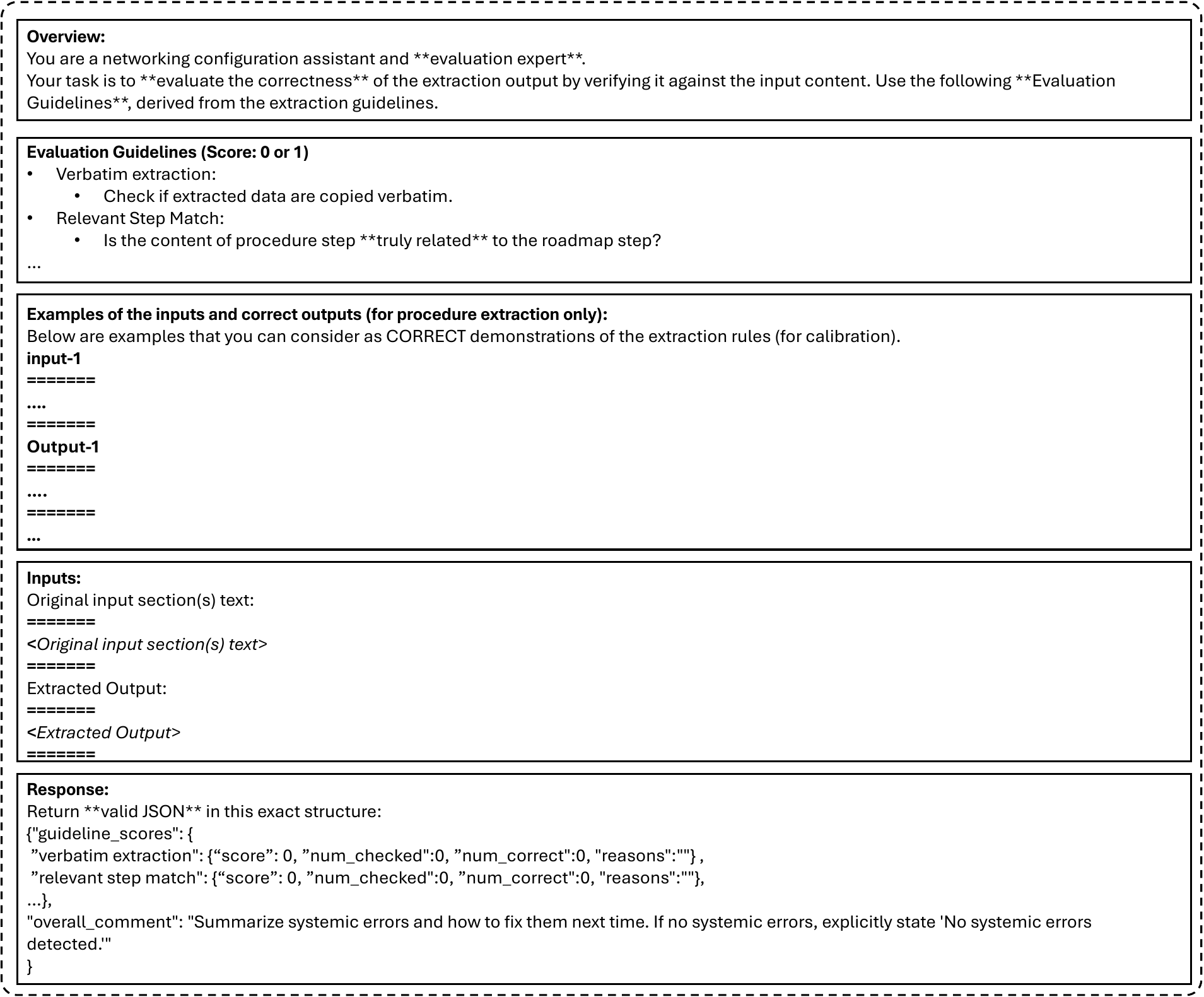}
\Description{}
\caption{The prompt for the EvalAgent. }
\label{fig:prompt_evaluation}
\end{figure*}

Figure~\ref{fig:prompt_evaluation} presents the simplified prompt used by the EvalAgent to assess extraction results for the three extraction tasks. The prompt is organized into six sections:
\begin{itemize}
  \item \textit{<Overview>} This section assigns the role of a networking configuration assistant and evaluation expert to LLM, and specifies the evaluation objective.
  \item \textit{<Evaluation Guidelines>} This section lists the criteria used to assess each evaluation guideline, derived from the original extraction guidelines provided to the ExtrAgent (e.g., verbatim copying, main-step boundary compliance, and step-numbering compliance). Since the three extraction tasks have different extraction objectives and output structures, their evaluation guidelines also differ accordingly. Detailed task-specific evaluation guidelines are provided below.
  \item \textit{<Examples of the inputs and correct outputs>} When including examples in the prompt of an ExtrAgent proved beneficial, the same set of examples is also included in the prompt for the Evalgent, this helps the LLM judge the outputs more accurately and consistently against the expected extraction behavior. Since only the procedure extraction agent benefited from the examples, we included them exclusively for procedure extraction evaluation, and not for the evaluations of roadmap extraction or roadmap–procedure mapping. We provide three example input–output pairs in the prompt as correct demonstrations of the extraction task, using the same three examples included in the corresponding procedure extraction prompt. Our preliminary study also confirms that adding these examples to the procedure evaluation prompt substantially improved the accuracy of the evaluation results.
  \item \textit{<Inputs>} This section provides the source text for each extraction task: Roadmap extraction uses the \emph{Configuration Roadmap} section text, roadmap--procedure mapping uses both the \emph{Configuration Roadmap} and \emph{Procedure} section texts, and procedure extraction uses the \emph{Procedure} section text, together with the extracted entities for that task.
  \item \textit{<Response>} This section defines the evaluation output format. The output follows a fixed JSON schema that reports, for each guideline, a binary compliance score (\textit{score}), two counting statistics (\textit{num\_checked} and \textit{num\_correct}), and a list of diagnostic explanations (\textit{reason}) when violations occur; it also includes an \textit{overall\_commment} summarizing any issues and suggested improvements. Here, \textit{num\_checked} denotes the total number of JSON entries evaluated for an evaluation guideline (including empty cases), whereas \textit{num\_correct} denotes the number of evaluated JSON entries that satisfy the corresponding rule.
\end{itemize}

Since the three extraction tasks differ in their objectives and expected outputs, the guidelines for the corresponding EvalAgent's prompt are task-specific. We derive the evaluation guidelines from the extraction guidelines for each task, as these explicitly define what knowledge should be extracted, how the output should be structured, and the task-specific constraints that should be satisfied. This alignment ensures that the EvalAgent evaluates each extracted KG entity using the same criteria that guided its extraction, making the evaluation more consistent and task-relevant.
The main guidelines for each task are summarized below.

\begin{itemize}
  \item \textbf{Roadmap extraction evaluation guidelines}
  \begin{itemize}
    \item \textbf{Step splitting:} Checks whether roadmap content is split into hierarchical steps only when explicit structural markers are present.
    \item \textbf{Context identification:} Checks whether only the introductory descriptive text before the first actual step is extracted as \textit{context}.
    \item \textbf{Goal extraction:} Checks whether explicitly stated purposes are correctly extracted into the \textit{goal} field.
    \item \textbf{Note extraction:} Checks whether clarifications, conditions, or background information are correctly extracted into the \textit{note} field.
    \item \textbf{Numbering:} Checks whether extracted steps follow a sequential and consistent hierarchical numbering scheme.
    \item \textbf{Verbatim copying:} Checks whether all extracted texts are copied exactly from the source roadmap text.
    \item \textbf{Format compliance:} Checks whether the output strictly follows the required JSON structure.
  \end{itemize}

  \item \textbf{Roadmap--procedure mapping evaluation guidelines}
  \begin{itemize}
    \item \textbf{Main-step boundary compliance:} Checks whether each roadmap or procedure step is treated as a complete top-level numbered block.
    \item \textbf{Step-numbering compliance:} Checks whether step numbers are sequential and accurately represented in the output.
    \item \textbf{Relevant step match:} Checks whether each matched procedure step is truly relevant to the roadmap step.
    \item \textbf{Multiple match inclusion:} Checks whether all relevant procedure steps are included when a roadmap step matches multiple procedure steps.
    \item \textbf{Device identifier consistency:} Checks whether matched procedure steps preserve the same device identifiers as those in the roadmap step.
    \item \textbf{Text completeness:} Checks whether the full texts of the roadmap and procedure steps are completely preserved.
    \item \textbf{Structural format:} Checks whether the output follows the required JSON schema.
  \end{itemize}

  \item \textbf{Procedure extraction evaluation guidelines}
  \begin{itemize}
    \item \textbf{Step coverage:} Checks whether all steps, sub-steps, and deeper-level steps are preserved without omission.
    \item \textbf{Step-numbering compliance:} Checks whether extracted steps are numbered hierarchically and sequentially.
    \item \textbf{Command extraction correctness:} Checks whether \textit{command} correctly captures the command lines or interactive inputs for each step.
    \item \textbf{Expected-output extraction correctness:} Checks whether \textit{expectedOutput} correctly captures execution outputs and related explanatory text.
    \item \textbf{Note classification \& attachment correctness:} Checks whether notes are correctly identified and attached to the appropriate preceding step.
    \item \textbf{Text completeness \& verbatim copying:} Checks whether all extracted texts are copied verbatim from the source procedure text.
    \item \textbf{Structural format \& schema compliance:} Checks whether the output strictly conforms to the required hierarchical JSON schema.
  \end{itemize}
\end{itemize}

The output of the EvalAgent contains two parts: (i) per-guideline scores and reasons for any errors under each guideline, and (ii) an overall comment summarizing the main issue observed and suggestions for improvements.
For example, in the output of a roadmap extraction evaluation, the Guideline scores section reports several compliance checks, including Step Splitting, Context Identification, Goal Extraction, Note Extraction, Numbering, Verbatim Copying, and Format Compliance. For each guideline, it provides a binary score, the counts of correct vs. checked JSON entries, and an optional error list that provides concrete evidence when violations occur.

For instance, if an error is detected in Note Extraction, the evaluation output specifies which note is incorrect and explains why it violates the note extraction guideline, such as by incorrectly placing essential action content into the note field. The overall comment further generalizes this issue, emphasizing that notes should only include clarifications or conditions, rather than repeating essential configuration actions.

Given an evaluation output, we calculate the \textbf{overall correctness score} as follows:

\begin{equation}
\label{eq:correctness-score}
\textit{correctness\_score}
=
\frac{\sum_{i=1}^{m}\textit{num\_correct}_i}
{\sum_{i=1}^{m}\textit{num\_checked}_i}.
\end{equation}

where \(\textit{num\_checked}_i\) denotes the number of checked JSON entries for guideline \(i\), and \(\textit{num\_correct}_i\) denotes the number of correct JSON entries for guideline \(i\). \(m\) is the total number of evaluation guidelines defined for the corresponding extraction task. Each JSON entry in the extraction output corresponds to a KG entity.
This metric measures the fraction of correctly extracted JSON entries among all evaluated JSON entries. 

Moreover, since the extracted KG entities are converted into triples by a deterministic script following the predefined KG schema, this conversion process does not introduce additional uncertainty or require further LLM-based inference. Therefore, evaluating the correctness of the extracted KG entities is equivalent to evaluating the correctness of the resulting KG.

\vspace{4pt}
\noindent\textbf{Extraction Improvement Agent.}
\begin{figure*}[htbp]
\centering
\includegraphics[width=0.85\textwidth]{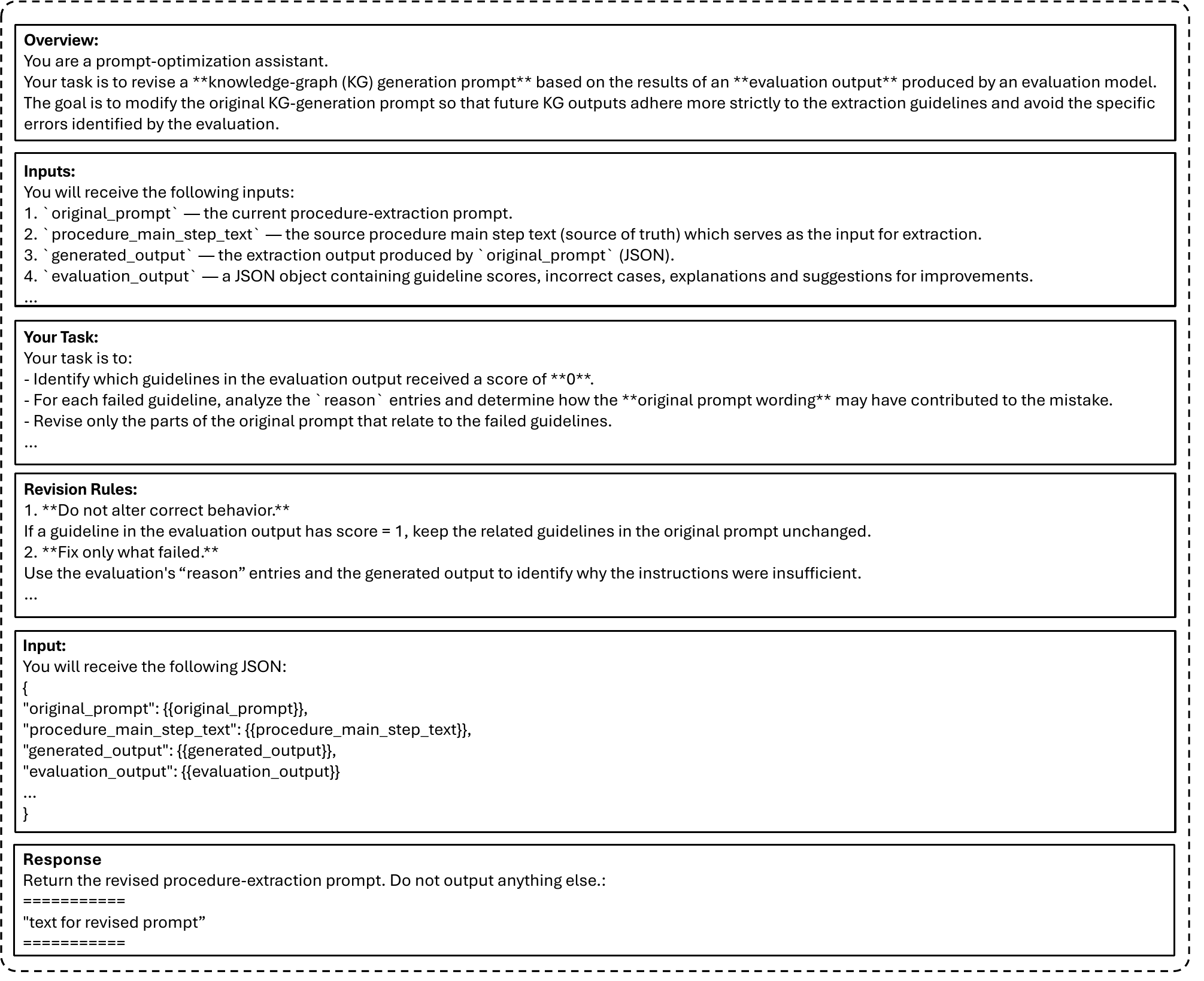}
\Description{}
\caption{The prompt for the ImprovAgent. }
\label{fig:prompt_improvement}
\end{figure*}
If the overall correctness score is below the predefined threshold, the ImprovAgent is triggered to refine the extraction prompt using the evaluation feedback and regenerate the KG entities.
Figure~\ref{fig:prompt_improvement} presents the simplified prompt used to guide the ImprovAgent to refine the original extraction prompt based on the evaluation feedback. The prompt is organized into seven sections:
\begin{itemize}
  \item \textit{<Overview>} This section assigns the LLM the role of a prompt-optimization assistant and defines the prompt improvement task.
  \item \textit{<Inputs>} This section lists and briefly describes the types of information provided to the ImprovAgent.
 \item \textit{<Your task>} This section defines the overall improvement task of the ImprovAgent, specifying how it should analyze the previous result and refine the prompt for the next iteration.
 \item \textit{<Revision Rules>} This section specifies the rules for refining the prompt, such as do not change the guidelines regarding the correct behavior (i.e, with score $=1$),  avoid vague instructions, and avoid modifying other prompt sections (e.g., overview, response format, etc.).
  \item \textit{<Input>} This section presents the actual input given to the ImprovAgent, such as the original extraction prompt, the input text (i.e, roadmap and/or procedure), the extraction output, and the evaluation feedback.
  \item \textit{<Response>} This section specifies the required output of the ImprovAgent.
\end{itemize}

\begin{figure*}[htbp]
\centering
\includegraphics[width=0.8\textwidth]{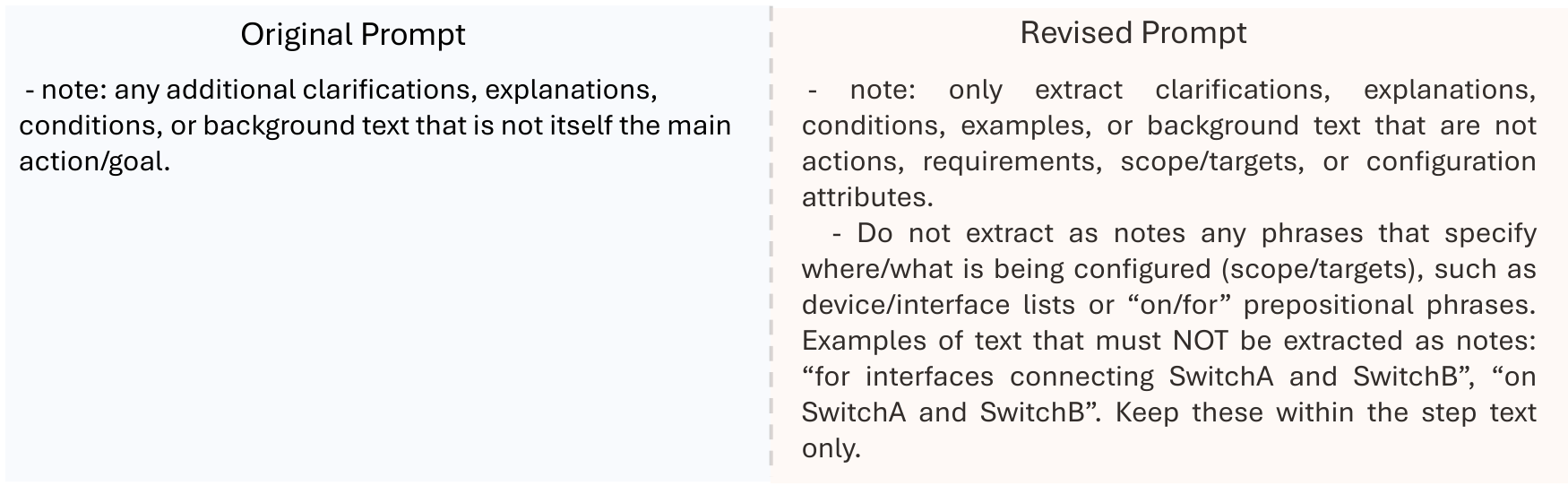}
\Description{}
\caption{Example of the revised prompt. }
\label{fig:revised_prompt}
\end{figure*}

Figure~\ref{fig:revised_prompt} shows an example of a revised prompt for the roadmap extraction task. In this case, the EvalAgent identified errors only in the Note Extraction evaluation guideline, while all other extraction guidelines were satisfied. Therefore, the ImprovAgent refined only the Note Extraction guideline in the extraction prompt, while leaving all other parts unchanged. Each ESCM may have its own writing style and formatting conventions, which may require manual-specific rules for accurate KG entity extraction. By refining only the guidelines associated with the observed errors, the LLM agent can effectively adapt the prompt to the characteristics of an individual ESCM.

\subsubsection{Knowledge Graph Enhancement}
According to the KG schema shown in Figure~\ref{fig:kg_schema}, the triples generated by roadmap extraction, roadmap–procedure mapping, and procedure extraction are connected to the \emph{Use Case Scenario} entity, which corresponds to the title of the ESCM, through the \textit{hasRoadmap} and \textit{hasProcedure} relations. 
To provide additional information that may support downstream tasks, particularly TCS generation, we further enhance the generated KG by incorporating supplementary sections that are not covered by the three main extraction tasks. 
Specifically, we include the \emph{Configuration Files} and \emph{Networking Requirements} sections, which contain substantial, critical supplementary information essential for correctly interpreting, executing, and reproducing the configuration task, as additional KG entities. The \emph{Configuration Files} section provides the complete device configurations to support deployment and reproducibility, while the \emph{Networking Requirements} section characterizes the assumed network environment, including topology, device roles, and operational objectives. 
These entities are linked to the \emph{Use Case Scenario} entity via the \textit{hasConfigurationFile} and \textit{hasNetworkingRequirements} relations. Together, these entities and relations form the final constructed KG for each ESCM.
This enhancement enriches the KG with supplementary configuration and contextual information, further supporting the generation of KG-derived TCSs. For other downstream tasks, additional sections can be incorporated in the same manner when they provide task-relevant information.

\subsection{Example Application Scenario: System-Level Test Case Specification Generation}
The final constructed KG captures rich semantics from the ESCM and provides structured domain knowledge for downstream tasks, particularly test automation. While the KG facilitates efficient retrieval of relevant information by enabling structured, relation-aware queries and lightweight graph traversals, and provides structured domain knowledge that can help reduce LLM hallucinations~\cite{su2024enhancing, su2025automated}, downstream automation tasks also require artifacts that are not only machine-consumable but also engineer-friendly and human-reviewable, enabling engineers to inspect, validate, and correct the extracted knowledge before execution~\cite{da2018test,milchevski2025multi,utting2010practical}. Moreover, recent LLM-based test generation studies increasingly adopt a two-stage workflow, where requirements or source documents are first transformed into intermediate test specifications or structured test descriptions, which are then used to guide executable test generation~\cite{milchevski2025multi, yang2025requirements, masuda2026generating, sarma2009automatic}. Therefore, to investigate how effectively the generated KGs can support such downstream testing tasks, we convert the KGs into system-level TCSs, which provide a precise description of the testing target~\cite{tcs1993automated, milchevski2025multi} and serve as structured input for automated downstream tasks, such as test case generation~\cite{milchevski2025multi,yang2025requirements,sarma2009automatic}. Specifically, we traverse the KG to identify the entities and relations associated with each configuration scenario, then map the extracted information to the predefined TCS template. 
The generated TCS is represented in JSON format and organizes the information in the KG triples into the following key elements, all of which are required for test generation:
\begin{itemize}
  \item \textbf{\textit{use\_case}}: The use case describes the configuration scenario and its objective, and is derived from the \emph{Use Case Scenario} entity.
  \item \textbf{\textit{preconditions}}: The preconditions specify the required initial environmental settings and configuration state that must be satisfied before executing the configuration steps, and are derived from the \emph{Networking Requirements} entity.
  \item \textbf{\textit{configuration\_steps}}: The configuration steps integrate information from both the \emph{Configuration Roadmap} and \emph{Procedure} entities. They include the roadmap context, roadmap steps with their IDs, contents, substeps, goals, notes, and mappings to the corresponding procedure steps, as well as the mapped procedure steps with their IDs, contents, commands, substeps, notes, and expected outputs. This field, therefore, preserves both the high-level configuration intent and the detailed executable actions required to implement and verify the configuration.
  \item \textbf{\textit{configuration\_file}}: The configuration file provides the complete device configuration needed for deployment and reproducibility, which is derived from the \emph{Configuration File} entity.
\end{itemize}

The generated TCS serves as a structured and human-reviewable specification for system-level Ethernet switch configuration testing. It summarizes the testing objective, required preconditions, configuration steps, and verification steps derived from the KG, thereby providing a clear basis for test engineers to review, refine, and convert into executable test cases.

Using a TCS offers several practical advantages for automated, reproducible downstream tasks. Compared to KG, TCS is easier to validate and debug. When the generated tests contain errors, diagnosing their source is substantially more difficult in KG than in a TCS, where preconditions, steps, and expected outputs are explicitly grouped and ordered in a structured way. Moreover, while KGs are useful for graph-level reasoning, system-level TCS presents information in a human-readable, tool-friendly form that can be reviewed and reused directly. Overall, the KG-to-TCS transformation converts the extracted knowledge into a structured, reviewable artifact, enabling systematic validation and reliable automation of downstream testing tasks.

\subsection{Adaptation to Other Documents and Systems}
\label{sec:adapt_framework}

Although our framework is evaluated in the context of ESCMs, it is designed for technical documents in general. Its modular design allows it to be adapted to other types of technical documents and industrial systems. Such adaptation, however, requires domain-specific customization.

The first step is to either reuse an existing KG schema or design one that reflects the relevant knowledge embedded in the target documents and system domain in a structured way. In this study, our KG schema is specifically tailored for ESCMs, capturing entities and relations that represent key configuration knowledge, including the attributes of configuration steps and their dependencies. This knowledge is valuable for downstream tasks such as test generation. For other domains, the entities and relations should be defined to reflect the domain’s specific concepts, terminology, and requirements. 

Second, the ExtrAgents should be adapted to the target domain by modifying their assigned roles and task definitions. In our experiments, the ExtrAgents are instructed to act as networking configuration assistants because the target documents describe Ethernet switch configurations. For other domains, this role should be changed accordingly. Such role customization helps guide the LLM toward the terminology, reasoning patterns, and information types that are relevant to the target domain.

Third, the extraction guidelines need to be redesigned to align with both the designed KG schema and the characteristics of the target documents. The guidelines should clearly specify which entities should be extracted and provide extraction rules. If the extraction guidelines are insufficient to guide the agent effectively, representative in-context examples should be provided. The examples should be carefully selected to cover representative extraction scenarios, such as straightforward and ambiguous cases, explicitly stated and implicitly inferred information, and simple and nested structures. These examples demonstrate how information should be extracted and represented across diverse document fragments, thereby helping reduce inconsistent or incomplete extractions.

Fourth, the EvalAgents should also be adapted to the new extraction tasks. Since EvalAgents assess the generated KGs against predefined evaluation guidelines, those guidelines must be revised to align with the new KG schema and extraction guidelines. 

Fifth, the stopping criteria for the EEI loop should be configured based on the target domain's quality requirements and resource constraints. In this study, we set the correctness threshold to 0.9 and the maximum number of EEI iterations to 3. When adapting the framework to other domains, these two parameters can be adjusted based on the desired level of extraction quality, the cost of LLM-based refinement, and the time budget of the application scenario. A higher threshold may lead to better extraction quality but trigger more refinement iterations, while a lower threshold may reduce computational cost but yield lower-quality KGs. Similarly, increasing the maximum number of iterations may allow the EEI loop to perform additional rounds of prompt refinement for outputs, but it also increases the overall refinement cost. Therefore, these parameters should be selected based on the trade-off between desired KG quality and available resources.

After the domain-specific KG schema, extraction prompts, evaluation prompts, and EEI stopping criteria are properly defined, the proposed framework can be reused to generate KGs from other technical documents. Therefore, the framework is not limited to ESCMs, but manual-specific components must be replaced or adapted to reflect the knowledge structure and documentation characteristics of the target domain. This adaptation process enables the same EEI loop to support KG generation for different systems and downstream engineering tasks.

\section{Study Design}

This section describes the design of our empirical study for evaluating the proposed approach. We first present the research questions addressed in this study, followed by the dataset used in the experiments. We then introduce the evaluation metric for assessing the correctness of the generated KG entities. Next, we describe the study process followed to answer each research question. Finally, we present the implementation details of our framework and experimental setup.

\subsection{Research Questions}

\noindent\textbf{RQ1: What is the extraction correctness score for each ESCM using the original prompt?}

This RQ aims to evaluate the correctness of KG entities generated by the three ExtrAgents (i.e., roadmap extraction agent, roadmap--procedure mapping agent, and procedure extraction agent) using the original prompts. The original prompts refer to the initial extraction prompts before any refinement by the EEI loop, allowing us to assess how well the approach performs without prompt tuning.

\noindent
\textbf{RQ2: How much does the EEI loop improve the extraction correctness score across ESCMs?}

This RQ aims to measure the improvement in correctness score achieved through the EEI loop, guided by LLM-as-a-Judge feedback and governed by predefined stopping criteria.

\noindent\textbf{RQ3: How consistent are extraction correctness scores assigned by the LLM-as-a-Judge with those assigned by human evaluators?} 

Our approach relies on an LLM-as-a-Judge approach to evaluate extraction quality and to guide improvements to the extraction process using its feedback. RQ3 aims to assess the alignment between the scores assigned by LLM-as-a-Judge and those assigned by human evaluators.

\noindent\textbf{RQ4: How effective are the generated KGs in supporting TCS generation?}

This RQ aims to evaluate the practical effectiveness of the generated KGs in a downstream testing task. Specifically, we use the generated KGs to derive system-level TCSs and ask testing experts to assess their quality. 

\subsection{Dataset}
Our dataset was drawn from the \emph{S300, S500, S2700, S5700, and S6700 Series Ethernet Switches Product Documentation} \footnote{\url{https://support.huawei.com/hedex/hdx.do?docid=EDOC1100333029&id=index}}, which contains approximately 1,500 ESCMs. 
These manuals are organized into four categories: (i) \emph{Typical Configuration Examples}, (ii) \emph{CLI-based Configuration Guide}, (iii) \emph{Web-based Configuration Guide}, and (iv) \emph{Security Hardening Guide}. We selected our dataset from the \emph{Typical Configuration Examples} because they provide the most representative configuration scenarios. Within this category, we further filtered ECSMs to those that include both a \emph{Configuration Roadmap} section and a \emph{Procedure} section, resulting in 208 ESCMs. The resulting ESCMs span 18 sub-categories (e.g., \emph{Typical Device Management Configuration}, \emph{Typical QoS Configuration}, and \emph{Typical Reliability Configuration}). Due to budget constraints (e.g., LLM inference cost and human evaluation effort), we selected a manageable subset of ESCMs for experimentation. Specifically, to ensure broad coverage across subcategories, we used proportional random sampling and selected 50 ESCMs, covering all 18 subcategories.

\subsection{Evaluation metrics}
We assess extraction quality both before and after the EEI loop using the \emph{overall correctness score}, as defined in Eq.~\ref{eq:correctness-score}.
This score is calculated as the proportion of correctly extracted JSON entries among all checked JSON entries across the $m$ evaluation guidelines (see Section~\ref{sec:eeiloop}), where each JSON entry in the extraction output corresponds to a KG entity.
It is important to note that, since the extracted KG entities are deterministically converted into triples by a script following the predefined KG schema, their correctness directly reflects the correctness of the resulting KG.

\subsection{Study Process}
\label{sec:study-process}

To answer the four research questions, we designed a study process comprising four main stages: KG extraction with the original prompts, prompt refinement through the EEI loop, human evaluation to validate the LLM-as-a-Judge results, and questionnaire-based evaluation of the TCSs derived from the generated KGs, as an intermediate artifact for downstream test case generation. 

\subsubsection{Process for RQ1}
To answer RQ1, we applied the three ExtrAgents to each ESCM using the original prompts, i.e., the prompts before any refinement by the EEI loop. 
For each ESCM and extraction task, the generated KG entities were evaluated by the EvalAgent using the task-specific evaluation guidelines. The EvalAgent checks whether the generated KG entities satisfy each guideline and, for each evaluation guideline, reports the number of correct and checked JSON entries in the generated JSON-format KG entities output. Based on this evaluation output, we calculated the overall correctness score for each ESCM and each extraction task. These scores were used to measure the extraction performance of the original prompts before any EEI improvement.

\subsubsection{Process for RQ2}
To answer RQ2, we applied the EEI loop to the ESCMs whose extraction correctness scores did not meet the predefined correctness threshold of 0.9. The loop terminated when the correctness score reached 0.9 or when the maximum number of iterations, set to 3, was reached.
To quantify the effectiveness of the EEI loop, we compared correctness scores from the original prompts with those from EEI-refined prompts. The improvement is measured as the difference between the final correctness score, obtained from the last EEI iteration, and the original correctness score for each ESCM and each extraction task.

\subsubsection{Process for RQ3}
To answer RQ3, we conducted a human evaluation on all 50 ESCMs across the three extraction tasks to assess the reliability of the LLM-as-a-Judge evaluation results. For each extraction task and each ECSM, two authors independently evaluated the quality of the generated KG entities. Specifically, for each extraction task, human evaluators assessed the generated KG entities against the same set of evaluation guidelines used by the LLM-as-a-Judge. For each guideline, the human evaluators recorded the evaluation result in the same JSON format used by the LLM-as-a-Judge, including the binary score, the number of checked and correct JSON entries, and the reasons for any detected errors. By using the same evaluation guidelines, scoring criteria, and output format across the three extraction tasks, we ensure that the human evaluation results are directly comparable to those produced by the LLM-as-a-Judge. Any discrepancies were discussed and resolved until full consensus was reached. We calculated Cohen's kappa statistic to assess agreement between the two human raters. The resulting scores ranged from 0.90 to 0.96 across three extraction tasks, indicating strong agreement between the two raters.

 We then compared the LLM-as-a-Judge judgments with the human judgments across ESCMs. To quantify agreement between the two evaluation methods, we also calculated Cohen's kappa for each extraction task by aggregating labels across all evaluation guidelines. This allows us to assess whether the LLM-as-a-Judge can produce evaluation results consistent with those of human evaluators. Moreover, we analyzed the discrepancies between the two evaluation methods to understand the types of mismatches.

\subsubsection{Process for RQ4}

To answer RQ4, we evaluated whether the generated KGs effectively support an important downstream testing task: test case specification (TCS) generation. We selected five representative ESCMs covering various sizes, structures, and levels of complexity and converted the corresponding generated KGs into TCSs. We then designed a questionnaire to collect feedback from five testers from our industry partner. All of the testers have sufficient levels of experience in testing. It is important to note that, to ensure the reliability of the questionnaire results, we specifically prioritized industry testers with relevant experience from our Ethernet switch industry partner. While this focus is challenging and limits the number of responses, it strengthens the credibility of our findings. Consequently, we collected five questionnaire responses. The respondents have 1 to 16 years of general engineering experience, with an average of 4.4 years. Their experience in Ethernet switch testing ranges from 0 to 4 years, with an average of 1.8 years. Notably, most respondents (4 out of 5) have direct experience in Ethernet switch testing. These respondents therefore provide a relevant industrial perspective for assessing the usefulness and quality of the KG-derived TCSs.
The questionnaire asked the testers to assess the quality and practical usefulness of the generated TCSs.

\noindent\textbf{Questionnaire Design.}
Table~\ref{tab:questionnaire} provides an overview of the questionnaire, including question categories and the answer types. At the beginning of the questionnaire, we first collected respondents’ background information, including their years of testing or engineering experience and their experience with Ethernet switch testing. 
The remaining questionnaire questions are organized into six categories. The practical usefulness questions (category A) examine whether the TCS can support test generation in practice. The clarity and understandability questions (category B) focus on whether the TCS is easy to understand and whether the test steps are clear and unambiguous. The completeness questions (category C) assess whether the TCS contains the necessary information for test generation. The correctness and technical accuracy questions (category D) evaluate whether the TCS, test steps, and expected results are technically correct. The overall quality question (category E) asks respondents to rate the extent to which they would recommend the TCS for supporting test generation in practice. The open-ended feedback question (category F) allows respondents to provide additional comments, suggested improvements, or identified issues.

Categories A--D are measured using a five-point Likert scale: \emph{strongly disagree}, \emph{disagree}, \emph{neither agree nor disagree}, \emph{agree}, and \emph{strongly agree}. Category E uses a five-point rating scale ranging from \emph{very low quality} (score = 1) to \emph{very high quality} (score = 5). Category F is an open-ended question used to collect qualitative feedback from respondents.

\begin{table*}[t]
\centering
\caption{Questionnaire designed for evaluating the generated test case specifications.}
\label{tab:questionnaire}
\small
\begin{tabular}{p{0.12\linewidth} p{0.08\linewidth} p{0.40\linewidth} p{0.14\linewidth}}
\hline
\textbf{Category} & \textbf{Question number} & \textbf{Question text} & \textbf{Answer type} \\
\hline
Practical usefulness (A) 
& A1 & The specification is useful for test generation. & 5-point Likert scale \\
& A2 & The specification reduces manual effort. & 5-point Likert scale \\
& A3 & The specification can be directly used or adapted. & 5-point Likert scale \\
& A4 & The specification captures dependencies between configuration steps. & 5-point Likert scale \\
\hline
Clarity and understandability (B) 
& B1 & The test specification is easy to understand. & 5-point Likert scale \\
& B2 & The steps are clear and unambiguous. & 5-point Likert scale \\
\hline
Completeness (C) 
& C1 & The test specification includes all necessary steps. & 5-point Likert scale \\
& C2 & Preconditions are sufficient. & 5-point Likert scale \\
& C3 & Expected results are clearly defined. & 5-point Likert scale \\
\hline
Correctness and technical accuracy (D) 
& D1 & The test specification is technically correct. & 5-point Likert scale \\
& D2 & The test steps are precisely described. & 5-point Likert scale \\
& D3 & Expected results are valid. & 5-point Likert scale \\
\hline
Overall quality (E) 
& E1 & To what extent would you recommend this test specification to support test generation in practice? & 5-point rating scale \\

\hline
Feedback (F) 
& F1 & Please provide any additional comments, suggested improvements, or identified issues, such as ambiguities, contradictions, or missing information. & Free text \\

\hline
\end{tabular}
\end{table*}

\noindent\textbf{Data Analysis.}
For the Likert-scale questions, we converted the ratings into numerical scores ranging from 1 to 5, where \emph{strongly disagree} was assigned a score of 1, \emph{disagree} a score of 2, \emph{neither agree nor disagree} a score of 3, \emph{agree} a score of 4, and \emph{strongly agree} a score of 5. Higher scores, therefore, indicate more positive evaluations. 
For each question category, we aggregated all ratings from the corresponding questions across the five evaluated TCSs and all five respondents, and then calculated the average, minimum, and maximum scores. We also calculated the positive rating rate as the proportion of ratings with a score of 4 or 5, corresponding to \emph{agree} or \emph{strongly agree} for the Likert-scale questions.
For the overall quality question, which was rated directly on a 1–5 scale, we aggregated the ratings across the five TCSs and all five respondents. For the open-ended responses, we reviewed the additional comments to supplement the quantitative results and to identify the strengths and possible areas for improvement in the generated TCSs. Together, these quantitative and qualitative results provide preliminary evidence of whether the KG-derived TCSs are perceived as useful, understandable, complete, technically correct, and practical for downstream test case generation.

\subsection{Implementation}

Our approach is implemented in Python, leveraging the LangChain~\cite{2025LangchainWebsite} library for LLM integration. Model inference is performed via the OpenAI API~\cite{OpenAI_API_Reference} using GPT-5 with the default configuration across all experiments for KG extraction, evaluation, and improvement. We choose GPT-5 for its strong instruction-following capabilities and ability to produce consistent, reliable outputs~\cite{openai2025gpt5}, which are critical for faithful extraction under strict formatting constraints. 
The pipeline is executed on a MacBook Pro equipped with an Apple M4 chip and 16 GB of RAM, running macOS Sequoia.

\section{Results}
In this section, we report and analyze the results related to our research questions and discuss their practical implications. 

\begin{table}[]
\centering
\caption{\textbf{Roadmap extraction evaluation results}}
\label{tab:task_1_results}
\begin{threeparttable}
\resizebox{0.99\linewidth}{!}{
\begin{tabular}{@{}*{5}{ccccc}@{}}
\toprule
\multirow{2}{*}{\textbf{No}} &
\multirow{2}{*}{\textbf{Original}} &
\multicolumn{2}{c}{\textbf{EEI iter}} &
\multirow{2}{*}{\textbf{$\Delta$ Corr.}} &
\multirow{2}{*}{\textbf{No}} &
\multirow{2}{*}{\textbf{Original}} &
\multicolumn{2}{c}{\textbf{EEI iter}} &
\multirow{2}{*}{\textbf{$\Delta$ Corr.}} &
\multirow{2}{*}{\textbf{No}} &
\multirow{2}{*}{\textbf{Original}} &
\multicolumn{2}{c}{\textbf{EEI iter}} &
\multirow{2}{*}{\textbf{$\Delta$ Corr.}} &
\multirow{2}{*}{\textbf{No}} &
\multirow{2}{*}{\textbf{Original}} &
\multicolumn{2}{c}{\textbf{EEI iter}} &
\multirow{2}{*}{\textbf{$\Delta$ Corr.}} &
\multirow{2}{*}{\textbf{No}} &
\multirow{2}{*}{\textbf{Original}} &
\multicolumn{2}{c}{\textbf{EEI iter}} &
\multirow{2}{*}{\textbf{$\Delta$ Corr.}} \\
\cmidrule(lr){3-4}\cmidrule(lr){8-9}\cmidrule(lr){13-14}\cmidrule(lr){18-19}\cmidrule(lr){23-24}
 &  & \textbf{1} & \textbf{2} &  &
   &  & \textbf{1} & \textbf{2} &  &
   &  & \textbf{1} & \textbf{2} &  &
   &  & \textbf{1} & \textbf{2} &  &
   &  & \textbf{1} & \textbf{2} &  \\
\midrule
1  & 0.91 &      &      & 0.00 & 11 & 0.93 &      &      & 0.00 & 21 & 1.00 &      &      & 0.00 & 31 & 1.00 &      &      & 0.00 & 41 & 1.00 &      &      & 0.00 \\
2  & 1.00 &      &      & 0.00 & 12 & 1.00 &      &      & 0.00 & 22 & 1.00 &      &      & 0.00 & 32 & 1.00 &      &      & 0.00 & 42 & 1.00 &      &      & 0.00 \\
3  & 1.00 &      &      & 0.00 & 13 & 1.00 &      &      & 0.00 & 23 & 1.00 &      &      & 0.00 & 33 & 1.00 &      &      & 0.00 & 43 & 1.00 &      &      & 0.00 \\
4  & 1.00 &      &      & 0.00 & 14 & 1.00 &      &      & 0.00 & 24 & 0.97 &      &      & 0.00 & 34 & 1.00 &      &      & 0.00 & 44 & 1.00 &      &      & 0.00 \\
5  & 0.92 &      &      & 0.00 & 15 & 0.88 & 1.00 &      & 0.12 & 25 & 1.00 &      &      & 0.00 & 35 & 1.00 &      &      & 0.00 & 45 & 1.00 &      &      & 0.00 \\
6  & 1.00 &      &      & 0.00 & 16 & 1.00 &      &      & 0.00 & 26 & 0.95 &      &      & 0.00 & 36 & 0.97 &      &      & 0.00 & 46 & 1.00 &      &      & 0.00 \\
7  & 1.00 &      &      & 0.00 & 17 & 1.00 &      &      & 0.00 & 27 & 1.00 &      &      & 0.00 & 37 & 1.00 &      &      & 0.00 & 47 & 0.94 &      &      & 0.00 \\
8  & 0.91 &      &      & 0.00 & 18 & 0.97 &      &      & 0.00 & 28 & 1.00 &      &      & 0.00 & 38 & 0.90 &      &      & 0.00 & 48 & 1.00 &      &      & 0.00 \\
9  & 1.00 &      &      & 0.00 & 19 & 1.00 &      &      & 0.00 & 29 & 1.00 &      &      & 0.00 & 39 & 1.00 &      &      & 0.00 & 49 & 1.00 &      &      & 0.00 \\
10 & 1.00 &      &      & 0.00 & 20 & 1.00 &      &      & 0.00 & 30 & 1.00 &      &      & 0.00 & 40 & 0.89 & 1.00 &      & 0.11 & 50 & 1.00 &      &      & 0.00 \\
\bottomrule
\end{tabular}
}
\begin{tablenotes}[flushleft]
\footnotesize
\item \textit{Original} denotes the correctness score obtained using the original extraction prompt (i.e., before applying any EEI iteration).
\item \textit{Iterations 1--2} denote the correctness scores after the 1st--2nd EEI iterations, respectively.
\item \textit{$\Delta$ Corr.} denotes the change in correctness score relative to ``Original''.
\end{tablenotes}
\end{threeparttable}
\end{table}

\begin{table}[]
\centering
\caption{\textbf{Roadmap–Procedure Mapping Evaluation Results}}
\label{tab:task_2_results}
\begin{threeparttable}
\resizebox{0.99\linewidth}{!}{
\begin{tabular}{@{}*{5}{ccccc}@{}}
\toprule
\multirow{2}{*}{\textbf{No}} &
\multirow{2}{*}{\textbf{Original}} &
\multicolumn{2}{c}{\textbf{EEI iter}} &
\multirow{2}{*}{\textbf{$\Delta$ Corr.}} &
\multirow{2}{*}{\textbf{No}} &
\multirow{2}{*}{\textbf{Original}} &
\multicolumn{2}{c}{\textbf{EEI iter}} &
\multirow{2}{*}{\textbf{$\Delta$ Corr.}} &
\multirow{2}{*}{\textbf{No}} &
\multirow{2}{*}{\textbf{Original}} &
\multicolumn{2}{c}{\textbf{EEI iter}} &
\multirow{2}{*}{\textbf{$\Delta$ Corr.}} &
\multirow{2}{*}{\textbf{No}} &
\multirow{2}{*}{\textbf{Original}} &
\multicolumn{2}{c}{\textbf{EEI iter}} &
\multirow{2}{*}{\textbf{$\Delta$ Corr.}} &
\multirow{2}{*}{\textbf{No}} &
\multirow{2}{*}{\textbf{Original}} &
\multicolumn{2}{c}{\textbf{EEI iter}} &
\multirow{2}{*}{\textbf{$\Delta$ Corr.}} \\
\cmidrule(lr){3-4}\cmidrule(lr){8-9}\cmidrule(lr){13-14}\cmidrule(lr){18-19}\cmidrule(lr){23-24}
 &  & \textbf{1} & \textbf{2} &  &
   &  & \textbf{1} & \textbf{2} &  &
   &  & \textbf{1} & \textbf{2} &  &
   &  & \textbf{1} & \textbf{2} &  &
   &  & \textbf{1} & \textbf{2} &  \\
\midrule
1  & 0.90 &      &      & 0.00 & 11 & 1.00 &      &      & 0.00 & 21 & 1.00 &      &      & 0.00 & 31 & 1.00 &      &      & 0.00 & 41 & 1.00 &      &      & 0.00 \\
2  & 1.00 &      &      & 0.00 & 12 & 0.97 &      &      & 0.00 & 22 & 1.00 &      &      & 0.00 & 32 & 1.00 &      &      & 0.00 & 42 & 1.00 &      &      & 0.00 \\
3  & 1.00 &      &      & 0.00 & 13 & 1.00 &      &      & 0.00 & 23 & 0.98 &      &      & 0.00 & 33 & 1.00 &      &      & 0.00 & 43 & 1.00 &      &      & 0.00 \\
4  & 1.00 &      &      & 0.00 & 14 & 1.00 &      &      & 0.00 & 24 & 0.98 &      &      & 0.00 & 34 & 0.97 &      &      & 0.00 & 44 & 1.00 &      &      & 0.00 \\
5  & 1.00 &      &      & 0.00 & 15 & 1.00 &      &      & 0.00 & 25 & 0.79 & 0.97 &      & 0.18 & 35 & 1.00 &      &      & 0.00 & 45 & 1.00 &      &      & 0.00 \\
6  & 1.00 &      &      & 0.00 & 16 & 1.00 &      &      & 0.00 & 26 & 0.95 &      &      & 0.00 & 36 & 1.00 &      &      & 0.00 & 46 & 1.00 &      &      & 0.00 \\
7  & 1.00 &      &      & 0.00 & 17 & 0.93 &      &      & 0.00 & 27 & 0.88 & 0.79 & 1.00 & 0.12 & 37 & 1.00 &      &      & 0.00 & 47 & 0.92 &      &      & 0.00 \\
8  & 1.00 &      &      & 0.00 & 18 & 1.00 &      &      & 0.00 & 28 & 1.00 &      &      & 0.00 & 38 & 0.89 & 1.00 &      & 0.11 & 48 & 0.97 &      &      & 0.00 \\
9  & 1.00 &      &      & 0.00 & 19 & 0.98 &      &      & 0.00 & 29 & 1.00 &      &      & 0.00 & 39 & 1.00 &      &      & 0.00 & 49 & 0.91 &      &      & 0.00 \\
10 & 1.00 &      &      & 0.00 & 20 & 0.95 &      &      & 0.00 & 30 & 0.81 & 1.00 &      & 0.19 & 40 & 1.00 &      &      & 0.00 & 50 & 0.84 & 0.83 & 0.97 & 0.13 \\
\bottomrule
\end{tabular}
}
\begin{tablenotes}[flushleft]
\footnotesize
\item \textit{Original} denotes the correctness score obtained using the original extraction prompt (i.e., before applying any EEI iteration).
\item \textit{Iterations 1--2} denote the correctness scores after the 1st--2nd EEI iterations, respectively.
\item \textit{$\Delta$ Corr.} denotes the change in correctness score relative to ``Original''.
\end{tablenotes}
\end{threeparttable}
\end{table}

\begin{table}[]
\centering
\caption{\textbf{Procedure Extraction Evaluation Results}}
\label{tab:task_3_results}
\begin{threeparttable}
\resizebox{0.99\linewidth}{!}{
\begin{tabular}{@{}*{5}{ccccc}@{}}
\toprule
\multirow{2}{*}{\textbf{No}} &
\multirow{2}{*}{\textbf{Original}} &
\multicolumn{2}{c}{\textbf{EEI iter}} &
\multirow{2}{*}{\textbf{$\Delta$ Corr.}} &
\multirow{2}{*}{\textbf{No}} &
\multirow{2}{*}{\textbf{Original}} &
\multicolumn{2}{c}{\textbf{EEI iter}} &
\multirow{2}{*}{\textbf{$\Delta$ Corr.}} &
\multirow{2}{*}{\textbf{No}} &
\multirow{2}{*}{\textbf{Original}} &
\multicolumn{2}{c}{\textbf{EEI iter}} &
\multirow{2}{*}{\textbf{$\Delta$ Corr.}} &
\multirow{2}{*}{\textbf{No}} &
\multirow{2}{*}{\textbf{Original}} &
\multicolumn{2}{c}{\textbf{EEI iter}} &
\multirow{2}{*}{\textbf{$\Delta$ Corr.}} &
\multirow{2}{*}{\textbf{No}} &
\multirow{2}{*}{\textbf{Original}} &
\multicolumn{2}{c}{\textbf{EEI iter}} &
\multirow{2}{*}{\textbf{$\Delta$ Corr.}} \\
\cmidrule(lr){3-4}\cmidrule(lr){8-9}\cmidrule(lr){13-14}\cmidrule(lr){18-19}\cmidrule(lr){23-24}
 &  & \textbf{1} & \textbf{2} &  &
   &  & \textbf{1} & \textbf{2} &  &
   &  & \textbf{1} & \textbf{2} &  &
   &  & \textbf{1} & \textbf{2} &  &
   &  & \textbf{1} & \textbf{2} &  \\
\midrule
\midrule
1  & 0.99 &      &      & 0.00 & 11 & 1.00 &      &      & 0.00 & 21 & 1.00 &      &      & 0.00 & 31 & 1.00 &      &      & 0.00 & 41 & 0.96 &      &      & 0.00 \\
2  & 1.00 &      &      & 0.00 & 12 & 1.00 &      &      & 0.00 & 22 & 0.99 &      &      & 0.00 & 32 & 1.00 &      &      & 0.00 & 42 & 0.98 &      &      & 0.00 \\
3  & 1.00 &      &      & 0.00 & 13 & 1.00 &      &      & 0.00 & 23 & 1.00 &      &      & 0.00 & 33 & 0.99 &      &      & 0.00 & 43 & 1.00 &      &      & 0.00 \\
4  & 1.00 &      &      & 0.00 & 14 & 1.00 &      &      & 0.00 & 24 & 1.00 &      &      & 0.00 & 34 & 1.00 &      &      & 0.00 & 44 & 0.98 &      &      & 0.00 \\
5  & 0.99 &      &      & 0.00 & 15 & 1.00 &      &      & 0.00 & 25 & 1.00 &      &      & 0.00 & 35 & 0.96 &      &      & 0.00 & 45 & 1.00 &      &      & 0.00 \\
6  & 1.00 &      &      & 0.00 & 16 & 1.00 &      &      & 0.00 & 26 & 1.00 &      &      & 0.00 & 36 & 1.00 &      &      & 0.00 & 46 & 1.00 &      &      & 0.00 \\
7  & 0.96 &      &      & 0.00 & 17 & 1.00 &      &      & 0.00 & 27 & 1.00 &      &      & 0.00 & 37 & 0.99 &      &      & 0.00 & 47 & 0.99 &      &      & 0.00 \\
8  & 1.00 &      &      & 0.00 & 18 & 0.98 &      &      & 0.00 & 28 & 1.00 &      &      & 0.00 & 38 & 0.99 &      &      & 0.00 & 48 & 1.00 &      &      & 0.00 \\
9  & 1.00 &      &      & 0.00 & 19 & 0.97 &      &      & 0.00 & 29 & 1.00 &      &      & 0.00 & 39 & 0.99 &      &      & 0.00 & 49 & 1.00 &      &      & 0.00 \\
10 & 1.00 &      &      & 0.00 & 20 & 0.96 &      &      & 0.00 & 30 & 0.99 &      &      & 0.00 & 40 & 1.00 &      &      & 0.00 & 50 & 1.00 &      &      & 0.00 \\
\bottomrule
\end{tabular}
}
\begin{tablenotes}[flushleft]
\footnotesize
\item \textit{Original} denotes the correctness score obtained using the original extraction prompt (i.e., before applying any EEI iteration).
\item \textit{Iterations 1--2} denote the correctness scores after the 1st--2nd EEI iterations, respectively.
\item \textit{$\Delta$ Corr.} denotes the change in correctness score relative to ``Original''.
\end{tablenotes}
\end{threeparttable}
\end{table}

\subsection{Extraction Correctness using the original prompts (RQ1)}
\label{sec:rq1}
Tables~\ref{tab:task_1_results}, \ref{tab:task_2_results}, and~\ref{tab:task_3_results} report the correctness scores for roadmap extraction, roadmap–procedure mapping, and procedure extraction across 50 ESCMs. The scores were assigned by an LLM-as-a-Judge based on the extraction outputs produced with the original prompts (i.e., the “\textit{Original}” column).

\noindent\textbf{Roadmap Extraction} Table~\ref{tab:task_1_results} shows that the original prompt already yields strong extraction performance for roadmap extraction, with correctness scores ranging from 0.88 to 1.00 (mean~=~0.98, median~=~1.00), and 38 out of 50 ESCMs achieve a perfect score of 1.00.

Among the roadmap extraction results that contain errors (with correctness scores below 1.00), most errors fall under two evaluation guidelines: Note Extraction and Goal Extraction. Specifically, Note Extraction violations occur in 11/50 ESCMs (22\%), while Goal Extraction violations are observed in 5/50 ESCMs (10\%). No violations are observed for the remaining evaluation guidelines (i.e., Context Identification, Format Compliance, Numbering, Step Splitting, and Verbatim Copying).

These errors likely arise because goals and notes are often embedded in roadmap step descriptions and expressed in diverse, sometimes ambiguous ways, which can confuse the LLM.
For example, in ESCMs, goals are often introduced by purpose clauses that begin with ``to'' (e.g., ``to implement interworking in the OSPF network''). However, the LLM may over-trigger on this pattern and incorrectly label non-purpose phrases as goals, such as ``to the inbound direction of GE0/0/3 on the Switch'', which actually describes an action detail. In addition, the LLM may misclassify step action details or execution scope as notes. For instance, extracting ``on SwitchA and SwitchB'' as a note, even though it specifies where the step should be applied. These cases suggest that the extraction prompt would benefit from a more precise definition of goals and notes (especially when they appear inside step descriptions), and certain ESCMs may require additional, manual-specific clarifications to reduce such ambiguities.

\noindent\textbf{Roadmap–Procedure Mapping} According to Table~\ref{tab:task_2_results}, the original extraction prompt for roadmap–procedure mapping also yields strong performance, with correctness scores ranging from 0.79 to 1.00 (mean~=~0.97, median~=~1.00), and 33 out of 50 ESCMs achieve a perfect score of 1.00.

Among the roadmap–procedure mapping extraction results that contain errors (i.e., with correctness scores below 1.00), the violations fall under six evaluation guidelines: Text Completeness (9/50 ESCMs, 18\%) and Relevant Step Match (6/50 ESCMs, 12\%), followed by Multiple Match Inclusion (4/50 ESCMs, 8\%) and Structural Format (3/50 ESCMs, 6\%). There are also a few violations of Step Numbering Compliance and Device Identifier Consistency, each occurring in only 1/50 ESCMs (2\%).  We found that most Text Completeness and Structural Format violations (9/50 ESCMs, 18\%) appear to be false positives from the EvalAgent. Specifically, the EvalAgent sometimes reports that there are minor copying mistakes, such as copying \textit{``**quit**''} as \textit{``**quit\textbackslash''}, which in turn can trigger Structural Format violations by introducing unescaped quotation marks (i.e., \textit{``\textbackslash''}) that break JSON string literals. However, upon careful manual inspection, we confirm that the extracted content is correctly copied. This indicates that the EvalAgent may occasionally hallucinate when recognizing strings containing special characters, such as Markdown markers (e.g., ``**'').
In addition, many violations of Relevant Step Match and Multiple Match Inclusion arise from incorrect mappings of roadmap steps and procedure verification steps. These errors suggest that the extraction prompt could be improved by providing more precise, rule-based guidance for mapping steps. Moreover, the evaluation results provide detailed analyses of mapping failures, for example, "Roadmap Step 3 was mapped to Procedure Step 4 (Verify the configuration). Step 4 does not verify edge-port status on SwitchC/SwitchD and is not directly related to configuring edge ports, so it is not a truly relevant match for Roadmap Step 3." Such feedback serves as valuable guidance for further refining the extraction prompt.

\noindent\textbf{Procedure Extraction}. 
Table~\ref{tab:task_3_results} shows that, using the original prompt, the procedure extraction agent produces consistently high-quality results across all 50 ESCMs, with the correctness scores ranging from 0.96 to 1.00 (mean~=~0.99, median~=~1.00), and 33 out of 50 ESCMs achieve a perfect score of 1.00.

Among the procedure extraction results that contain errors (with correctness scores below 1.00), most errors fall under four evaluation guidelines: Expected Output Extraction Correctness (11/50 ESCMs, 22\%), Text Completeness \& Verbatim Copy (10/50 ESCMs, 20\%), Command Extraction Correctness (8/50 ESCMs, 16\%), Note Classification \& Attachment Correctness (3/50 ESCMs, 6\%). No violations are observed for other evaluation guidelines (i.e., Step Coverage, Structural Format \& Schema Compliance, and Step Numbering Compliance). 
We observed that most errors occur when commands or expected outputs are embedded within step descriptions, rather than appearing as isolated command/output blocks. For example, when the command ``Run the **display ip routing-table** command on PE1 and PE2'' appears within the step description, it was misclassified as expected output. Moreover, the text ``If the status is **effective** , the static entry has taken effect.'' is embedded in the step description and should be extracted as the expected output, but the LLM failed to recognize it. However, these errors do not lead to information loss, since both the command and the expected output information are still preserved in the step content in the extracted output. 
We also observe minor copying and formatting errors. For example, the LLM may omit leading spaces or line breaks during extraction, and it may occasionally introduce small spelling mistakes (e.g., misspelling ``behavior'' as ``bhehavior''). While such issues reduce strict textual alignment with the source, they generally do not remove critical content.

\begin{tcolorbox}
\textbf{Answer to RQ1: The original prompts used in our approach already achieve strong performance for KG entity extraction, with mean correctness scores ranging from 0.97 to 0.99 across three extraction tasks on 50 ESCMs. Certain ESCMs require additional manual-specific refinements in the prompts to further improve the quality of extracted KG entities. }   
\end{tcolorbox}

\subsection{Extraction Correctness after the EEI Loop (RQ2)}
Tables~\ref{tab:task_1_results}, \ref{tab:task_2_results}, and~\ref{tab:task_3_results} report the correctness scores across EEI iterations, together with the corresponding improvement in correctness score (\(\Delta\)Corr) for roadmap extraction, roadmap–procedure mapping, and procedure extraction over 50 ESCMs.

Across the three tasks, EEI was triggered for only a small subset of ESCMs. Specifically, 2 out of 50 ESCMs required EEI for roadmap extraction, 5 out of 50 for roadmap–procedure mapping, and none for procedure extraction. Among the ESCMs that triggered EEI, the average correctness gain (\(\Delta\)Corr) was 0.12 for roadmap extraction and 0.15 for roadmap–procedure mapping. These results suggest that the original prompts already perform strongly in most cases, while EEI yields substantial improvements for the small subset of ESCMs that require additional refinement.

\noindent\textbf{Roadmap Extraction} Table~\ref{tab:task_1_results} shows that, given the threshold of 0.9, EEI is applied to only 2 out of 50 ESCMs (i.e., No.~15 and No.~40), indicating that the EEI loop is rarely needed for this task and is sufficient to fully correct all roadmap extraction cases whose initial correctness scores fall below the threshold. For ESCM 15, the correctness score increases from 0.88 to 1.00 after the first EEI iteration (\(\Delta\)Corr~=~0.12). For ESCM 40, the score increases from 0.89 to 1.00 after the first EEI iteration (\(\Delta\)Corr~=~0.11). 
The errors in both the ESCMs 15 and 40 fall under Note Extraction. The EEI loop leverages feedback from the EvalAgent to refine the extraction prompt. For example, the evaluation agent reports: ``Step 1 note misclassified: `for interfaces connecting SwitchA and SwitchB' is part of the main action target, not a clarification/note.''
Based on this feedback, the ImprovAgent refines the extraction prompt by adding explicit note-extraction rules, such as: ``Do not extract as notes any phrases that specify where/what is being configured (scope/targets), such as device/interface lists.'' This case suggests that the EEI loop can adapt the prompt to manual-specific error patterns, thereby improving extraction quality.

\noindent\textbf{Roadmap–Procedure Mapping} Table~\ref{tab:task_2_results} shows that the EEI loop is only applied to a subset of ESCMs (i.e., 5 out of 50) for roadmap--procedure mapping task. Among these five ESCMs, three (Nos.~25, 30, and 38) are corrected after the first EEI iteration, while the remaining two (Nos.~27 and 50) require a second iteration to achieve further improvement.
Most errors in these ESCMs fall under the Text Completeness guideline. However, as discussed in Section~\ref{sec:rq1}, many of these violations appear to be false positives caused by occasional hallucinations of the evaluation agent.

Moreover, for ESCMs 27 and 50, the correctness score decreases after the first EEI iteration. This drop is caused by new errors introduced during regeneration, which are then corrected in the second EEI iteration, resulting in higher correctness scores. This indicates the need for multi-iteration EEI, which can detect and fix regression errors introduced in earlier iterations, thereby improving stability.

\noindent\textbf{Procedure Extraction}
Table~\ref{tab:task_3_results} shows that the original prompt already achieves consistently strong performance for procedure extraction, with all ESCMs exceeding the threshold of 0.9, and therefore none of them triggers the EEI loop. 
One plausible reason is that procedure extraction usually involves a larger number of KG entities than roadmap extraction and roadmap–procedure mapping. As a result, one or two incorrect entities have a relatively smaller impact on the overall correctness score, whereas the same number of errors can cause a larger score decrease in tasks with fewer extracted entities. 
Moreover, procedure extraction has 17 ESCMs with a correctness score below 1.00, indicating that some ESCMs still contain small issues. 
However, as discussed in Section~\ref{sec:rq1}, these errors are mostly minor, such as slight misclassification or small formatting/copying errors. The relevant knowledge is still extracted into the KG extraction output, but it may be assigned to a less appropriate field or represented with minor formatting inconsistencies. Therefore, these errors typically do not lead to missing key information from the generated KG. As a result, even without further improvement from EEI, the original prompt already yields sufficiently high-quality procedure-extraction outputs for downstream use.

\begin{tcolorbox}
\textbf{Answer to RQ2: The EEI loop is triggered for the few most challenging ESCMs that require manual-specific refinements to the extraction prompts, leading to substantial improvements in their correctness scores.}   
\end{tcolorbox}

\subsection{Consistency of LLM and Human Judgment on Extraction Correctness (RQ3)}

The results show that, for the three extraction tasks, Cohen’s kappa scores range from 0.72 to 0.80, indicating substantial agreement between human evaluation and the LLM-as-a-Judge.

\noindent\textbf{Roadmap Extraction}
For roadmap extraction, the agreement between human evaluation and the LLM-as-a-Judge is 0.80 (Cohen's kappa), indicating substantial agreement. The main source of disagreement lies in the extraction of notes and goals. In some cases, the LLM treats explanatory phrases within the main action, such as “in delay mode” or “including AC system parameters and AP management”, as notes. Similarly, it sometimes interprets phrases such as “based on which the switch permits access of the NMS” as goals. However, from the human evaluators' perspective, these phrases are better understood as part of the explanatory content embedded in the main action itself, rather than as standalone notes or explicit goals.

Importantly, the impact of these disagreements on the quality of the extracted KG entities is limited. Even when such phrases are misclassified as notes or goals, the underlying information is still preserved in the extracted output. 
Moreover, some phrases are semantically close to both the main action and the associated goal or note fields, making their categorization less clear.
Therefore, many of these disagreements are better interpreted as differences in categorization criteria, rather than as substantial extraction errors.

\noindent\textbf{Roadmap–Procedure Mapping}
For roadmap–procedure mapping, the agreement between human evaluation and the LLM-as-a-Judge is 0.76 (Cohen’s kappa), indicating substantial agreement. The main source of disagreement is that some procedure steps are partially or indirectly related to the corresponding roadmap step. In these cases, the original KG extraction is considered correct by the human evaluator because the selected procedure steps still contributed to the implementation or validation of the roadmap step. However, the LLM-as-a-Judge sometimes treats such matches as irrelevant and marks them as incorrect.

In addition, a small number of disagreements come from very minor issues, such as extra copied spaces, as well as hallucinated error reports from the LLM-as-a-Judge. For example, in some cases, symbols such as \textit{**} are actually copied correctly, but the LLM-as-a-Judge incorrectly judges them as missing.

Overall, these disagreement cases have a limited impact on the validity of the extracted KG entities. A partially or indirectly linked procedure step still contributes useful contextual or operational information for understanding how a roadmap step is executed in practice. Even when the LLM-as-a-Judge marks these mappings as incorrect, the extracted KG entities often still retains the correct high-level correspondence between roadmap and procedure content. Therefore, many of these disagreements are better interpreted as the result of a stricter or more conservative evaluation criterion used by the LLM-as-a-Judge, rather than as evidence of real mapping errors.

\noindent\textbf{Procedure Extraction}
For procedure extraction, the agreement between human evaluation and the LLM-as-a-Judge is 0.72 according to Cohen’s kappa, indicating substantial agreement. Most disagreements involve relatively minor issues rather than major extraction failures, such as minor deviations in verbatim copying and formatting inconsistencies. For instance, some discrepancies arise when the extracted KG entities omits spaces or line breaks, or when command fields include additional words such as “run” or “or”. The LLM-as-a-Judge tends to regard these additions as deviations from the original command lines and thus marks them as errors, whereas the human evaluators consider them minor variations that preserve the original technical meaning and do not affect the practical usefulness of the extracted KG.

Another source of disagreement comes from hallucinations by the LLM-as-a-Judge. In some cases, it incorrectly reports missing symbols (e.g., **) or spelling errors that are not actually present in the extracted output. These false error reports explain part of the inconsistency between the two evaluation methods.

Only two disagreement cases are related to real extraction errors that are overlooked by the LLM-as-a-Judge. These cases occur when the ESCMs contain parameter-setting information that the extraction agent sometimes mistakenly treats as substeps. Although this led to incorrect KG extraction, the LLM-as-a-Judge did not identify the problem.

\begin{tcolorbox}
\textbf{Answer to RQ3: Across all three tasks, human evaluation and the LLM-as-a-Judge show substantial agreement, with Cohen’s kappa scores above 0.72. Most disagreements stem from differences in categorization criteria or minor copying and formatting issues, rather than major errors that substantially affect the correctness or usefulness of the extracted KG.}   
\end{tcolorbox}

\subsection{Effectiveness of Generated KGs in Supporting Test Case Specification Generation (RQ4)}

\begin{table}[t]
\centering
\caption{Questionnaire results for evaluating KG-derived TCSs.}
\label{tab:rq4_questionnaire_results}
\begin{tabular}{lcccc}
\hline
\textbf{Evaluation Categories} & \textbf{Avg. Score} & \textbf{Min} & \textbf{Max} & \textbf{Positive Ratings} \\
\hline
Practical usefulness & 4.17 & 3 & 5 & 99.0\% \\
Clarity and understandability & 4.24 & 3 & 5 & 96.0\% \\
Completeness & 4.12 & 2 & 5 & 88.0\% \\
Correctness and technical accuracy & 4.23 & 3 & 5 & 97.3\% \\
Overall quality & 4.36 & 4 & 5 & 100.0\% \\
\hline
\textbf{Overall} & \textbf{4.18} & -- & -- & \textbf{95.3\%} \\
\hline
\end{tabular}
\vspace{1mm}
\begin{flushleft}
\footnotesize
Note: The overall row is calculated by aggregating ratings across the four evaluation categories: practical usefulness, clarity and understandability, completeness, and correctness and technical accuracy. The overall quality rating is reported separately and is not included in the overall row calculation.
\end{flushleft}
\end{table}

Table~\ref{tab:rq4_questionnaire_results} summarizes the questionnaire results from the five respondents for evaluating the KG-derived TCSs. Overall, the generated TCSs received positive feedback from the respondents, with an overall score of 4.18 out of 5 and a positive rating rate of 95.3\% (i.e., ratings of `agree' or `strongly agree'). 
Note that this overall result was calculated by aggregating ratings from the first four evaluation categories, namely practical usefulness, clarity and understandability, completeness, and correctness and technical accuracy. 
The overall quality category, which reflects respondents’ impressions of the generated TCSs, received consistently high ratings, ranging from 4 to 5. These results indicate that the respondents strongly agree that the generated TCSs are useful, understandable, complete, technically accurate, and practical for supporting downstream test generation. 
Given that the respondents are industry testers from our Ethernet switch industry partner and most of them have direct experience in Ethernet switch testing, their evaluations provide credible, contextually relevant evidence of the practical value of the KG-derived TCSs.

The generated TCSs received high positive ratings of at least 96\% across three evaluation categories: practical usefulness, clarity and understandability, and correctness and technical accuracy. Most ratings for these three categories were either `agree' or `strongly agree', with only a small number of neutral (`neither agree nor disagree') responses and no negative responses (`disagree' or `strongly disagree'). These results suggest that the generated TCSs were generally perceived as useful for supporting test generation, easy to understand, and technically accurate.

Completeness received an 88\% positive rating, slightly lower than the other three categories. 
A more detailed analysis shows that the generated TCSs generally cover the necessary test steps, as the question ``The test specification includes all necessary steps'' received only one neutral response, while all remaining ratings were positive. However, respondents were more cautious about the sufficiency of preconditions and the clarity of expected results. Specifically, ``Preconditions are sufficient'' received four neutral responses, the ``Expected results are clearly defined'' received one `disagree’ response and three neutral responses, indicating that the preconditions and expected results may not always be specific or explicit enough for direct test execution. Overall, these results suggest that the generated TCSs are largely aligned with testers’ needs for downstream test automation, but they may still require further refinement before being used as executable TCSs.

The open-ended feedback further explains why respondents were more cautious about the sufficiency of preconditions and the clarity of expected results. In one comment, a respondent suggested moving the goal of the first configuration step into the preconditions and explicitly distinguishing which IP addresses should be allowed or denied access. This indicates that although the generated TCS captures the relevant configuration intent, some information may need to be reorganized into the appropriate TCS fields to better support direct test execution. In another comment, a respondent suggested refining the purpose of “managing and maintaining the switch” in the preconditions to better align with the expected results. This suggests that some generated preconditions are still relatively high-level and may require further refinement to make the testing objective and expected behavior more explicit. Overall, these comments indicate that the generated TCSs are close to testers’ needs, as they capture the key testing knowledge and necessary steps. However, making them fully ready for direct execution may require further refinement, especially by explicitly specifying preconditions and expected results. Therefore, LLMs can play an important role in downstream tasks by reasoning over the KG and source information to reorganize and enrich the generated TCSs.

\begin{tcolorbox}
\textbf{Answer to RQ4: The TCSs generated from KGs received positive feedback from experienced respondents, with an overall average score of 4.18 and a positive rating rate of 95.3\% across the evaluation categories. The results suggest that the generated TCSs are generally useful, understandable, complete, technically accurate, and practical for supporting downstream test case generation.}
\end{tcolorbox}

\section{Threats to Validity}
In this section, we discuss potential threats to the validity of our study and the steps we have taken to mitigate them.

\noindent\textbf{Construct Validity}

The primary potential construct threat arises from our use of an LLM-as-a-Judge as the primary evaluation mechanism. While the LLM-based evaluator enables scalable, systematic assessment, its judgments may not always align with human expert judgment, especially in ambiguous cases such as implicit goals and notes. This threat is particularly important as it may cause the system to optimize toward the LLM-based evaluator criteria rather than toward a ground truth. To mitigate this threat, we also conducted a manual evaluation in which two authors independently assessed all the generated KGs. Any disagreements were discussed until a consensus was reached. To further quantify the agreement between the LLM-as-a-Judge and human evaluators, we measured Cohen's kappa. The results show that Cohen's kappa is at least 0.72 across all extraction tasks, indicating substantial agreement between human evaluation and the LLM-as-a-Judge. This provides evidence that the LLM-based evaluation is reasonably reliable.

\noindent\textbf{Conclusion Validity}

A potential threat to conclusion validity concerns the strength of the conclusions that can be drawn from the observed improvements produced by the EEI loop. Although our results show that the EEI loop improved extraction quality when triggered, it was executed only a limited number of times. This is because most original extraction outputs already achieved high correctness scores and therefore did not require further improvement. As a result, while the observed improvements suggest that the EEI loop is effective for improving low-quality extractions, the limited number of triggered cases restricts the strength of our conclusions regarding its overall effectiveness. In future work, we plan to evaluate the EEI loop on more ESCMs or larger datasets to provide stronger evidence of its general effectiveness.

Another potential threat to conclusion validity concerns the questionnaire-based evaluation of TCSs. The number of testers who participated in the questionnaire was limited, which may affect the statistical strength and representativeness of the conclusions drawn from their feedback. To mitigate this threat, we sought to enhance the feedback's credibility by prioritizing qualified testing experts with relevant experience in Ethernet switch testing, and we designed the questionnaire to collect structured feedback across multiple aspects. 

\noindent\textbf{Internal Validity}

A potential threat to internal validity lies in the configuration of the employed LLM, since different parameter settings may affect the quality and stability of the generated results. To mitigate this threat, all experiments in this study were conducted using GPT-5 within the LangChain framework under the same default parameter settings. This consistent setup helps reduce the influence of configuration differences across experiments.

Another potential threat to internal validity arises from the inherent variability of LLM-based generation. In our experiments, we used GPT-5 with the default configuration provided through the API. Since the selected GPT-5 model does not support setting the temperature parameter, we were unable to set temperature=0 to make the output more deterministic. To reduce potential variability, we kept the model and all other experimental settings consistent across all experiments. In addition, running our experiments multiple times would be costly in terms of time, API fees, and human evaluation effort. Therefore, although we controlled the experimental configuration consistently across experiments, some run-to-run variability may still exist. Nevertheless, our experiments were conducted on 50 ESCMs rather than a single manual, and the extraction performance remained consistently strong across all of them. Moreover, GPT-5 is a reasoning model designed for harder and more complex tasks, which may mitigate this threat by improving the accuracy, consistency, and predictability of the generated outputs.~\cite{openai2025gpt5}.

Another potential threat to internal validity lies in the use of non-uniform prompting strategies across the three extraction tasks. Specifically, few-shot examples were used only for the procedure extraction task, while the other two tasks used prompts without examples. This design choice was based on a preliminary study on the first 20 of the 50 ESCMs, which showed that roadmap extraction and roadmap–procedure mapping already achieved very high correctness scores without examples, leaving little room for improvement. In contrast, including examples substantially improved the correctness of the procedure extraction task. 
As a result, this task-specific design may limit the direct comparability of results across the three tasks. However, the decision was made to better align the prompting strategy with each task's characteristics.

\noindent\textbf{External Validity}

The generalizability of our approach is limited by the dataset's scope, which consists of 50 real-world ESCMs from industry sources. Although these manuals provide realistic and diverse scenarios within this domain, they may not represent other networking products or other forms of technical documentation. The effectiveness of our approach may vary when applied to documents with different structures or different writing styles. However, the overall multi-agent framework was designed to be modular and may be adapted to other types of technical documents by revising the corresponding KG schema and the prompts used by the agents, as discussed in Section~\ref{sec:adapt_framework}. In future work, we plan to investigate methods for adapting prompts and KG schemas to different categories of technical documents.

\section{Related Work}
In this section, we review prior work related to our proposed LLM-based framework for KG generation, evaluation, and improvement from ESCMs. We first provide a general overview of the existing research on knowledge extraction and KG generation from technical documents. We then focus on LLM-based KG generation, as the proposed KG approach is built upon LLMs. Finally, we discuss prompting strategies and the use of LLM-as-a-Judge in existing LLM-based KG generation and verification methods.

\subsection{Knowledge Extraction and KG Generation from Technical Documents}

Knowledge extraction and KG generation from technical documents have been increasingly investigated and evolved in recent years due to their potential to support automation of downstream engineering tasks. Early approaches primarily relied on rule-based techniques by creating task-specific parsers that exploit recurring patterns, predefined syntax, and documentation structures to extract entities and their relationships~\cite{li2014configuration}. 
However, these methods can be sensitive to document structure, and minor variations can affect their performance~\cite{su2024enhancing}. In addition, these methods are primarily limited to syntactic analysis and often fail to capture the underlying semantics of technical documents, such as implicit dependencies and preconditions~\cite{su2024enhancing}. To address this limitation, semantic-based approaches have been proposed that leverage domain ontologies to model technical knowledge~\cite{rizvi2018ontology}. While these approaches improve interpretability and support reasoning over technical instructions, they lack generalizability and adaptability to new domains. 
As technical documents evolve over time or are applied to new domains, these approaches require substantial manual effort from domain experts to update existing ontologies or even construct new ones, limiting their scalability and adaptability.

Traditional NLP approaches~\cite{otomo2021towards, su2022constructing} have also been explored for capturing both the semantics and syntactic structure of technical documents. These approaches primarily rely on document or sentence embeddings combined with similarity-based analysis to capture the semantics of technical instructions. However, these approaches often struggle with long and complex technical sentences~\cite{su2022constructing}. Moreover, they tend to extract only high-level semantics from technical instructions~\cite{su2024enhancing} and are generally unable to capture implicit dependencies between them, which, in our context, are essential for accurately representing configuration behavior. As a result, the knowledge representations extracted by these approaches often lack the precision required for downstream tasks such as test generation~\cite{su2024enhancing}. In contrast, when carefully leveraged within a well-designed framework, LLMs can interpret complex technical documents and extract implicit attributes, dependencies, and relationships within technical instructions. Motivated by this potential, we designed and developed an LLM-based framework to extract knowledge from semi-structured product documentation, particularly ESCMs, and represent it as KGs to support the automation of downstream engineering tasks.

\subsection{LLM-based Knowledge Graph Generation}

The emergence of LLMs has significantly advanced KG generation approaches in recent years. However, most existing LLM-based KG generation approaches are designed for general text rather than technical documentation. For instance, Zhang et al.~\cite{zhang2024extract} proposed EDC (Extract-Define-Canonicalize), an LLM-based framework for KG generation. Since EDC is designed for general text, it follows the Open Information Extraction (OIE) paradigm, using few-shot prompting to directly extract entity-relation triples from the input text without relying on a predefined schema. This process is followed by a schema definition step based on the extracted triples, and then a canonicalization step that removes redundant and semantically equivalent relations from the resulting schema. The latter is intended to reduce ambiguity in the resulting KG and to make it more suitable for downstream applications. Their evaluation on general text datasets with diverse relation types shows that EDC can effectively construct concise, non-redundant schemas and generate canonicalized KGs.

Similarly, Kommineni et al.~\cite{kommineni2024human} proposed a semi-automatic KG construction pipeline for general text, in which LLMs are employed both to generate an initial ontology and to construct the KG based on that ontology. Their proposed approach starts with generating a set of competency questions (CQs) using LLMs, which are then refined and extended by domain experts. CQs are high-level, abstract questions that capture the underlying domain and serve as requirement specifications in ontology development.
These verified CQs are subsequently used in few-shot prompting to extract relevant concepts and relationships for ontology construction. They further employ the Retrieval-Augmented Generation (RAG) technique, using five selected documents as references to help the LLM answer the verified CQs, followed by simple text processing that refines the LLM-generated answers. Finally, the verified CQs, their corresponding answers, and the LLM-generated ontology are provided to an LLM to extract entities and relationships from the answers and map them onto the ontology, resulting in the final KG.

In the context of general text, where entities and relationships are highly diverse, the existing approaches primarily focus on the flexibility and dynamic construction of the KG schema. In contrast, KG generation from technical documents poses a different set of challenges, primarily focused on the correctness and precision of extracted knowledge. This is particularly critical in the context of Ethernet switches, where enabling automation of downstream tasks using KGs generated from ESCMs requires the accurate capture of configuration steps and their attributes and dependencies. While existing LLM-based KG generation approaches offer flexibility in handling diverse input texts and dynamically constructing schemas, they are generally not designed to achieve the level of correctness and granularity required for ESCMs. This difficulty stems from the inherent complexity of such technical documents, which require highly accurate and fine-grained KG generation approaches. To address this difficulty, we design a fine-grained schema tailored to the structure and characteristics of ESCMs, enabling accurate representation of various configuration step attributes and the dependencies embedded in configuration documents. In addition, our approach incorporates an EEI loop that iteratively refines knowledge extraction by identifying inaccurate or incomplete outputs and regenerating them using targeted, refined prompts. This iterative process improves the correctness of extracted knowledge, particularly the implicit dependencies between configuration steps, enhancing the overall quality and reliability of the generated KGs.

Su et al.~\cite{su2024enhancing} also leveraged LLMs with few-shot prompting to generate KGs from bug reports. Although their approach also targets technical documentation, ESCMs are typically more complex, making KG generation more challenging, particularly due to implicit dependencies between configuration steps.
Their manual evaluation shows that their approach can achieve strong performance in KG generation from bug reports without requiring iterative refinement of extraction prompts.
In contrast, our results show that, for ESCMs, omitting the EEI loop can lead to suboptimal performance in some cases. In particular, extracting the Roadmap–Procedure mappings from ESCMs often necessitates improving the initial prompts, requiring one or, in some cases, even two rounds of EEI loop to achieve a high correctness score.

In another relevant study, Bi et al.~\cite{bi2024codekgc} proposed CodeKGC, a method that leverages code-language models by reformulating KG generation as a code-generation task. The core idea is to transform input text into a code-like format with predefined syntactic and structural characteristics, enabling code language models to generate accurate KGs.
CodeKGC first transforms both the original natural-language input and the defined KG schema into a code-like representation using a predefined Python script that adheres to Python grammar. Specifically, the input sentences, as well as each entity and relation type in the KG schema, are converted into predefined Python classes and incorporated into the prompt. While their experimental results, conducted on datasets with a limited number of entity and relation types, demonstrate the effectiveness of CodeKGC, the LLM input token limitations constrain how much of the transformed input and the full KG schema can be incorporated into the prompt. This limitation becomes more pronounced as the complexity of both the schema and input increases. In the subsequent section, we provide a detailed discussion of the impact of prompt design and prompt engineering strategies on LLM-based KG generation.

\subsubsection{Prompt Engineering for KG generation}

Prompt engineering plays a critical role in LLM-based knowledge extraction and KG generation~\cite{pan2023large}, as prompt design directly affects the quality of the resulting KG. Moreover, since the extracted knowledge must conform to the predefined KG schema, incorporating schema information into the prompt is essential to guide the LLM toward producing structured and valid outputs. Existing approaches incorporate schema information in different ways. EDC~\cite{zhang2024extract}, which is based on open information extraction, dynamically derives the schema using LLMs and subsequently feeds both the extracted triples and the generated schema back to the LLM for canonicalization. Similarly, Kommineni et al.~\cite{kommineni2024human} directly provide the LLM with an ontology generated from competency questions to guide KG generation. In another study, Bi et al.~\cite{bi2024codekgc} transform both the original natural language input and the predefined KG schema into Python code, and incorporate these representations into the prompt to guide KG generation.

However, including the entire KG schema in the prompt, especially when it is complex and lengthy, may not be feasible due to LLM input token limitations. As a result, few-shot learning has been adopted in LLM-based KG generation~\cite{su2024enhancing, bi2024codekgc}. Given the LLM's input limitations, selecting a representative set of examples to include in the prompts is crucial. To address this, our approach employs an agentic framework comprising three specialized agents for entity extraction. Each agent is guided by a carefully designed, task-specific prompt. Where beneficial, we incorporate a minimal yet representative set of examples that captures the complexity of the task while providing sufficient guidance to the model.

Nevertheless, prior research suggests that prompt-based approaches can be sensitive to prompt design and may introduce biases~\cite{pan2023large}. To address this, both Bi et al.~\cite{bi2024codekgc} and Su et al.~\cite{su2024enhancing} improved their approaches using Chain-of-Thought (CoT) prompting, where intermediate reasoning steps are included to improve extraction performance. In contrast, our approach introduces an EEI loop that automatically refines extraction prompts based on the evaluation of extracted entities for each specific ESCM. Within this loop, an ImprovAgent revises the extraction prompt based on detailed feedback from an EvalAgent. Our results demonstrate that such an improvement loop is often necessary to ensure a high correctness score for extracted knowledge, particularly for challenging tasks such as extracting Roadmap-Procedure mappings from ESCMs.

\subsubsection{LLM-as-a-Judge}

The strong performance of LLMs across a wide range of domains has led to their adoption as evaluators, commonly referred to as LLM-as-a-Judge~\cite{gu2024survey}. This paradigm leverages LLMs’ ability to approximate human-like reasoning, enabling them to perform evaluation tasks traditionally carried out by human experts while providing a highly scalable, cost-effective alternative.
LLM-as-a-Judge has also been explored in the context of KG generation by Kommineni et al.~\cite{kommineni2024human}. They employed LLM-as-a-Judge to assess the quality of both the generated competency question answers and the extracted KG. They further refine the LLM judge's outputs using automatic text processing techniques. Their results show that such refinement minimizes disagreements between LLM and human judgments and confirm the reliability of LLM-as-a-Judge in KG evaluation.
Similarly, we leverage an LLM-as-a-Judge, adopting a guideline-based evaluation strategy in which the LLM assesses the generated KGs against predefined evaluation criteria. 
Our empirical analysis of the agreement between LLM and human judgments across 50 ESCMs shows substantial consistency, with a Cohen's kappa score of at least 0.72, further supporting the reliability of LLM-as-a-Judge in our approach.

\section{Conclusion}
In this paper, we introduce a multi-agent LLM-based framework for KG generation, evaluation, and improvement from technical documents to support system testing. We evaluate our approach using Ethernet switches as a case study by extracting KGs from ESCMs, semi-structured documents that contain highly detailed, complex technical information. Our results show that this framework has the potential to effectively support the automation of Ethernet switch testing by accurately capturing semantic knowledge, including the key attributes of configuration steps and the implicit dependencies among them, while ensuring high correctness scores in the generated KGs.
We design a fine-grained KG schema specifically tailored to ESCMs, enabling a precise and highly granular representation of the configuration knowledge contained in ESCMs. Based on this schema, the framework combines specialized agents, each guided by carefully designed prompts, with an iterative Extract–Evaluate–Improve (EEI) mechanism to ensure high correctness for the generated KGs. After initial KG generation using original prompts, the EEI loop employs an LLM-as-a-Judge with task-specific evaluation guidelines to assess the generated KGs and produce targeted feedback when needed. This feedback is used to refine the extraction prompts, which are then applied to generate new KGs, resulting in progressively more accurate KGs. 
The framework’s modular, agent-based design supports strong generalizability, enabling it to be easily adapted to configuration manuals in new domains and extended to other types of technical documents.
Our empirical evaluation on real-world ESCMs from industry demonstrates that the generated KGs consistently achieve high average correctness scores. 
Additionally, our investigation into the usefulness of generated KGs for deriving system-level TCSs, based on a structured questionnaire with experienced industry testers, shows consistently high Likert-scale ratings, indicating that the generated and verified KGs can effectively support automated test generation.

We plan to further improve the robustness of our framework across a wider range of technical documents and domains. Although our approach is designed to be generalizable, adapting it to domains with different writing styles or domain-specific terminology may require prompt adaptation and schema evolution.
Moreover, we aim to further assess the practical utility of the generated KGs in downstream tasks by integrating them into end-to-end pipelines, particularly for automated test case generation and validation. This includes developing methods to automatically generate executable test cases using the generated KGs and to assess whether these test cases are effective in real testing scenarios.

\begin{acks}
This work was supported by a research grant from Huawei Canada, the Discovery Grant and Canada Research Chair programs of the Natural Sciences and Engineering Research Council of Canada (NSERC), and a Research Ireland grant 13/RC/2094-2.
\end{acks}

\bibliographystyle{ACM-Reference-Format}
\bibliography{main}


\begin{thebibliography}{39}


\ifx \showCODEN    \undefined \def \showCODEN     #1{\unskip}     \fi
\ifx \showISBNx    \undefined \def \showISBNx     #1{\unskip}     \fi
\ifx \showISBNxiii \undefined \def \showISBNxiii  #1{\unskip}     \fi
\ifx \showISSN     \undefined \def \showISSN      #1{\unskip}     \fi
\ifx \showLCCN     \undefined \def \showLCCN      #1{\unskip}     \fi
\ifx \shownote     \undefined \def \shownote      #1{#1}          \fi
\ifx \showarticletitle \undefined \def \showarticletitle #1{#1}   \fi
\ifx \showURL      \undefined \def \showURL       {\relax}        \fi
\providecommand\bibfield[2]{#2}
\providecommand\bibinfo[2]{#2}
\providecommand\natexlab[1]{#1}
\providecommand\showeprint[2][]{arXiv:#2}

\bibitem[202(2025)]%
        {2025LangchainWebsite}
 \bibinfo{year}{Accessed: 2025}\natexlab{}.
\newblock \bibinfo{title}{Langchain Official Website}.
\newblock
\urldef\tempurl%
\url{https://www.langchain.com/}
\showURL{%
\tempurl}


\bibitem[Banerjee and Lavie(2005)]%
        {banerjee2005meteor}
\bibfield{author}{\bibinfo{person}{Satanjeev Banerjee} {and} \bibinfo{person}{Alon Lavie}.} \bibinfo{year}{2005}\natexlab{}.
\newblock \showarticletitle{METEOR: An automatic metric for MT evaluation with improved correlation with human judgments}. In \bibinfo{booktitle}{\emph{Proceedings of the acl workshop on intrinsic and extrinsic evaluation measures for machine translation and/or summarization}}. \bibinfo{pages}{65--72}.
\newblock


\bibitem[Bi et~al\mbox{.}(2024)]%
        {bi2024codekgc}
\bibfield{author}{\bibinfo{person}{Zhen Bi}, \bibinfo{person}{Jing Chen}, \bibinfo{person}{Yinuo Jiang}, \bibinfo{person}{Feiyu Xiong}, \bibinfo{person}{Wei Guo}, \bibinfo{person}{Huajun Chen}, {and} \bibinfo{person}{Ningyu Zhang}.} \bibinfo{year}{2024}\natexlab{}.
\newblock \showarticletitle{Codekgc: Code language model for generative knowledge graph construction}.
\newblock \bibinfo{journal}{\emph{ACM Transactions on Asian and Low-Resource Language Information Processing}} \bibinfo{volume}{23}, \bibinfo{number}{3} (\bibinfo{year}{2024}), \bibinfo{pages}{1--16}.
\newblock


\bibitem[Chen et~al\mbox{.}(2025)]%
        {chen2025prometheus}
\bibfield{author}{\bibinfo{person}{Zimin Chen}, \bibinfo{person}{Yue Pan}, \bibinfo{person}{Siyu Lu}, \bibinfo{person}{Jiayi Xu}, \bibinfo{person}{Claire~Le Goues}, \bibinfo{person}{Martin Monperrus}, {and} \bibinfo{person}{He Ye}.} \bibinfo{year}{2025}\natexlab{}.
\newblock \showarticletitle{Prometheus: Unified knowledge graphs for issue resolution in multilingual codebases}.
\newblock \bibinfo{journal}{\emph{arXiv preprint arXiv:2507.19942}} (\bibinfo{year}{2025}).
\newblock


\bibitem[Da~Silva et~al\mbox{.}(2018)]%
        {da2018test}
\bibfield{author}{\bibinfo{person}{Alberto~Rodrigues Da~Silva}, \bibinfo{person}{Ana~CR Paiva}, {and} \bibinfo{person}{Valter~ER Da~Silva}.} \bibinfo{year}{2018}\natexlab{}.
\newblock \showarticletitle{A test specification language for information systems based on data entities, use cases and state machines}. In \bibinfo{booktitle}{\emph{International Conference on Model-Driven Engineering and Software Development}}. Springer, \bibinfo{pages}{455--474}.
\newblock


\bibitem[Fu et~al\mbox{.}(2024)]%
        {fu2024gptscore}
\bibfield{author}{\bibinfo{person}{Jinlan Fu}, \bibinfo{person}{See~Kiong Ng}, \bibinfo{person}{Zhengbao Jiang}, {and} \bibinfo{person}{Pengfei Liu}.} \bibinfo{year}{2024}\natexlab{}.
\newblock \showarticletitle{Gptscore: Evaluate as you desire}. In \bibinfo{booktitle}{\emph{Proceedings of the 2024 Conference of the North American Chapter of the Association for Computational Linguistics: Human Language Technologies (Volume 1: Long Papers)}}. \bibinfo{pages}{6556--6576}.
\newblock


\bibitem[Gu et~al\mbox{.}(2024)]%
        {gu2024survey}
\bibfield{author}{\bibinfo{person}{Jiawei Gu}, \bibinfo{person}{Xuhui Jiang}, \bibinfo{person}{Zhichao Shi}, \bibinfo{person}{Hexiang Tan}, \bibinfo{person}{Xuehao Zhai}, \bibinfo{person}{Chengjin Xu}, \bibinfo{person}{Wei Li}, \bibinfo{person}{Yinghan Shen}, \bibinfo{person}{Shengjie Ma}, \bibinfo{person}{Honghao Liu}, {et~al\mbox{.}}} \bibinfo{year}{2024}\natexlab{}.
\newblock \showarticletitle{{A survey on llm-as-a-judge}}.
\newblock \bibinfo{journal}{\emph{{The Innovation}}} (\bibinfo{year}{2024}).
\newblock


\bibitem[Guo et~al\mbox{.}(2021)]%
        {guo2021automatic}
\bibfield{author}{\bibinfo{person}{Liang Guo}, \bibinfo{person}{Fu Yan}, \bibinfo{person}{Yuqian Lu}, \bibinfo{person}{Ming Zhou}, {and} \bibinfo{person}{Tao Yang}.} \bibinfo{year}{2021}\natexlab{}.
\newblock \showarticletitle{An automatic machining process decision-making system based on knowledge graph}.
\newblock \bibinfo{journal}{\emph{International journal of computer integrated manufacturing}} \bibinfo{volume}{34}, \bibinfo{number}{12} (\bibinfo{year}{2021}), \bibinfo{pages}{1348--1369}.
\newblock


\bibitem[Hogan et~al\mbox{.}(2021)]%
        {hogan2021knowledge}
\bibfield{author}{\bibinfo{person}{Aidan Hogan}, \bibinfo{person}{Eva Blomqvist}, \bibinfo{person}{Michael Cochez}, \bibinfo{person}{Claudia d’Amato}, \bibinfo{person}{Gerard~De Melo}, \bibinfo{person}{Claudio Gutierrez}, \bibinfo{person}{Sabrina Kirrane}, \bibinfo{person}{Jos{\'e} Emilio~Labra Gayo}, \bibinfo{person}{Roberto Navigli}, \bibinfo{person}{Sebastian Neumaier}, {et~al\mbox{.}}} \bibinfo{year}{2021}\natexlab{}.
\newblock \showarticletitle{{Knowledge graphs}}.
\newblock \bibinfo{journal}{\emph{{ACM Computing Surveys (Csur)}}} \bibinfo{volume}{54}, \bibinfo{number}{4} (\bibinfo{year}{2021}), \bibinfo{pages}{1--37}.
\newblock


\bibitem[Ji et~al\mbox{.}(2023)]%
        {ji2023survey}
\bibfield{author}{\bibinfo{person}{Ziwei Ji}, \bibinfo{person}{Nayeon Lee}, \bibinfo{person}{Rita Frieske}, \bibinfo{person}{Tiezheng Yu}, \bibinfo{person}{Dan Su}, \bibinfo{person}{Yan Xu}, \bibinfo{person}{Etsuko Ishii}, \bibinfo{person}{Ye~Jin Bang}, \bibinfo{person}{Andrea Madotto}, {and} \bibinfo{person}{Pascale Fung}.} \bibinfo{year}{2023}\natexlab{}.
\newblock \showarticletitle{{Survey of hallucination in natural language generation}}.
\newblock \bibinfo{journal}{\emph{{ACM computing surveys}}} \bibinfo{volume}{55}, \bibinfo{number}{12} (\bibinfo{year}{2023}), \bibinfo{pages}{1--38}.
\newblock


\bibitem[Jung et~al\mbox{.}(2024)]%
        {jung2024trust}
\bibfield{author}{\bibinfo{person}{Jaehun Jung}, \bibinfo{person}{Faeze Brahman}, {and} \bibinfo{person}{Yejin Choi}.} \bibinfo{year}{2024}\natexlab{}.
\newblock \showarticletitle{{Trust or escalate: Llm judges with provable guarantees for human agreement}}.
\newblock \bibinfo{journal}{\emph{{arXiv preprint arXiv:2407.18370}}} (\bibinfo{year}{2024}).
\newblock


\bibitem[Kommineni et~al\mbox{.}(2024)]%
        {kommineni2024human}
\bibfield{author}{\bibinfo{person}{Vamsi~Krishna Kommineni}, \bibinfo{person}{Birgitta K{\"o}nig-Ries}, {and} \bibinfo{person}{Sheeba Samuel}.} \bibinfo{year}{2024}\natexlab{}.
\newblock \showarticletitle{From human experts to machines: An LLM supported approach to ontology and knowledge graph construction}.
\newblock \bibinfo{journal}{\emph{arXiv preprint arXiv:2403.08345}} (\bibinfo{year}{2024}).
\newblock


\bibitem[Li et~al\mbox{.}(2014)]%
        {li2014configuration}
\bibfield{author}{\bibinfo{person}{Fuliang Li}, \bibinfo{person}{Jiahai Yang}, \bibinfo{person}{Jianping Wu}, \bibinfo{person}{Zhiyan Zheng}, \bibinfo{person}{Huijing Zhang}, {and} \bibinfo{person}{Xingwei Wang}.} \bibinfo{year}{2014}\natexlab{}.
\newblock \showarticletitle{Configuration analysis and recommendation: Case studies in IPv6 networks}.
\newblock \bibinfo{journal}{\emph{Computer Communications}}  \bibinfo{volume}{53} (\bibinfo{year}{2014}), \bibinfo{pages}{37--51}.
\newblock


\bibitem[Lin(2004)]%
        {lin2004rouge}
\bibfield{author}{\bibinfo{person}{Chin-Yew Lin}.} \bibinfo{year}{2004}\natexlab{}.
\newblock \showarticletitle{Rouge: A package for automatic evaluation of summaries}. In \bibinfo{booktitle}{\emph{Text summarization branches out}}. \bibinfo{pages}{74--81}.
\newblock


\bibitem[Liu et~al\mbox{.}(2024)]%
        {liu2024lost}
\bibfield{author}{\bibinfo{person}{Nelson~F Liu}, \bibinfo{person}{Kevin Lin}, \bibinfo{person}{John Hewitt}, \bibinfo{person}{Ashwin Paranjape}, \bibinfo{person}{Michele Bevilacqua}, \bibinfo{person}{Fabio Petroni}, {and} \bibinfo{person}{Percy Liang}.} \bibinfo{year}{2024}\natexlab{}.
\newblock \showarticletitle{{Lost in the middle: How language models use long contexts}}.
\newblock \bibinfo{journal}{\emph{{Transactions of the Association for Computational Linguistics}}}  \bibinfo{volume}{12} (\bibinfo{year}{2024}), \bibinfo{pages}{157--173}.
\newblock


\bibitem[Liu et~al\mbox{.}(2023)]%
        {liu2023g}
\bibfield{author}{\bibinfo{person}{Yang Liu}, \bibinfo{person}{Dan Iter}, \bibinfo{person}{Yichong Xu}, \bibinfo{person}{Shuohang Wang}, \bibinfo{person}{Ruochen Xu}, {and} \bibinfo{person}{Chenguang Zhu}.} \bibinfo{year}{2023}\natexlab{}.
\newblock \showarticletitle{{G-eval: NLG evaluation using gpt-4 with better human alignment}}.
\newblock \bibinfo{journal}{\emph{{arXiv preprint arXiv:2303.16634}}} (\bibinfo{year}{2023}).
\newblock


\bibitem[Masuda et~al\mbox{.}(2026)]%
        {masuda2026generating}
\bibfield{author}{\bibinfo{person}{Satoshi Masuda}, \bibinfo{person}{Satoshi Kouzawa}, \bibinfo{person}{Kyousuke Sezai}, \bibinfo{person}{Hidetoshi Suhara}, \bibinfo{person}{Yasuaki Hiruta}, {and} \bibinfo{person}{Kunihiro Kudou}.} \bibinfo{year}{2026}\natexlab{}.
\newblock \showarticletitle{Generating high-level test cases from requirements using LLM: An industry study}. In \bibinfo{booktitle}{\emph{2026 International Conference on Artificial Intelligence, Computer, Data Sciences and Applications (ACDSA)}}. IEEE, \bibinfo{pages}{1--9}.
\newblock


\bibitem[Milchevski et~al\mbox{.}(2025)]%
        {milchevski2025multi}
\bibfield{author}{\bibinfo{person}{Dragan Milchevski}, \bibinfo{person}{Gordon Frank}, \bibinfo{person}{Anna H{\"a}tty}, \bibinfo{person}{Bingqing Wang}, \bibinfo{person}{Xiaowei Zhou}, {and} \bibinfo{person}{Zhe Feng}.} \bibinfo{year}{2025}\natexlab{}.
\newblock \showarticletitle{Multi-Step Generation of Test Specifications using Large Language Models for System-Level Requirements}. In \bibinfo{booktitle}{\emph{Proceedings of the 63rd Annual Meeting of the Association for Computational Linguistics (Volume 6: Industry Track)}}. \bibinfo{pages}{132--146}.
\newblock


\bibitem[{OpenAI}({[n.\,d.]})]%
        {OpenAI_API_Reference}
\bibfield{author}{\bibinfo{person}{{OpenAI}}.} \bibinfo{year}{[n.\,d.]}\natexlab{}.
\newblock \bibinfo{title}{OpenAI {API} Reference}.
\newblock \bibinfo{howpublished}{\url{https://platform.openai.com/docs/api-reference/introduction}}.
\newblock
\newblock
\shownote{Accessed: 2026-02-04}.


\bibitem[{OpenAI}(2025)]%
        {openai2025gpt5}
\bibfield{author}{\bibinfo{person}{{OpenAI}}.} \bibinfo{year}{2025}\natexlab{}.
\newblock \bibinfo{title}{Introducing GPT-5}.
\newblock \bibinfo{howpublished}{\url{https://openai.com/index/introducing-gpt-5/}}.
\newblock
\newblock
\shownote{Accessed: 2026-04-23}.


\bibitem[Otomo et~al\mbox{.}(2021)]%
        {otomo2021towards}
\bibfield{author}{\bibinfo{person}{Kazuki Otomo}, \bibinfo{person}{Satoru Kobayashi}, \bibinfo{person}{Kensuke Fukuda}, \bibinfo{person}{Osamu Akashi}, \bibinfo{person}{Kimihiro Mizutani}, {and} \bibinfo{person}{Hiroshi Esaki}.} \bibinfo{year}{2021}\natexlab{}.
\newblock \showarticletitle{Towards extracting semantics of network config blocks}. In \bibinfo{booktitle}{\emph{2021 IEEE 45th Annual Computers, Software, and Applications Conference (COMPSAC)}}. IEEE, \bibinfo{pages}{1443--1448}.
\newblock


\bibitem[Pan et~al\mbox{.}(2023)]%
        {pan2023large}
\bibfield{author}{\bibinfo{person}{Jeff~Z Pan}, \bibinfo{person}{Simon Razniewski}, \bibinfo{person}{Jan-Christoph Kalo}, \bibinfo{person}{Sneha Singhania}, \bibinfo{person}{Jiaoyan Chen}, \bibinfo{person}{Stefan Dietze}, \bibinfo{person}{Hajira Jabeen}, \bibinfo{person}{Janna Omeliyanenko}, \bibinfo{person}{Wen Zhang}, \bibinfo{person}{Matteo Lissandrini}, {et~al\mbox{.}}} \bibinfo{year}{2023}\natexlab{}.
\newblock \showarticletitle{Large language models and knowledge graphs: Opportunities and challenges}.
\newblock \bibinfo{journal}{\emph{arXiv preprint arXiv:2308.06374}} (\bibinfo{year}{2023}).
\newblock


\bibitem[Papineni et~al\mbox{.}(2002)]%
        {papineni2002bleu}
\bibfield{author}{\bibinfo{person}{Kishore Papineni}, \bibinfo{person}{Salim Roukos}, \bibinfo{person}{Todd Ward}, {and} \bibinfo{person}{Wei-Jing Zhu}.} \bibinfo{year}{2002}\natexlab{}.
\newblock \showarticletitle{Bleu: a method for automatic evaluation of machine translation}. In \bibinfo{booktitle}{\emph{Proceedings of the 40th annual meeting of the Association for Computational Linguistics}}. \bibinfo{pages}{311--318}.
\newblock


\bibitem[Peng et~al\mbox{.}(2023)]%
        {peng2023knowledge}
\bibfield{author}{\bibinfo{person}{Ciyuan Peng}, \bibinfo{person}{Feng Xia}, \bibinfo{person}{Mehdi Naseriparsa}, {and} \bibinfo{person}{Francesco Osborne}.} \bibinfo{year}{2023}\natexlab{}.
\newblock \showarticletitle{{Knowledge graphs: Opportunities and challenges}}.
\newblock \bibinfo{journal}{\emph{{Artificial intelligence review}}} \bibinfo{volume}{56}, \bibinfo{number}{11} (\bibinfo{year}{2023}), \bibinfo{pages}{13071--13102}.
\newblock


\bibitem[Press et~al\mbox{.}(2021)]%
        {press2021train}
\bibfield{author}{\bibinfo{person}{Ofir Press}, \bibinfo{person}{Noah~A Smith}, {and} \bibinfo{person}{Mike Lewis}.} \bibinfo{year}{2021}\natexlab{}.
\newblock \showarticletitle{{Train short, test long: Attention with linear biases enables input length extrapolation}}.
\newblock \bibinfo{journal}{\emph{{arXiv preprint arXiv:2108.12409}}} (\bibinfo{year}{2021}).
\newblock


\bibitem[Rizvi et~al\mbox{.}(2018)]%
        {rizvi2018ontology}
\bibfield{author}{\bibinfo{person}{Syed Tahseen~Raza Rizvi}, \bibinfo{person}{Dominique Mercier}, \bibinfo{person}{Stefan Agne}, \bibinfo{person}{Steffen Erkel}, \bibinfo{person}{Andreas Dengel}, {and} \bibinfo{person}{Sheraz Ahmed}.} \bibinfo{year}{2018}\natexlab{}.
\newblock \showarticletitle{Ontology-based Information Extraction from Technical Documents.}. In \bibinfo{booktitle}{\emph{ICAART (2)}}. \bibinfo{pages}{493--500}.
\newblock


\bibitem[Sarma and Mall(2009)]%
        {sarma2009automatic}
\bibfield{author}{\bibinfo{person}{Monalisa Sarma} {and} \bibinfo{person}{Rajib Mall}.} \bibinfo{year}{2009}\natexlab{}.
\newblock \showarticletitle{Automatic generation of test specifications for coverage of system state transitions}.
\newblock \bibinfo{journal}{\emph{Information and Software Technology}} \bibinfo{volume}{51}, \bibinfo{number}{2} (\bibinfo{year}{2009}), \bibinfo{pages}{418--432}.
\newblock


\bibitem[Sneed(1993)]%
        {tcs1993automated}
\bibfield{author}{\bibinfo{person}{HM Sneed}.} \bibinfo{year}{1993}\natexlab{}.
\newblock \showarticletitle{{Automated tool support for ANSI/IEEE STD: 829-1983 software test documentation}}. In \bibinfo{booktitle}{\emph{{Proceedings 1993 Software Engineering Standards Symposium}}}. {IEEE}, \bibinfo{pages}{308--316}.
\newblock


\bibitem[Su et~al\mbox{.}(2022)]%
        {su2022constructing}
\bibfield{author}{\bibinfo{person}{Yanqi Su}, \bibinfo{person}{Zheming Han}, \bibinfo{person}{Zhenchang Xing}, \bibinfo{person}{Xin Xia}, \bibinfo{person}{Xiwei Xu}, \bibinfo{person}{Liming Zhu}, {and} \bibinfo{person}{Qinghua Lu}.} \bibinfo{year}{2022}\natexlab{}.
\newblock \showarticletitle{Constructing a system knowledge graph of user tasks and failures from bug reports to support soap opera testing}. In \bibinfo{booktitle}{\emph{Proceedings of the 37th IEEE/ACM International Conference on Automated Software Engineering}}. \bibinfo{pages}{1--13}.
\newblock


\bibitem[Su et~al\mbox{.}(2024)]%
        {su2024enhancing}
\bibfield{author}{\bibinfo{person}{Yanqi Su}, \bibinfo{person}{Dianshu Liao}, \bibinfo{person}{Zhenchang Xing}, \bibinfo{person}{Qing Huang}, \bibinfo{person}{Mulong Xie}, \bibinfo{person}{Qinghua Lu}, {and} \bibinfo{person}{Xiwei Xu}.} \bibinfo{year}{2024}\natexlab{}.
\newblock \showarticletitle{Enhancing exploratory testing by large language model and knowledge graph}. In \bibinfo{booktitle}{\emph{Proceedings of the IEEE/ACM 46th International Conference on Software Engineering}}. \bibinfo{pages}{1--12}.
\newblock


\bibitem[Su et~al\mbox{.}(2025)]%
        {su2025automated}
\bibfield{author}{\bibinfo{person}{Yanqi Su}, \bibinfo{person}{Zhenchang Xing}, \bibinfo{person}{Chong Wang}, \bibinfo{person}{Chunyang Chen}, \bibinfo{person}{Sherry Xu}, \bibinfo{person}{Qinghua Lu}, {and} \bibinfo{person}{Liming Zhu}.} \bibinfo{year}{2025}\natexlab{}.
\newblock \showarticletitle{Automated soap opera testing directed by llms and scenario knowledge: Feasibility, challenges, and road ahead}.
\newblock \bibinfo{journal}{\emph{Proceedings of the ACM on Software Engineering}} \bibinfo{volume}{2}, \bibinfo{number}{FSE} (\bibinfo{year}{2025}), \bibinfo{pages}{757--778}.
\newblock


\bibitem[Utting and Legeard(2010)]%
        {utting2010practical}
\bibfield{author}{\bibinfo{person}{Mark Utting} {and} \bibinfo{person}{Bruno Legeard}.} \bibinfo{year}{2010}\natexlab{}.
\newblock \bibinfo{booktitle}{\emph{Practical model-based testing: a tools approach}}.
\newblock \bibinfo{publisher}{Elsevier}.
\newblock


\bibitem[Wang et~al\mbox{.}(2023)]%
        {wang2023chatgpt}
\bibfield{author}{\bibinfo{person}{Jiaan Wang}, \bibinfo{person}{Yunlong Liang}, \bibinfo{person}{Fandong Meng}, \bibinfo{person}{Zengkui Sun}, \bibinfo{person}{Haoxiang Shi}, \bibinfo{person}{Zhixu Li}, \bibinfo{person}{Jinan Xu}, \bibinfo{person}{Jianfeng Qu}, {and} \bibinfo{person}{Jie Zhou}.} \bibinfo{year}{2023}\natexlab{}.
\newblock \showarticletitle{Is chatgpt a good nlg evaluator? a preliminary study}.
\newblock \bibinfo{journal}{\emph{arXiv preprint arXiv:2303.04048}} (\bibinfo{year}{2023}).
\newblock


\bibitem[Yang et~al\mbox{.}(2025)]%
        {yang2025requirements}
\bibfield{author}{\bibinfo{person}{Zhenzhen Yang}, \bibinfo{person}{Rubing Huang}, \bibinfo{person}{Chenhui Cui}, \bibinfo{person}{Nan Niu}, {and} \bibinfo{person}{Dave Towey}.} \bibinfo{year}{2025}\natexlab{}.
\newblock \showarticletitle{Requirements-based test generation: A comprehensive survey}.
\newblock \bibinfo{journal}{\emph{ACM Transactions on Software Engineering and Methodology}} (\bibinfo{year}{2025}).
\newblock


\bibitem[Ye et~al\mbox{.}(2022)]%
        {ye2022generative}
\bibfield{author}{\bibinfo{person}{Hongbin Ye}, \bibinfo{person}{Ningyu Zhang}, \bibinfo{person}{Hui Chen}, {and} \bibinfo{person}{Huajun Chen}.} \bibinfo{year}{2022}\natexlab{}.
\newblock \showarticletitle{Generative knowledge graph construction: A review}. In \bibinfo{booktitle}{\emph{Proceedings of the 2022 conference on empirical methods in natural language processing}}. \bibinfo{pages}{1--17}.
\newblock


\bibitem[Zhang and Soh(2024)]%
        {zhang2024extract}
\bibfield{author}{\bibinfo{person}{Bowen Zhang} {and} \bibinfo{person}{Harold Soh}.} \bibinfo{year}{2024}\natexlab{}.
\newblock \showarticletitle{Extract, define, canonicalize: An llm-based framework for knowledge graph construction}. In \bibinfo{booktitle}{\emph{Proceedings of the 2024 conference on empirical methods in natural language processing}}. \bibinfo{pages}{9820--9836}.
\newblock


\bibitem[Zhang et~al\mbox{.}(2025)]%
        {zhang2025siren}
\bibfield{author}{\bibinfo{person}{Yue Zhang}, \bibinfo{person}{Yafu Li}, \bibinfo{person}{Leyang Cui}, \bibinfo{person}{Deng Cai}, \bibinfo{person}{Lemao Liu}, \bibinfo{person}{Tingchen Fu}, \bibinfo{person}{Xinting Huang}, \bibinfo{person}{Enbo Zhao}, \bibinfo{person}{Yu Zhang}, \bibinfo{person}{Yulong Chen}, {et~al\mbox{.}}} \bibinfo{year}{2025}\natexlab{}.
\newblock \showarticletitle{{Siren’s Song in the AI Ocean: A Survey on Hallucination in Large Language Models}}.
\newblock \bibinfo{journal}{\emph{{Computational Linguistics}}} (\bibinfo{year}{2025}), \bibinfo{pages}{1--46}.
\newblock


\bibitem[Zheng et~al\mbox{.}(2023)]%
        {zheng2023judging}
\bibfield{author}{\bibinfo{person}{Lianmin Zheng}, \bibinfo{person}{Wei-Lin Chiang}, \bibinfo{person}{Ying Sheng}, \bibinfo{person}{Siyuan Zhuang}, \bibinfo{person}{Zhanghao Wu}, \bibinfo{person}{Yonghao Zhuang}, \bibinfo{person}{Zi Lin}, \bibinfo{person}{Zhuohan Li}, \bibinfo{person}{Dacheng Li}, \bibinfo{person}{Eric Xing}, {et~al\mbox{.}}} \bibinfo{year}{2023}\natexlab{}.
\newblock \showarticletitle{{Judging llm-as-a-judge with mt-bench and chatbot arena}}.
\newblock \bibinfo{journal}{\emph{{Advances in neural information processing systems}}}  \bibinfo{volume}{36} (\bibinfo{year}{2023}), \bibinfo{pages}{46595--46623}.
\newblock


\bibitem[Zhu et~al\mbox{.}(2023)]%
        {zhu2023judgelm}
\bibfield{author}{\bibinfo{person}{Lianghui Zhu}, \bibinfo{person}{Xinggang Wang}, {and} \bibinfo{person}{Xinlong Wang}.} \bibinfo{year}{2023}\natexlab{}.
\newblock \showarticletitle{{Judgelm: Fine-tuned large language models are scalable judges}}.
\newblock \bibinfo{journal}{\emph{{arXiv preprint arXiv:2310.17631}}} (\bibinfo{year}{2023}).
\newblock


\end{thebibliography}

\end{document}